\documentclass{aa} 
\usepackage[varg]{txfonts}

\usepackage[fleqn]{amsmath}
\usepackage{epstopdf}
\usepackage{multirow}
\usepackage{graphicx}
\usepackage{expdlist}
\usepackage{color}
\usepackage{epsfig}
\usepackage{pdflscape}
\usepackage{lscape}
\usepackage{rotating}
\usepackage{arydshln}

\usepackage{float}

\usepackage{pifont}

\usepackage{flafter}

\usepackage[utf8]{inputenc}
\usepackage{natbib,twoopt}
\usepackage[breaklinks=true]{hyperref} 
\bibpunct{(}{)}{;}{a}{}{,} 
\makeatletter
\newcommandtwoopt{\citeads}[3][][]{\href{http://adsabs.harvard.edu/abs/#3}%
{\def\hyper@linkstart##1##2{}%
\let\hyper@linkend\@empty\citealp[#1][#2]{#3}}}
\newcommandtwoopt{\citepads}[3][][]{\href{http://adsabs.harvard.edu/abs/#3}%
{\def\hyper@linkstart##1##2{}%
\let\hyper@linkend\@empty\citep[#1][#2]{#3}}}
\newcommandtwoopt{\citetads}[3][][]{\href{http://adsabs.harvard.edu/abs/#3}%
{\def\hyper@linkstart##1##2{}%
\let\hyper@linkend\@empty\citet[#1][#2]{#3}}}
\newcommandtwoopt{\citeyearads}[3][][]%
{\href{http://adsabs.harvard.edu/abs/#3}
{\def\hyper@linkstart##1##2{}%
\let\hyper@linkend\@empty\citeyear[#1][#2]{#3}}}
\makeatother

\newcommand{\grau}      {$^{\circ}$}

\newcommand{\kms}       {km\,s$^{-1}$}
\newcommand{\mJy}       {mJy\,beam$^{-1}$}
\newcommand{\Jy}        {Jy\,beam$^{-1}$}
\newcommand{\msun}      {M$_{\sun}$\xspace}
\newcommand{\lsun}      {L$_{\sun}$\xspace}

\newcommand{\ie}        {i.\,e.,}
\newcommand{\eg}        {e.\,g.,}

\newcommand{\phs}       {\phantom{$<$}}
\newcommand{\phn}       {\phantom{0}}
\newcommand{\phnn}      {\phantom{00}}

\newcommand{\phns}      {\phantom{0.}}

\newcommand{\nh}        {NH$_3$}
\newcommand{\mum}       {$\mu$m}
\newcommand{\hii}       {\ion{H}{ii}}

\begin{document}

\title{Radio survey of the stellar population in the infrared dark cloud G14.225-0.506}

\author{Elena Díaz-Márquez\inst{1,2,3}
\and
Roger Grau \inst{4}
\and Gemma Busquet \inst{1,2,5}
\and Josep Miquel Girart \inst{3,5}
\and Álvaro Sánchez-Monge\inst{3,5}
\and Aina Palau\inst{6}
\and Matthew S. Povich\inst{7}
\and Nacho Añez-López\inst{8}
\and Hauyu Baobab Liu\inst{9}
\and Qizhou Zhang\inst{10}
\and Robert Estalella\inst{1,2,5}
} 

\institute{Departament de Física Quàntica i Astrofísica (FQA), Universitat de Barcelona (UB), Martí i Franquès 1, E-08028 Barcelona, Catalonia, Spain
\and
Institut de Ciències del Cosmos (ICCUB), Universitat de Barcelona, Martí i
Franquès 1, E-08028 Barcelona, Catalonia, Spain
\and
Institut de Ci\`encies de l'Espai (ICE, CSIC), Can Magrans, s/n, E-08193 Cerdanyola del Vall\`es, Catalonia, Spain
\and 
Institut de Física d'Altes Energies (IFAE), The Barcelona Institute of Science and Technology (BIST), 08193 Bellaterra, Catalonia, Spain
\and
Institut d'Estudis Espacials de Catalunya (IEEC), Gran Capità, 2-4, E-08034, Barcelona, Catalonia, Spain
\and 
Instituto de Radioastronom\'ia y Astrof\'isica, Universidad Nacional Aut\'onoma de M\'exico, Antigua Carretera a P\'atzcuaro 8701, Ex-Hda. San Jos\'e de la Huerta, 58089 Morelia, Michoac\'an, M\'exico
\and 
Department of Physics and Astronomy, California State Polytechnic University, 3801 West Temple Avenue, Pomona, CA 91768, USA
\and 
Université Paris-Saclay, Université Paris Cité, CEA, CNRS, AIM, 91191, Gif-sur-Yvette, France
\and Department of Physics, National Sun Yat-Sen University, No. 70, Lien-Hai Road, Kaohsiung City 80424, Taiwan, R.O.C.
  \and Harvard-Smithsonian Center for Astrophysics, Cambridge, MA 02138, USA
} 

\date{Received 2023 September 27 / Accepted 2023 November 18}

\abstract 
{
The infrared dark cloud (IRDC) G14.225-0.506, is part of the extended and massive molecular cloud located
at the south west of the \hii\ region M17. The cloud is associated with a network of filaments, which result
in two different dense hubs, as well as with several signposts of star formation activity and a rich population
of protostars and YSOs.
} {The aim of this work is to study the centimeter continuum emission in order to characterize the stellar population in both regions, as well as to study the evolutionary sequence across the IRDC G14.225-0.506.
} {We performed deep ($\sim1.5$--3~$\mu$\Jy) radio continuum observations at 6 and 3.6~cm toward the IRDC G14.225-0.506 using the Karl G.\ Jansky Very Large Array (VLA) in its most extended A configuration ($\sim$ 0.3$^{\prime\prime}$). Data at both C and X bands were imaged using the same \textit{(u,v)} range in order to derive spectral indices. We have also made use of observations taken during different days to study the presence of variability at short timescales towards the detected sources.} 
{We detected a total of 66 sources, 32 in the northern region G14.2-N and 34 in the southern region G14.2-S.
Ten of the sources are found to be variable, 3 located in G14.2-N and 7 in G14.2-S. Based on their spectral index, the emission in G14.2-N is mainly dominated by non-thermal sources while G14.2-S contains more thermal emitters. Approximately 75\% of the sources present a counterpart at other wavelengths. When considering the inner 0.4~pc region around the center of each hub, the number of IR sources in G14.2-N is larger than in G14.2-S by a factor of 4. We also studied the relation between the radio luminosity and the bolometric luminosity, finding that the thermal emission of the studied sources is compatible with thermal radio jets. For our sources with X-ray counterparts, the non-thermal emitters follow a Güdel-Benz relation with $\kappa = 0.03$, as previously suggested for other similar regions.
}
{We found similar levels of fragmentation between G14.2-N and G14.2-S, suggesting that both regions are most likely twin hubs. The non-thermal emission found in the less evolved objects, mainly coming from G14.2-N, suggests that G14.2-N may be composed of more massive YSOs as well as being in a more advanced evolutionary stage, consistent with the filament-halo gradient in age and mass from previous works. Overall, our results confirm a wider evolutionary sequence from southwest to northeast starting in G14.2-S as the youngest part, followed by G14.2-N, and ending with the most evolved region M17.} 

\maketitle

\section{Introduction}\label{s:intro}

The formation of intermediate and high-mass stars is a complex process that involves several evolutionary stages, and it is usually found associated with clusters of lower-mass stars \citep[\eg][]{pudritz2002,lada2003}. However, when and in what stage massive stars form relative to their low-mass cluster members remains an open question \citep[\eg][]{vazquez-semadeni2017,motte2018}. Moreover, it is interesting to probe the earliest stages at which ionization might be present to understand the implications of stellar feedback that could soon disrupt the star-forming cores and limit their further growth. The lack of observational data  characterizing these phenomena is due to the distances involved, typically larger than 2~kpc, and the clustered nature of high-mass star-forming regions, which call for high angular resolution and high sensitivity observations.

The radio continuum emission at centimeter wavelengths is found in association with young stellar objects (YSOs) in all stages of the star formation processes (from Class~0 to Class~III). The origin of radio continuum emission can be distinguished through the spectral index $\alpha$, defined as $S_\nu\propto\nu^\alpha$.
A \textit{non-thermal} origin for the radio continuum emission results in a spectral index $\alpha<-0.1$ (in the frequency range 4--12~GHz, of interest for the current work), and it is generally the result of electrons in presence of magnetic fields. In star-forming regions, this type of emission is commonly detected towards YSOs with an active magnetosphere (gyrosynchrotron radiation) corresponding to Class~II/III YSOs \citep[\eg][]{Feigelson1985, Gudel2002, deller2013}. Low-mass Class~0/I objects also have an active magnetosphere and sometimes present synchrotron flares due to magnetic reconnections in the protostellar surface \citep{Liu2014}. Synchrotron emission can also be generated in very strong magnetized shock spots within jet lobes interacting with the ambient medium \citep[\eg][]{carrasco-gonzalez2010, Ainsworth2014, rodriguez-kamenetzky2016, rodriguez-kamenetzk2017, rodriguez-kamenetzky2019, osorio2017}, towards high-mass binary stars producing synchrotron radiation in the region where their winds collide \citep[\eg][]{rodriguez2012}, towards some \hii\ regions \citep[\eg][]{padovani2019, meng2019}, and finally, as contaminating background extragalactic sources.

On the other hand, the radio continuum emission can have a \textit{thermal} origin (\textit{thermal bremsstrahlung}) which is characterized by spectral indices between $-0.1$ and $+2$, and it is interpreted as emission from free-free electron encounters. This emission can arise from shocks in jets powered by low, intermediate and high-mass YSOs \citep[see][for a review]{anglada2018} with typical values for the spectral index $\alpha\sim+0.6$. Thermal emission is also commonly detected in the surroundings of massive stars, whose UV photons can ionize the gas and generate an \hii\ region. Depending on the evolutionary phase and the surrounding environment these \hii\ regions can be small ($<0.1$~pc) and dense ($>10^5$~cm$^{-3}$), referred to as hypercompact and ultracompact \hii\ regions, and be associated with both optically thin ($\alpha=-0.1$) and partially thick ($\alpha\sim+0.6$--+2) emission, or they can be large ($>1$~pc) and more diffuse ($<10^3$~cm$^{-3}$), referred to as classical or giant \hii\ regions, and preferentially associated with optically thin emission \citep[\eg][]{Kurtz1994, sanchez-monge2013, sanchez-monge2013b}. Other regions, such as the Orion Nebula Cluster (ONC), present protoplanetary disks under the influence of external photoevaporation by the cluster’s intense UV field. These formations, known as \textit{proplyds}, consist of a disk surrounded by an ionization front and present strong thermal radio emission \citep{ballering2023}. Finally, Class~0/I objects may exhibit strong winds that result in optically thick thermal emission, particularly in the dense region surrounding the protostar \citep[\eg][]{rodriguez1999}. Consequently, even if the protostar were to emit non-thermal radiation, it would likely be hidden by the optically thick free-free emission from the surrounding material and remain undetectable to an observer \citep[\eg][]{dzib2013, dzib2015}.

Previous radio continuum studies conducted toward massive star-forming regions and infrared dark clouds have been constrained by sensitivity limitations, since the noise level is typically in the order of mJy/beam. This restricts the detection to only the most massive objects \citep[\eg][]{Kurtz1994,sanchez-monge2013,Purcell_2013,depree2014,moscadelli2016,rosero2016,rosero2019,hofner2017,medina2018,Kavak2021,purser2021,irabor2023,dzib2023}, thereby missing a significant fraction of the stellar population.
The new capabilities of the Karl G.\ Jansky Very Large Array (VLA), reaching $\approx\mu$Jy/beam sensitivities, offer a unique opportunity to extend radio continuum studies in nearby molecular clouds to more distant regions, providing insights into the formation of massive stars, their associated clusters and their implications on the surrounding medium. 

Deep radio continuum surveys toward star-forming complexes in the solar neighbourhood, 
such as Ophiuchus, Taurus-Auriga, Serpens, Perseus, R Coronae Australis and Orion have shown the potential to characterize the population of YSOs within the radio frequency range along with their characteristics at other wavelengths \citep[\eg][]{dzib2013,dzib2015, Liu2014, kounkel2014,ortiz-leon2015, pech2016,forbrich2016,coutens2019,vargas-gonzalez2021}. Excluding Orion, the mean flux density at 7.5~GHz of the low-mass YSOs in the Gould's Belt VLA survey ranges from $\sim0.15$--0.8~mJy in Class~0/I protostars to $\sim0.2$--4~mJy in T\,Tauri stars \citep{dzib2015,pech2016}. Detecting such a low-mass stellar population in regions lying more than 10 times further away is challenging, since a 1~mJy source translates into a flux density of 10~$\mu$Jy, requiring, hence, a sensitivity of $\sim2~\mu$Jy\,beam$^{-1}$ to be detectable at a 5$\sigma$ level. The only high-mass star-forming complex that has been studied with such a sensitivity is the Orion Nebula Cluster \citep{forbrich2016,vargas-gonzalez2021}, showing an increase in the number of known compact radio sources compared to previous, more shallow surveys \citep{zapata2004,rivilla2015}. We aim at extending this kind of studies to other massive star-forming complexes.


The infrared dark cloud (IRDC) G14.22$-$0.506 (hereafter G14.2), also called M17\,SWex, is part of the extended ($77\times15$~pc) and massive ($>10^5$~\msun) molecular cloud first reported by \citet{elmegreen1976}, and located at the southwest of the \hii\ region M17, which contains the rich cluster NGC\,6618 with at least 16 O-type stars and over 100 B-type stars \citep{chini1980,hoffmeister2008}.  
Based on parallax and proper motions of 12~GHz CH$_3$OH masers, the distance to the cloud is estimated to be 1.98 $^{+0.14} _{-0.12}$ ~kpc \citep{xu2011,wu2014}. More recently, \citet{Zucker2020} obtained a distance range of 1488--1574~pc by combining stellar photometric data with \textit{Gaia} DR2 parallax measurements.  
For the present work, we use $1.6 ^{+0.3} _{-0.1}$~kpc as the distance to the cloud.

\begin{figure}[!t]
    \centering
    \includegraphics[width=0.5\textwidth]{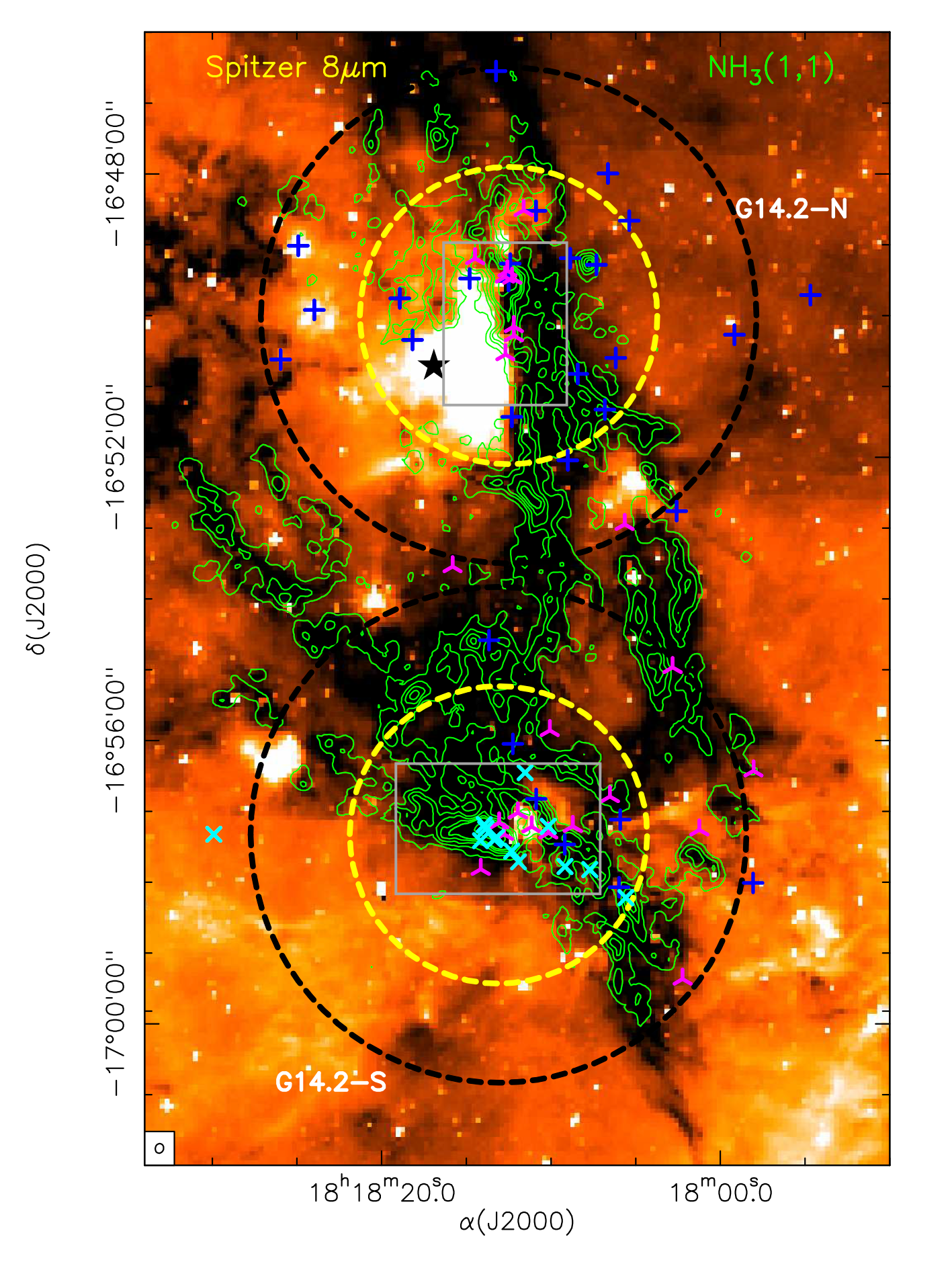}
    \caption{\textit{Spitzer} image at 8~\mum\ (color scale) overlaid on the \nh\,(1,1) integrated intensity (green contours) from \citet{busquet2013}. The contour levels range from 3 to 27 in steps of 6, and from 27 to 67 in steps of 20 times the RMS noise of the map, 9~\mJy\ \kms. The \nh\,(1,1) synthesized beam is shown in the bottom left corner. The black star depicts the position of IRAS\,18153$-$1651. Blue crosses and cyan four-point stars depict radio sources detected in this study only at 6~cm and 3.6~cm, respectively. Pink three-point stars indicate radio sources detected at both frequency bands. The black and yellow dashed circles represent the field of view at 6~cm ($\sim7'$ at 6~GHz) and 3.6~cm ($\sim4.2'$ at 10~GHz), respectively. The two observed fields in this work, G14.2-N and G14.2-S are labeled. 
   The grey rectangles indicate the close-up images presented in Figs. \ref{fig:SMA-hubN} and \ref{fig:SMA-hubS}.
   }
    \label{fig:radiosources}
\end{figure}

High angular resolution observations of the dense gas (\nh\ and N$_2$H$^{+}$) and submillimeter dust continuum emission \citep{lin2017} unveil a network of filaments comprising two hub-filament systems \citep[see Fig.~\ref{fig:radiosources};][]{busquet2013,chen2019}. The cloud is associated with several signposts of star formation activity, such as H$_2$O and CH$_3$OH masers \citep{jaffe1981,palagi1993,wang2006,Green_2010,sugiyama2017}, a rich population of protostars and YSOs detected with \textit{Spitzer}, and a population of intermediate-mass pre-main sequence stars emitting X-rays detected with the \textit{Chandra X-ray Observatory}, some of them lacking infrared excess emission from circumstellar disks \citep{povich2010,povich2016}. The cloud has also been observed with the Atacama Large Millimeter/submillimeter Array (ALMA) at 3~mm \citep{ohashi2016} and with the Submillimeter Array (SMA) at 1.2~mm \citep{busquet2016}. The embedded population consists of 48 dust cores, with masses ranging from 0.7~\msun\ up to 78~\msun.

One of the most prominent results found by \citet{povich2016} is that, despite the mass of the cloud ($>10^5$~\msun) and its high star formation rate ($\geqslant0.007$~\msun~yr$^{-1}$),  
there is a lack of O-type protostars. The brightest IRAS source in the field, IRAS\,18153$-$1651, is associated with an \hii\ region hosting two stars with spectral types B1 and B3 \citep{Gvaramadze2017}.
This absence suggests that either the IRDC G14.2 is only producing up to intermediate-mass stars but does not form massive O-type stars, or the massive clumps are still in the process of accreting enough material to form later the high-mass stars. Interestingly, \citet{povich2016} observed a large-scale `filament-halo' age gradient and mass segregation of the stellar population. The less-obscured population, which corresponds to diskless stars, is distributed across an extended halo of lower-density molecular gas surrounding the IRDC filaments. In contrast, in the more obscured core regions of the filaments ($A_{\mathrm{V}}>50$~mag), the more-obscured objects cluster together, containing all the youngest and most massive YSOs. Thus, the spatial distribution is accompanied by an apparent age spread. Diskless X-ray population is more evolved, less obscured, and less clustered with respect to the filaments in comparison to the YSOs, exhibiting an actual evolutionary effect.
These findings suggest that G14.2 is a complex and dynamic environment with ongoing star formation activity. 
However, infrared and X-ray data suffer from extinction limitations. In order to overcome these limitation, sensitive and high-angular resolution centimeter continuum observations can fill the missing piece of information by helping to identify which dust cores may actually be associated with centimeter continuum emission. This will provide a more representative 
sample of the protostellar population that might not be detected at other wavelengths.

\begin{figure*}[!ht]
    \centering
    \includegraphics[width=0.9\textwidth]{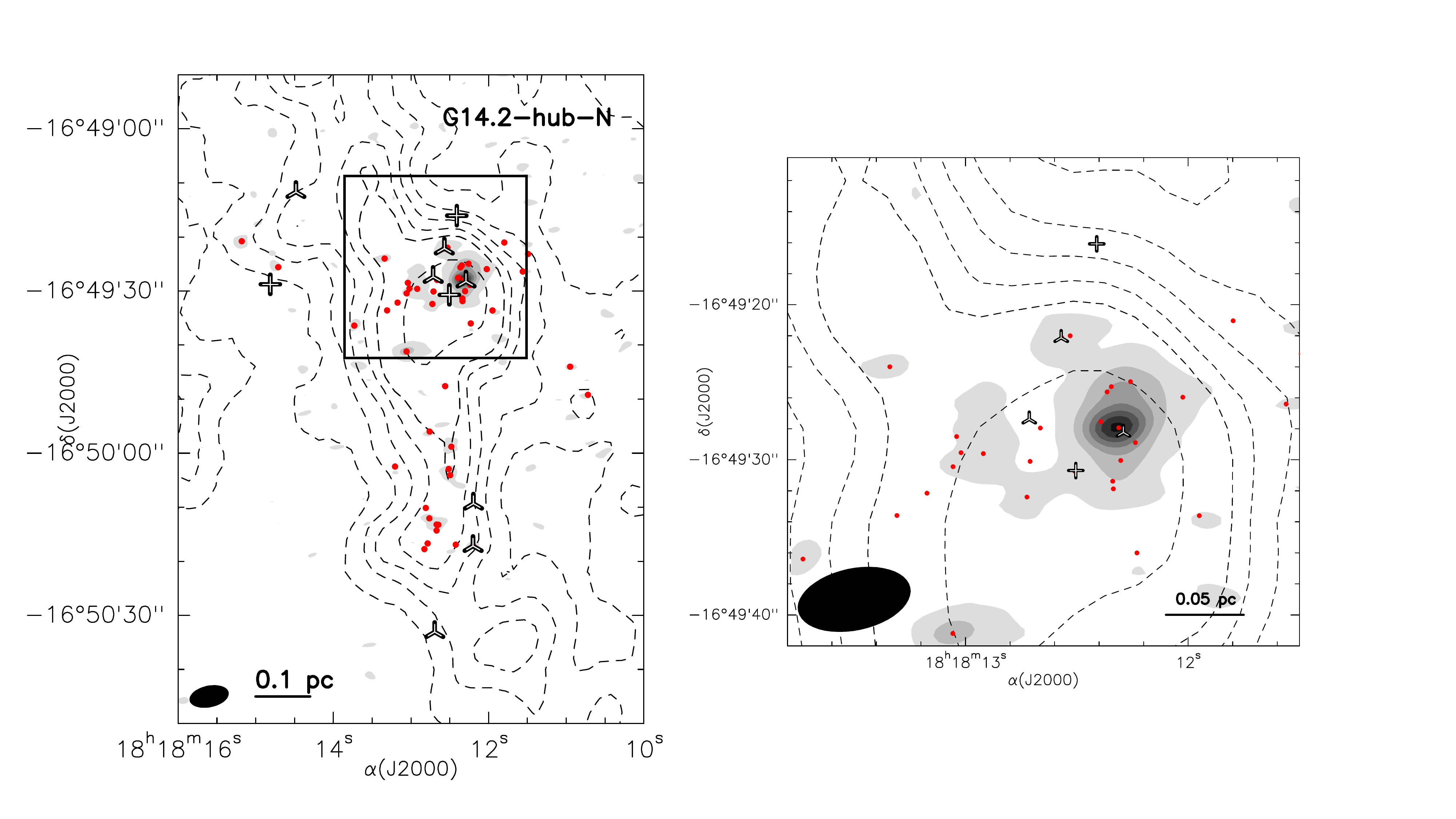}
    \caption{ALMA image (grey) at 3~mm \citep{ohashi2016} of G14.2-hub-N overlaid on the \nh\,(1,1) integrated intensity (black dashed contours) from \citet{busquet2013}. The left panel corresponds to the grey rectangle marked in Figure~\ref{fig:radiosources} while the right panel shows a close -up of the central region around G14.2-hub-N. In both panels contour levels of the grayscale image start at 3$\sigma$ and increase in steps of 15$\sigma$, where $\sigma$ is the rms of the map (0.2~mJy beam$^{-1}$). Red dots depict the dust continuum sources detected with ALMA at 1.3~mm (Zhang et al. private communication). The synthesized beam is shown in the bottom left corner of both images. Symbols are the same as in Figure~\ref{fig:radiosources}.
    }  
    \label{fig:SMA-hubN}  
\end{figure*}

\begin{figure}[!ht]
    \centering
    \includegraphics[width=0.5\textwidth]{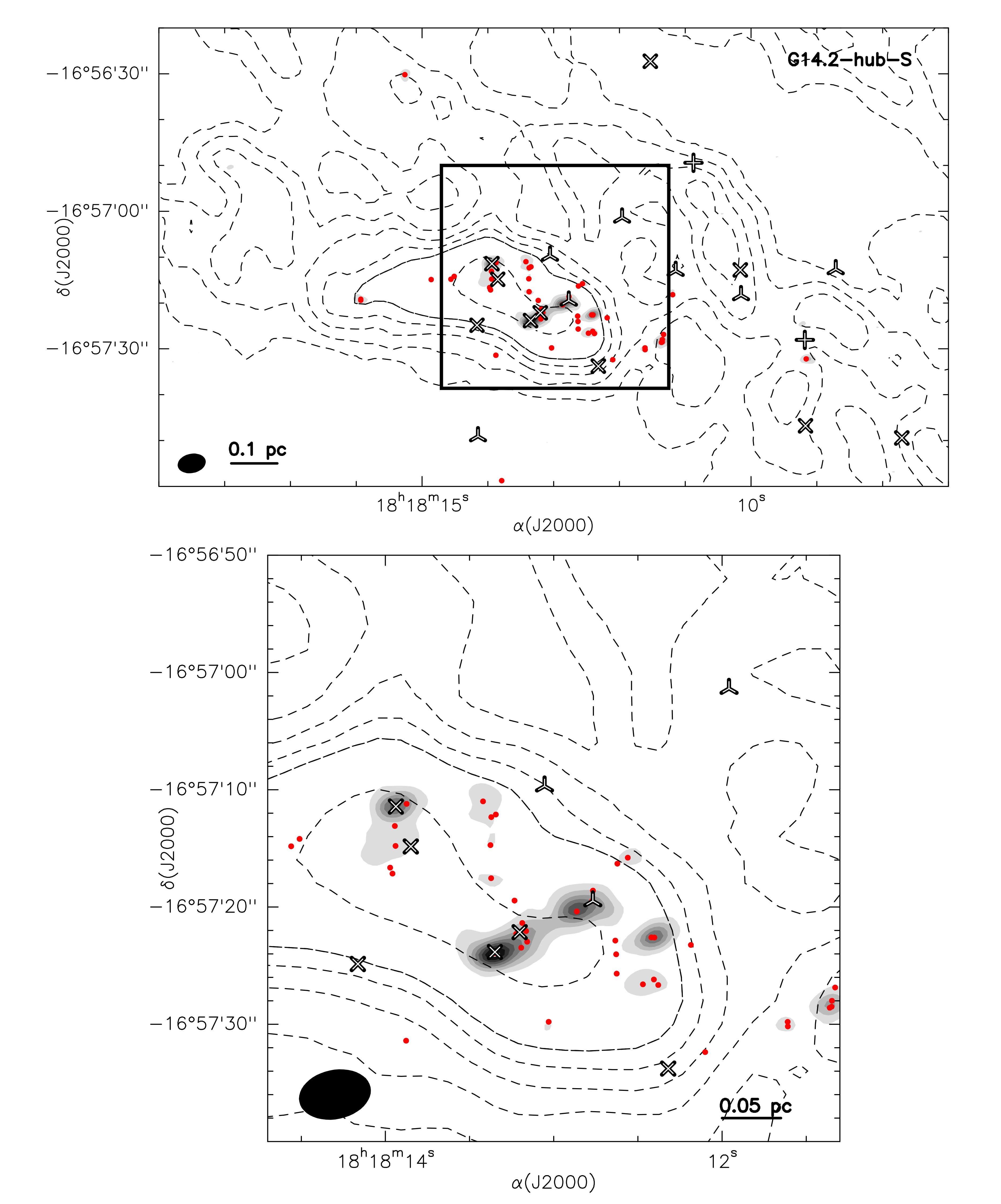}
    \caption{ALMA image (grey) at 3~mm \citep{ohashi2016} of G14.2-hub-S overlaid on the \nh\,(1,1) integrated intensity (black dashed contours) from \citet{busquet2013}. The top panel corresponds to the grey rectangle marked in Figure~\ref{fig:radiosources} while the bottom panel shows a close -up of the central region around G14.2-hub-N. In both panels contour levels of the grayscale image start at 3$\sigma$ and increase in steps of 6$\sigma$, where $\sigma$ is the rms of the map (0.2~mJy beam$^{-1}$). Red dots depict the dust continuum sources detected with ALMA at 1.3~mm (Zhang et al. private communication). The synthesized beam is shown in the bottom left corner of both images. Symbols are the same as in Figure~\ref{fig:radiosources}.
    }  
    \label{fig:SMA-hubS}  
\end{figure}


\begin{figure}[!ht]
    \centering
    \includegraphics[width=0.5\textwidth]{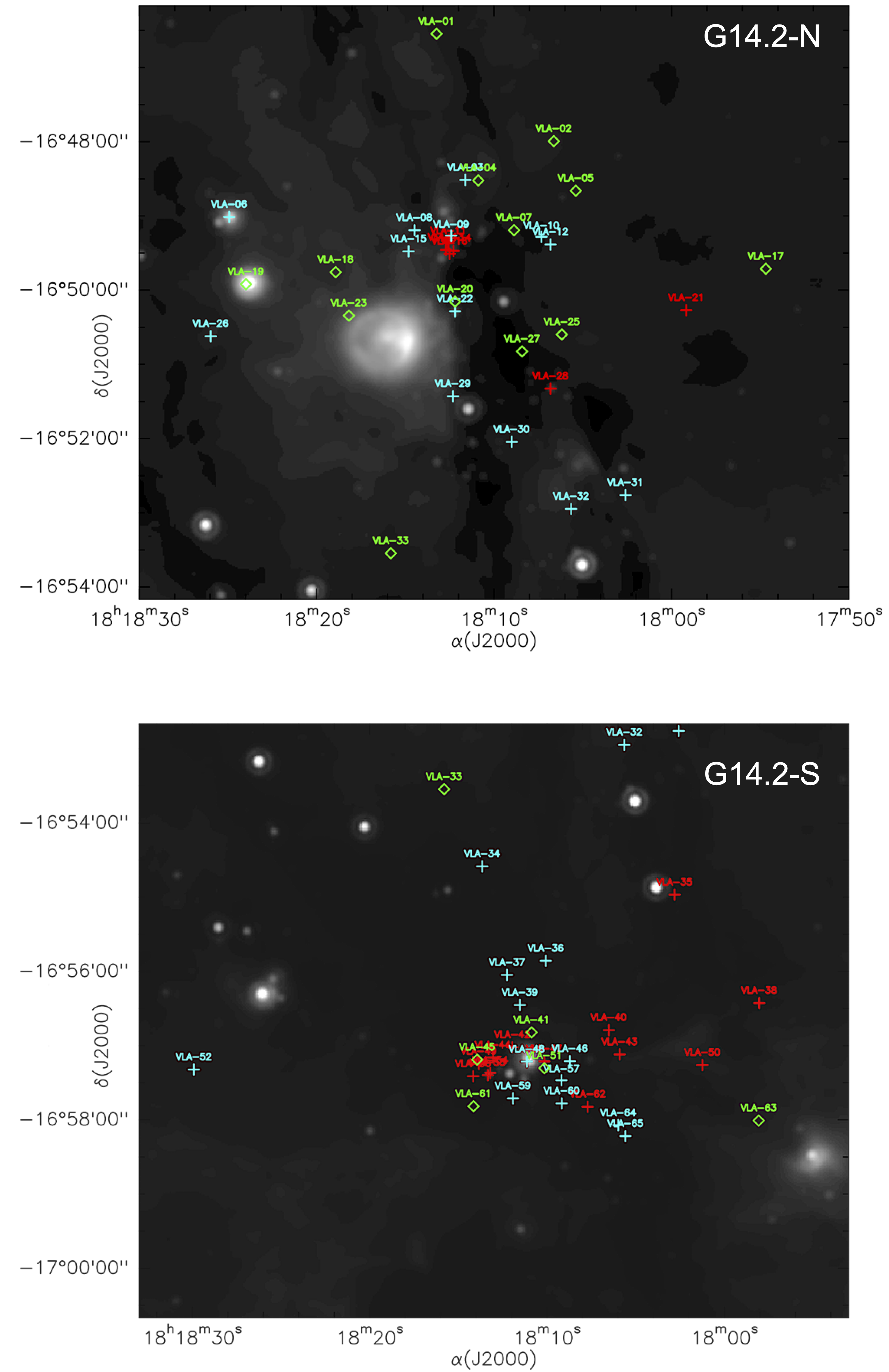}
    \caption{MIPSGAL image \citep{carey2009} at 24~\mum\ overlaid on the centimeter sources detected in this work in G14.2-N (top panel) and G14.2-S (bottom panel). Red crosses depict radio sources with an IR counterpart \citep{povich2016}. Cyan crosses depict radio sources with an X-ray and IR counterpart \citep{povich2016}. Green diamonds depict radio sources with no IR or X-ray counterpart.}
    \label{fig:24m}
\end{figure}


In this work, we present deep, large-scale radio continuum observations obtained with the VLA toward the IRDC G14.2.
The paper is structured as follows. In Sect.~\ref{s:obs} we describe the VLA observations, the data reduction and imaging processes. The results are presented in Sect.~\ref{s:res}. We analyze the radio properties (thermal/non-thermal emission) of the detected sources in Sect.~\ref{s:analysis} and discuss the characteristics of the stellar population in G14.2 
in Sect.~\ref{s:dis}.
Finally, in Sect.~\ref{s:conc}, we present the summary and main conclusions of this work.

\section{Observations}\label{s:obs}

\begin{table*}[!t]
\centering
\caption{\label{t:observations}Parameters of the observations at 6~cm and 3.6~cm with the VLA.}
\begin{scriptsize}
\begin{tabular}{lcccccccccccc}
\hline\hline \noalign{\smallskip}
&$\lambda^{\mathrm{a}}$
&Project
&\multicolumn{2}{c}{Phase center$^{\mathrm{c}}$} 
&Beam
&P.A.
&robust
&rms
&$t_{\mathrm{on-source}}$ 
&Beam$^{\dagger}$
&P.A.$^{\dagger}$
\\
\cline{4-5}
Field 
&(cm)
&code 
&$\alpha$(J2000)
&$\delta$(J2000)
&(arcsec)
&(\grau)
&weighting
&($\mu$Jy\,beam$^{-1}$)
&(hours)
&(arcsec)
&(\grau)
\\
\hline \noalign{\smallskip}
G14.2-N &\phns6  &VLA19A-147 &18 18 12.50  &$-$16 50 00.0  &0.55$\times$0.32 &\phn7.5 &$+1$ &2.2 &1.87 &0.55$\times$0.32 &\phn7.5 \\
        &3.6     &VLA19A-147 &18 18 12.50  &$-$16 50 00.0  &0.42$\times$0.21 &23.9    &$+2$ &1.5 &6.15 &0.41$\times$0.22 &22.4 \\
G14.2-S &\phns6  &VLA19A-147 &18 18 13.10  &$-$16 57 20.0  &0.55$\times$0.33 &\phn6.3 &$+1$ &2.9 &1.99 &0.55$\times$0.33 &\phn6.5 \\
        &3.6$^{\mathrm{b}}$     &VLA17B-236 &18 18 13.10  &$-$16 57 20.0  &0.56$\times$0.42 &87.2    &$+2$ &1.4 &7.07 &0.61$\times$0.45 &81.6 \\
\hline

\end{tabular}
\end{scriptsize}
\tablefoot{
{Columns marked with $^{\dagger}$ next to their names correspond to the parameters of the observations using the common $(u,v)$ range. \em $^{(a)}$}3.6~cm observations were performed in six runs while 6~cm observations were performed in two runs, with the array in the A configuration. {\em $^{(b)}$}
During the first three runs the array was in the BnA configuration while the array was being re-configured to its high-resolution A configuration for the last three runs. 
{\em $^{(c)}$} Units of right ascension ($\alpha$) are hours, minutes, and seconds, and units of declination ($\delta$) are degrees, arcminutes, and arcseconds.}
\end{table*}

The radio continuum observations of G14.2 (project codes 17B-236 and 19A-147) were carried out with the NSF's VLA of the National Radio Astronomy Observatory\footnote{The National Radio Astronomy Observatory is a facility of the National Science Foundation operated under cooperative agreement by Associated Universities, Inc.} at C-band (4--8~GHz; \ie\ $\sim$6~cm) and X-band (8--12~GHz; \ie\ $\sim$3.6~cm). We performed two pointings, labeled G14.2-N and G14.2-S, to encompass most of the cloud complex (see Fig.~\ref{fig:radiosources}). The phase center of each pointing was at $\alpha$(J2000)$=$18$
^\mathrm{h}$18$^{\mathrm{m}}$12.50$^{\mathrm{s}}$ and $\delta$(J2000)$=-16$\grau$50'00.0''$ for G14.2-N and $\alpha$(J2000)$=$18$
^\mathrm{h}$18$^{\mathrm{m}}$13.10$^{\mathrm{s}}$ and $\delta$(J2000)$=-16$\grau$57'20.0''$ for G14.2-S.

The observations were conducted in two different epochs. First, X-band observations toward G14.2-S were performed  
during February 2018 (project 17B-236). 
In the second epoch, we observed G14.2-S in the C-band in two runs 
(2019 September 24 and 27). 
For G14.2-N, we followed the same strategy (\ie\ two runs in C-band during 2019 September 25 and 26, and six runs in X-band during 2019 August 26, 28, and September 3, 6, 9, and 16). The duration of these individuals runs were 1.7~hours for C-band and 1.8~hours for X-band, yielding a total observing time of 3.4~hours and 10.8~hours at C- and X-bands, respectively. A summary of the VLA observational parameters is given in Table~\ref{t:observations}.

Data at both C- and X-bands were taken using two 2~GHz wide basebands (3-bit samplers) and in full polarization mode. The total 4~GHz bandwidth was split into 48 spectral windows, each with a bandwidth of 128~MHz, which were divided into 64 channels with a channel width of 2~MHz. 3C286 was used as the primary flux density and bandpass calibrator, and J1820$-$2528 was observed to calibrate the complex gains. The FWHM of the primary beam  (\ie\ the field of view) of the VLA has a diameter of $7'$ at 6~GHz and $4.2'$ at 10~GHz.

The data were processed using the VLA Calibration Pipeline\footnote{https://science.nrao.edu/facilities/vla/data-processing/pipeline} within the Common Astronomy Software Applications (CASA) environment, specifically the CASA~5.4.2 release.
Once the data were calibrated, they were imaged using the CASA task \textit{tclean} at each frequency band. Each epoch was analyzed separately to look for potential source variability.
During this analysis, two bright and highly variable sources that interfered in the final image were found outside the field of view, one at C-band in G14.2-N and another one at X-band in G14.2-S. The G14.2-N source peak showed approximately 
a factor 6 of variability, and the peak intensity of the G14.2-S source has a variability of more than one order of magnitude. In order to facilitate the cleaning process, these two variable sources were subtracted (see Appendix \ref{appendix:AppA}).  Additionally, there was a discrepancy in the positions and fluxes for all the sources in the X-band, which required the recentering of the data for the different observed days (see Appendix \ref{appendix:AppB}).

For the purpose of creating the final images, once these variables sources were subtracted and the recentering of the X-band was done, the visibilities of all observations were inspected to establish the $(u,v)$ plane coverage. For each region, images were created with the common $(u,v)$ range between the C and the X band (10.8 to 969.7~k$\lambda$ for G14.2-N and 6.8 to 791.1~k$\lambda$ for G14.2-S) in order to ensure that similar spatial-scale structures are recovered and detected in the images at both frequencies. Finally, we performed the imaging including all epochs. All images have been corrected for the primary beam attenuation. In Table~\ref{t:observations} we list the robust parameter used for the imaging, the synthesized beam, the position angle (P.A.) and the rms noise level of the combined image. As we can see, the beams in Table~\ref{t:observations} are slightly different at each frequency. In order to ensure a proper comparison between the images, we also created another set of images for each field, frequency and day of observation with a common beam of $0\farcs6\times0\farcs4$ for G14.2-N and $1\farcs0\times0\farcs9$ for G14.2-S. The position angle for the common beam images was set to zero. These images have been used when a comparison of sources between different frequencies was needed, \eg\ when studying the variability or estimating the spectral indices.

\section{Results}\label{s:res}
\subsection{Source identification}

The 3.6 and 6 cm continuum images towards the IRDC G14.2 reveal a rich population of compact radio continuum sources, whilst no extended emission is detected. This is likely due to the interferometric filtering which resolves out
structures with sizes greater than 3$^{\prime\prime}$.
In order to identify compact sources we used 
the Python Blob Detector and Source Finder package (PyBDSF\footnote{\url{https://pybdsf.readthedocs.io/en/latest/}}),  
which is a tool designed to decompose radio interferometry images into sources. The identification of the radio sources has been done by using the images without the primary beam correction.
By default, the PyBDSF module recognizes a source as those with a peak intensity larger than a certain threshold above the rms of the image ($\sigma$).  This tool allows us to get the positions, integrated flux, peak intensity, sizes from an elliptical fit and position angle of each identified source. Nevertheless, for homogeneity, we  
calculated the fluxes for each source by defining a polygon at the 3$\sigma$ level based on the positions identified as sources by PyBDSF. The same region defined has been used to get the fluxes on the images for the individual days when studying the variability (see Sect.~\ref{s:variability}) and when estimating the spectral index (see Sect.~\ref{s:spec}), using the images with the same (\textit{uv}) range and synthesized beam.

In our study, we adopted two different criteria to consider a firm detection: \textit{(i)} sources with a peak intensity larger than 6$\sigma$, where $\sigma$ is the rms of the image, or \textit{(ii)} sources with a reported counterpart at other wavelenghts and a peak intensity larger than 3$\sigma$. In order to find the counterparts at other wavelengths, we used the list of millimeter sources identified by \cite{busquet2016} and \cite{ohashi2016}, and the catalog of infrared and X-ray sources from \cite{povich2016}. We established a radius of 2$^{\prime\prime}$ ($\sim3200$~au) around every source and considered as counterpart the closest source inside that radius. This search radius is reasonable given that many of the VLA sources appear to be jets that ought to be offset from the driving stars.

\begin{figure*}[!ht]
    \centering
    \includegraphics[width=0.8\linewidth]{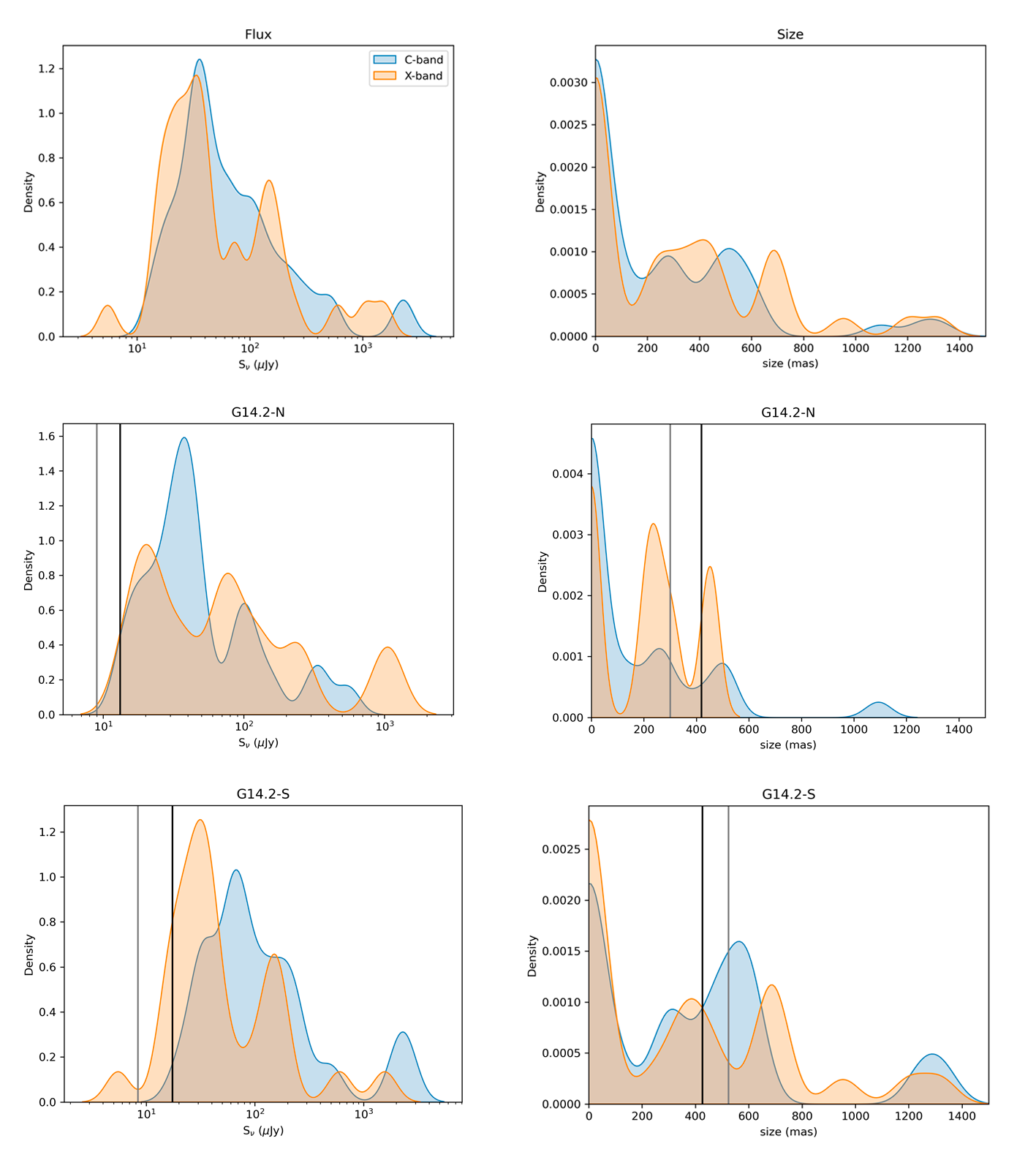}
    \caption{Kernel density estimation (KDE) of integrated fluxes (left) and sizes (right) of the detected sources in G14.2. The middle and bottom panels show the distribution in G14.2-N and G14.2-S, respectively, at C- (blue) and X-band (orange). Vertical lines in the left panels show the 6$\sigma$ value without primary beam correction for the C- (black) and X-band (grey), while vertical lines in the right panels show the FWHM of the synthesized beams for the C- (black) and X-band (grey).}
   \label{fig:panelKDE}
\end{figure*}

By combining these two criteria, a total of 66 sources were detected in the IRDC complex G14.2. Only $\sim10$\% of the sources present a peak intensity between 3 and 6$\sigma$, and only 18 out of the 66 sources do not present a reported counterpart at other wavelengths. A total of 52 sources were detected at 6~cm and 36 at 3.6~cm. Of all these, 22 sources were detected at both bands. Regarding their spatial distribution, 32 sources were located at the G14.2-N field and 34 sources at the G14.2-S. Two sources were detected in both regions because of the overlap in the field of view of the two pointings. Fig.~\ref{fig:radiosources} presents the location of the centimeter continuum sources detected in this work overlaid on dense gas emission traced by the NH$_3$\,(1,1) from \citet{busquet2013}, while a close-up view of the two hubs is presented in Figs.~\ref{fig:SMA-hubN} and \ref{fig:SMA-hubS}, showing also the ALMA 3~mm image from \citet{ohashi2016}.  Fig.~\ref{fig:24m} present the MIPSGAL 24~\mum\ image \citep{carey2009} overlaid on the centimeter sources detected in this work with their IR and/or X-ray counterparts.

\begin{table}[t]
\centering
\caption{Median fluxes and sizes for the radio sources in G14.2.}
\begin{tabular}{cccc}
\hline\hline \noalign{\smallskip}
    Region & Band & Flux ($\mu$Jy) & Size (mas) \\
    \hline \noalign{\smallskip}
    G14.2-N & C-band & 40 & $<$200  \\
            & X-band & 70 & \phs230 \\
    \hline \noalign{\smallskip}
    G14.2-S & C-band & 72 & \phs323 \\
            & X-band & 34 & \phs323 \\
\hline
\end{tabular}
\label{tab:median}
\end{table}

The parameters of the radio sources detected can be found in Appendix~\ref{app_fluxes}, where the primary beam correction has been applied. Appendix~\ref{app:comments} presents some of the sources that have been studied in more depth. The individual images of each source are presented in Appendix~\ref{app_figures}. In Fig.~\ref{fig:panelKDE} we show the distribution of sizes and intensities for the identified radio sources. 
We split the sample into sources detected at different frequency bands (top panels) as well as sources detected in the two fields (middle and bottom panels). As can be seen in the left column of Fig.~\ref{fig:panelKDE}, most of the radio sources detected in the IRDC G14.2 have flux densities between 30 to 70~$\mu$Jy. This kind of sources would have remained undetected in typical previous surveys of star-forming regions, which typically reach sensitivities of 0.1--1~mJy. In G14.2-N, 15 sources (10 detected at C-band and 5 at X-band) are above 50~$\mu$Jy. Only one of them, detected at X-band, has a flux density larger than 1~mJy. Regarding G14.2-S, there are three sources above 1~mJy (two at C-band and one at X-band). We have 25 sources with flux densities larger than 50~$\mu$Jy, mostly detected at C-band. G14.2-S presents a wider range of fluxes, although there are more differences between the values detected at the different frequencies. At the X-band, most of the sources are weaker than for the C-band, with 10 of them having flux densities below 30~$\mu$Jy. The median fluxes per field and band are reported in Table~\ref{tab:median}.

The right column of Fig.~\ref{fig:panelKDE} shows the distribution of source sizes. Most of the radio continuum sources in G14.2 detected in this work are compact ($<200$~mas, corresponding to $\approx300$~au at the G14.2 distance), with 35 sources (19 in G14.2-N and 16 in G14.2-S) remaining unresolved at our current angular resolution. Interestingly, the sources in G14.2-N seem to be slightly more compact than in G14.2-S. There are only 4 sources with sizes above 500~mas in G14.2-N, including a very extended and clumpy source ($\sim1100$~mas, or $\approx1700$~au) in VLA-19 (see Fig.~\ref{fig:vla19}), whereas for G14.2-S there are 13 sources over 500~mas. The median sizes of the sources per field and frequency are listed in Table~\ref{tab:median}.

\subsection{Background sources}\label{s:bg}

\begin{table}[!t]
\centering
\caption{Number of background sources}
\label{t:bkg}
\begin{tabular}{lccc}
\hline\hline \noalign{\smallskip}
Region
& Band
& $<N_{\rm FWHM}\tablefootmark{$a$} >$
& $<N_{\rm 0.4~pc}\tablefootmark{$b$} >$
\\
\hline \noalign{\smallskip}
G14.2-N & C-band & 8 & 0.11  \\
        & X-band & 3 & 0.11  \\
\hline \noalign{\smallskip}
G14.2-S & C-band & 7 & 0.09  \\
        & X-band & 3 & 0.11  \\
\hline 
\end{tabular}
\tablefoot{
\tablefoottext{$a$}{Considering the field of view of the C-band ($\sim7'$) and X-band ($\sim4.2'$).}
\tablefoottext{$b$}{Considering a diameter of 0.4~pc around the center of each hub.}
}
\end{table}

After identifying all the compact radio continuum sources in both fields (\ie\ in G14.2-N and G14.2-S), an estimation of the number of background sources for the VLA can be calculated using the formula from \citet{Anglada_1998}:
\begin{equation}
    \begin{aligned}
        & \langle N \rangle = 1.4\left\{1- \exp \left[-0.0066\left(\frac{\theta_F}{\mathrm{arcmin}}\right)^2\left(\frac{\nu}{5\, \mathrm{\mathrm{GHz}}}\right)^2 \right]\right\} \times \\ 
        & \left(\frac{S_0}{\mathrm{mJy}} \right)^{-0.75}\left(\frac{\nu}{5\,\mathrm{GHz}} \right)^{-2.52}
   \end{aligned}
\end{equation}
where $\theta_F$ is the field of view, $\nu$ is the frequency and $S_0$ is the flux density.

We used this expression and considered two different field of view sizes for each region and a flux density of 6$\sigma$ to estimate the number of background sources in our observations (see Table~\ref{t:bkg}). When considering only a region of 0.4~pc around the center of each hub ($\sim42''$ at a distance of 1.6~kpc), we obtain values below 1 for the number of background sources. Therefore, the probability of detecting an object not associated with the star-forming hub is small and we can assume that all the sources detected within the 0.4~pc inner region of each hub are indeed associated with G14.2-N and G14.2-S.
Although the level of background contamination is low in the inner region of the cluster-hubs, this may be an important factor when considering the whole field of view at both frequency bands, since we have about 6 to 15 sources being potential background sources (see Table~\ref{t:bkg}). Identifying counterparts at other wavelengths (see Sect.~\ref{s:count}) will ensure the membership of the object to the G14.2 complex.

\subsection{Variability}\label{s:variability}

\begin{figure}[!t]
    \centering
    \includegraphics[width=0.8\linewidth]{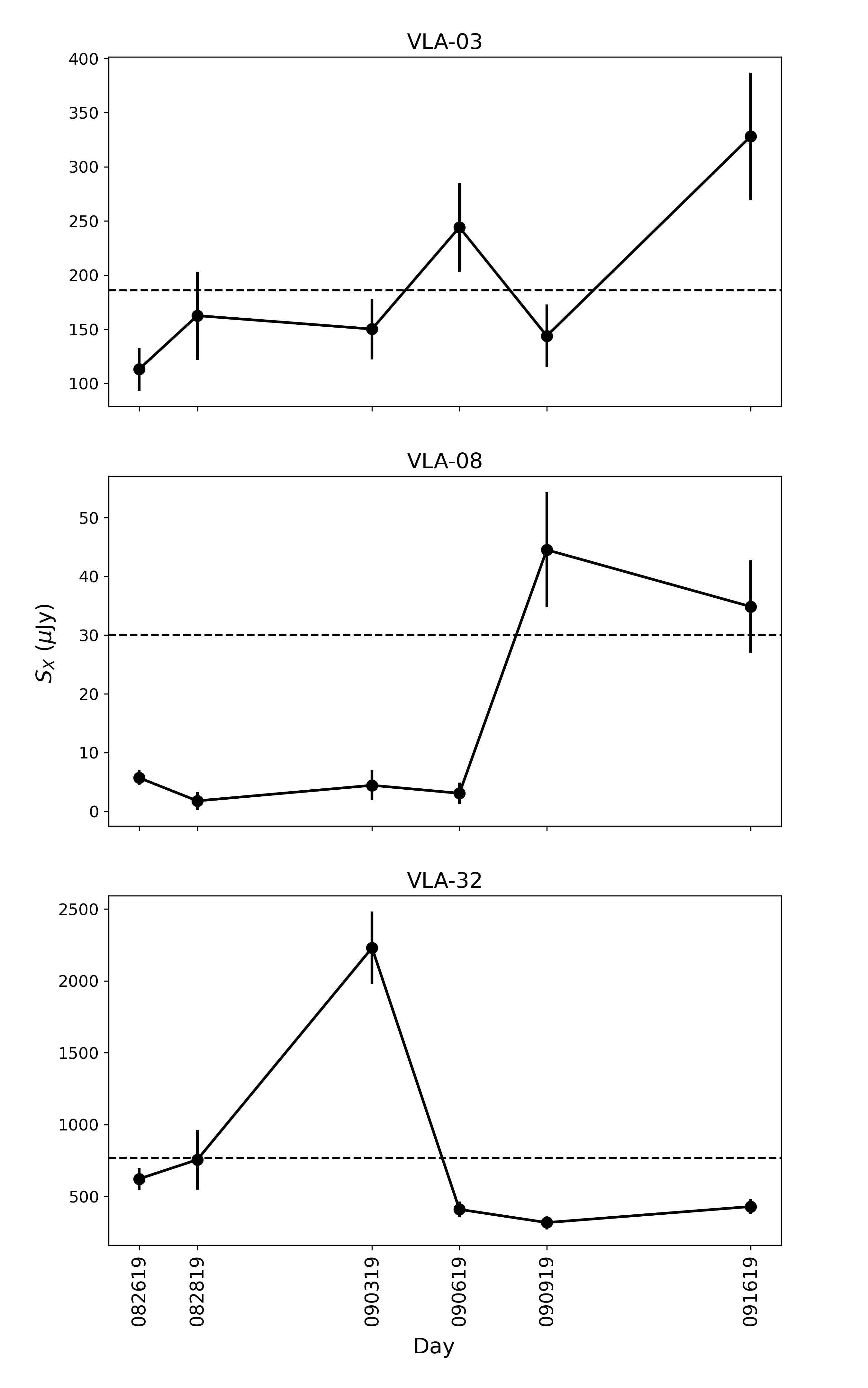}
    \caption{Integrated flux of the variable sources detected in G14.2-N at X-band during the observed days. The sources were observed during 2019 August 26, 28, and September 3, 6, 9 and 16. The black dashed line corresponds to the established cutoff for each source shown in Table~\ref{VarX}.}
    \label{fig:var-g14n-xband}
\end{figure}

\begin{figure}[!t]
    \centering
    \includegraphics[width=0.8\linewidth]{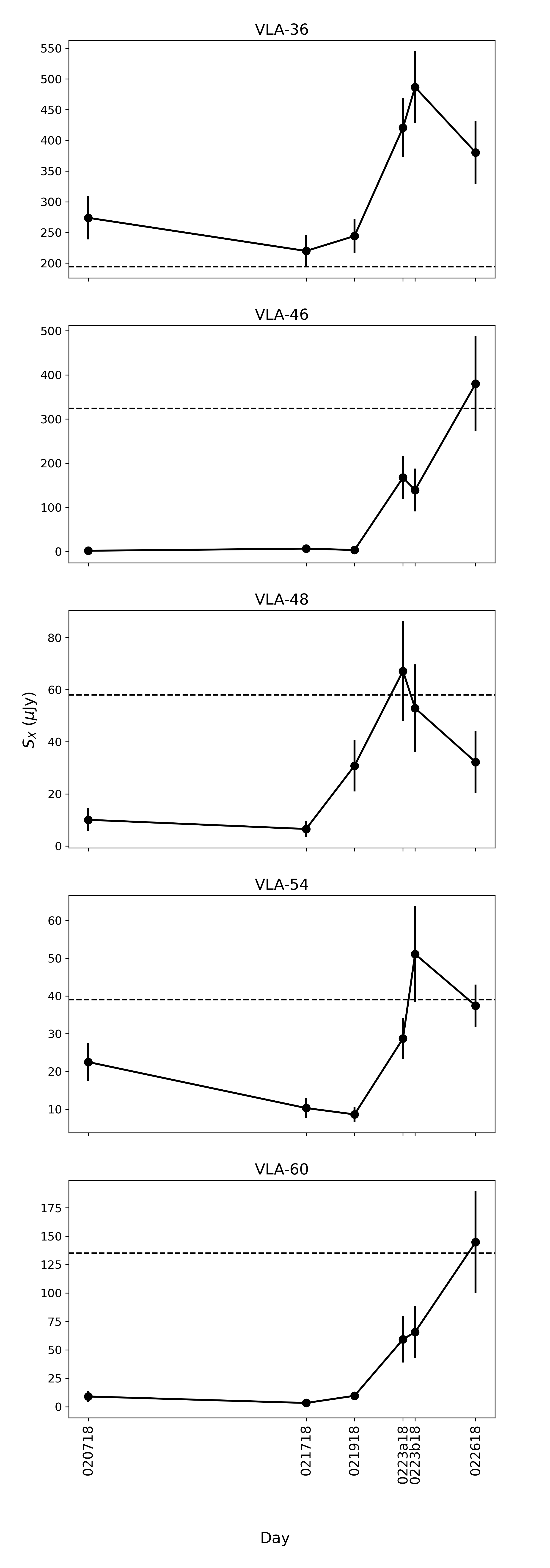}
    \caption{Integrated flux of the variable sources detected in G14.2-S at X-band during the observed days. The sources were observed during 2018 February. The black dashed line corresponds to the established cutoff for each source shown in Table~\ref{VarX}.}
    \label{fig:var-g14s-xband}
\end{figure}

We searched for variability in the radio continuum emission of the detected sources by extracting the flux of each source in each of the different observing days (see Sect.~\ref{s:obs}). We note that the synthesized beam of these images vary slightly from day to day. In order to avoid possible biases, we convolved all the images to a common beam of $0\farcs6\times0\farcs4$ for G14.2-N and $1\farcs0\times0\farcs9$ for G14.2-S, and then evaluated the flux for each source and day. After that, we calculated the difference in flux between the maximum and minimum value, establishing a cutoff at 3$\sigma$ level for variability detection. This cutoff was computed as $3\sqrt{\sigma_\text{max}^2+\sigma_\text{min}^2}$ where $\sigma_\text{max}$ and $\sigma_\text{min}$ are the uncertainties of the maximum and minimum flux, respectively. As indicated in Appendix~\ref{app_fluxes}, this uncertainty takes into account the uncertainty in the rms and the uncertainty in the flux calibration. Sources whose uncertainty in the measurement exceeded this cutoff were considered as variable.

Tables~\ref{VarC} and \ref{VarX} list the variable sources that have been detected at C-band and X-band, respectively. 
Notably, certain sources exhibit variability in one band but not in the other. It is important to note that this disparity does not necessarily indicate exclusive variability in a specific band. It may be due to variations in the observing days and in the duration of each observation, potentially preventing some sources from achieving the established cutoff. 

Figs.~\ref{fig:var-g14n-xband} and ~\ref{fig:var-g14s-xband}  
show in more detail the evolution of the variable sources at X-band over the six days of observation. The sources vary on short timescales, as the difference between consecutive observations ranges from hours to weeks. In G14.2-S, all sources present higher fluxes during the last three days of observation, although their behavior differ from source to source. For the rest of them, we cannot see any specific trend on the variability of the sources. Since at C-band we only have two days of observations, we cannot infer any specific trend on the variability from 
the fluxes reported in Table~\ref{VarC}.

\begin{table*}[!t]
\centering
\caption{Variability of radio sources at C-band.}
\label{VarC}
\begin{tabular}{l c c c c c}
\hline\hline \noalign{\smallskip}
& $S_{\text{C,max}}$ ($\mu$Jy)
& $S_{\text{C,min}}$ ($\mu$Jy)
& $S_{\text{C,diff}}$ ($\mu$Jy)
& $S_{\text{C,cutoff}}$ ($\mu$Jy)
& \% difference
\\
\hline \noalign{\smallskip}
\multicolumn{4}{l}{Southern region G14.2-S} \\
\hline \noalign{\smallskip}
VLA-37 & 47.5 $\pm$ 7.4 & 21.3 $\pm$ 4.7 & 26 $\pm$ 12 & 26 & 55 $\pm$ 12 \\
VLA-41 & 38.6 $\pm$ 5.9 & 17.2 $\pm$ 4.0 & 21 $\pm$ 10 & 21 & 56 $\pm$ 12 \\
\hline
\end{tabular}
\tablefoot{
The columns list: $S_{\text{C,max}}$ and $S_{\text{C,min}}$: maximum and minimum integrated fluxes of the variable sources detected at the C-band; $S_{\text{C,diff}}$: difference between those two fluxes in $\mu$Jy; $S_{\text{C,cutoff}}$: cutoff at which the flux difference is considered as variability; and the relative difference with respect to the maximum value.
}
\end{table*}

\begin{table*}[!h]
\centering
\caption{Variability of radio sources at X-band.}
\label{VarX}
\begin{tabular}{l c c c c c}
\hline\hline \noalign{\smallskip}
& $S_{\text{X,max}}$ ($\mu$Jy)
& $S_{\text{X,min}}$ ($\mu$Jy)
& $S_{\text{X,diff}}$ ($\mu$Jy)
& $S_{\text{X,cutoff}}$ ($\mu$Jy)
& \% difference
\\
\hline \noalign{\smallskip}
\multicolumn{4}{l}{Northern region G14.2-N} \\
\hline \noalign{\smallskip}
VLA-03 & \phn328.1 $\pm$ \phn58.7 & 113.1 $\pm$ 19.7       & \phn215 $\pm$ \phn78 & 186 & 66 $\pm$ 9 \\
VLA-08 & \phnn44.5 $\pm$ \phnn9.8 & \phnn1.8 $\pm$ \phn1.5 & \phnn43 $\pm$ \phn11 & 30  & 96 $\pm$ 4 \\
VLA-32 & 2229.9 $\pm$ 252.1       & 318.4 $\pm$ 47.2       & 1912 $\pm$ 299       & 770 & 86 $\pm$ 3 \\
\hline \noalign{\smallskip}
\multicolumn{4}{l}{Southern region G14.2-S} \\
\hline \noalign{\smallskip}
VLA-36 & 486.6 $\pm$ \phn58.6     & 244.1 $\pm$ 27.5       & 242 $\pm$ \phn86     & 194 & 50 $\pm$ 8 \\
VLA-46 & 379.9 $ \pm$ 107.9       & \phnn1.6 $\pm$ \phn1.6 & 378 $\pm$ 110        & 324 & 99 $\pm$ 1 \\
VLA-48 & \phn67.2 $\pm$ \phn19.1  & \phnn6.5 $\pm$ \phn3.1 & \phn61 $\pm$ \phn22  & 58  & 90 $\pm$ 5 \\
VLA-54 & \phn51.1 $\pm$ \phn12.7  & \phnn8.6 $\pm$ \phn2.0 & \phn42 $\pm$ \phn15  & 39  & 83 $\pm$ 6 \\
VLA-60 & 144.7 $\pm$ \phn44.9     & \phnn3.3 $\pm$ \phn3.5 & 141 $\pm$ \phn48     & 135 & 98 $\pm$ 3 \\
\hline
\end{tabular}
\tablefoot{
The columns list: $S_{\text{X,max}}$ and $S_{\text{X,min}}$: maximum and minimum integrated fluxes of the variable sources detected at the C-band; $S_{\text{X,diff}}$: difference between those two fluxes in $\mu$Jy; $S_{\text{X,cutoff}}$: cutoff at which the flux difference is considered as variability; and the relative difference with respect to the maximum value.
}
\end{table*}

As discussed in Sect.~\ref{s:obs}, two sources in the outer parts of the observing fields were found to be very bright and highly variable, and were subtracted to produce cleaner images. These sources are not shown in Tables~\ref{VarC} and \ref{VarX}, and their details can be found in Appendix~\ref{appendix:AppA}.

\subsection{Spectral indices}\label{s:spec}

In order to determine the origin of the radio continuum emission, we have calculated the spectral indices for the 66 sources detected in G14.2. To do this, we have used the set of images created with the common \textit{(u,v)} range and convolved to the same beam (see Sect.~\ref{s:obs}). Since the field of view between the C- and X-band differs, the calculation is limited to those sources within the common field of view. For example, source VLA-17 is located outside the field of view at X-band, which prevents us from deriving a reliable spectral index. For the sources that have been detected in only one of the bands, we assumed a 6$\sigma$ upper limit for the flux density at the non-detected band since the in-band spectral indices present very large uncertainties. Tables~\ref{t:IndexN} and \ref{t:IndexS} report the spectral indices for sources in G14.2-N and G14.2-S, respectively. After that, and taking into account the uncertainties, we classify the sources as thermal (if $\alpha > -0.1$) or non-thermal (if $\alpha < -0.1$) radio emitters based on their spectral index (see Sect.~\ref{s:analysis}).

\begin{table}[!ht]
\caption{Spectral indices of sources detected in G14.2-N.}
\label{t:IndexN}
\begin{small}
\begin{tabular}{l c c c}
\hline\hline \noalign{\smallskip}
       & $S_{\mathrm{C-band}}$  &$S_{\mathrm{X-band}}$    & Spectral \\
Source & ($\mu$Jy)              & ($\mu$Jy)               & index ($\alpha$) \\
\hline \noalign{\smallskip}

VLA-01     & \phn162.0 $\pm$ 16.4\phn & < 79.0             & < $-$1.4 \\
VLA-02     & \phnn21.9 $\pm$ 2.8\phnn & < 44.1             & < $+$1.4 \\
VLA-03$^*$ & \phn305.2 $\pm$ 32.0\phn & 243.3 $\pm$ 25.0   & $-$0.44 $\pm$ 0.29 \\
VLA-04     & \phnn17.3 $\pm$ 1.9\phnn & < 53.7             & < $+$2.2 \\
VLA-05     & \phnn19.3 $\pm$ 2.9\phnn & < 10.4             & < $-$1.2 \\
VLA-06     & \phnn31.7 $\pm$ 3.6\phnn & < 62.4             & < $+$1.3 \\
VLA-07     & \phnn31.1 $\pm$ 3.6\phnn & < 18.9             & < $-$1 \\
VLA-08$^*$ & \phnn42.1 $\pm$ 4.6\phnn & 79.3 $\pm$ 8.1     & $+$1.24 $\pm$ 0.29 \\
VLA-09     & \phnn16.2 $\pm$ 1.7\phnn & < 22.6             & < $+$0.7 \\
VLA-10     & \phnn45.3 $\pm$ 6.6\phnn & < 41.7             & < $-$0.2 \\
VLA-11     & \phn125.5 $\pm$ 12.7\phn & 129.6 $\pm$ 13.2   & $+$0.06 $\pm$ 0.28 \\
VLA-12     & \phnn57.4 $\pm$ 6.8\phnn & < 19.0             & < $-$2.2 \\
VLA-13     & \phnn98.0 $\pm$ 15.7\phn & 70.0 $\pm$ 9.7     & $-$0.66 $\pm$ 0.42 \\
VLA-14     & \phnn40.4 $\pm$ 14.4\phn & 22.9 $\pm$ 7.8     & $-$1.11 $\pm$ 0.97 \\
VLA-15     & \phnn38.7 $\pm$ 2.5\phnn & < 23.9             & < $-$0.9 \\
VLA-16     & \phnn14.6 $\pm$ 1.6\phnn & < 17.1             & < $+$0.3 \\
VLA-18     & \phnn29.3 $\pm$ 3.0\phnn & < 34.0             & < $+$0.3 \\
VLA-19     & \phn107.5 $\pm$ 27.8\phn & < 54.4             & < $-$1.3 \\
VLA-20     & \phnn21.7 $\pm$ 2.4\phnn & 15.2 $\pm$ 1.6     & $-$0.7 $\pm$ 0.30 \\
VLA-21     & \phnn30.4 $\pm$ 3.1\phnn & < 45.0             & < $+$0.8 \\
VLA-22     & \phnn40.4 $\pm$ 4.0\phnn & 37.2 $\pm$ 3.7     & $-$0.16 $\pm$ 0.28 \\
VLA-23     & \phnn12.9 $\pm$ 1.5\phnn & < 30.7             & < $+$1.7 \\
VLA-24     & \phnn38.1 $\pm$ 4.1\phnn & 20.3 $\pm$ 2.2     & $-$1.23 $\pm$ 0.30 \\
VLA-25     & \phnn30.6 $\pm$ 3.3\phnn & < 42.6             & < $+$0.7 \\
VLA-26     & \phnn86.4 $\pm$ 8.9\phnn & < 75.4             & < $-$0.3 \\
VLA-27     & \phnn37.2 $\pm$ 4.0\phnn & < 47.8             & < $+$0.5 \\
VLA-28     & \phnn48.8 $\pm$ 7.1\phnn & < 24.1             & < $-$1.4 \\
VLA-29     & \phnn24.7 $\pm$ 2.5\phnn & < 32.0             & < $+$0.5 \\
VLA-30     & \phnn42.7 $\pm$ 5.0\phnn & < 51.8             & < $+$0.4 \\
VLA-31     & \phnn96.0 $\pm$ 10.3\phn & < 92.3             & < $-$0.1 \\
VLA-32$^*$ & 2192.3 $\pm$ 225.7       & 1050.0 $\pm$ 195.0 & $-$1.44 $\pm$ 0.42 \\
VLA-33     & \phn555.4 $\pm$ 15.2\phn & 600.2 $\pm$ 60.6   & $+$0.15 $\pm$ 0.20 \\
\hline
\end{tabular}
\tablefoot{
VLA-03, VLA-08 and VLA-32 are variable sources and therefore their spectral index may be inaccurate.
}
\end{small}
\end{table}

\begin{table}[ht]
\caption{Spectral indices of sources detected in G14.2-S.}
\label{t:IndexS}
\begin{small}
\begin{tabular}{l c c c}
\hline\hline \noalign{\smallskip}
       & $S_{\mathrm{C-band}}$  &$S_{\mathrm{X-band}}$    & Spectral \\
Source & ($\mu$Jy)              & ($\mu$Jy)               & index ($\alpha$) \\
\hline \noalign{\smallskip}
VLA-32$^*$ & 2192.3 $\pm$ 225.7 & 1050.0 $\pm$ 195.0      & $-$1.44 $\pm$ 0.42 \\
VLA-33     & 555.4 $\pm$ 15.2   & 600.2 $\pm$ 60.6        & $+$0.15 $\pm$ 0.2 \\
VLA-34     & 33.1 $\pm$ 4.0     & < 59.4                  & < $+$1.1 \\
VLA-35     & 114.2 $\pm$ 14.1   & 161.0 $\pm$ 16.6        & $+$0.67 $\pm$ 0.32 \\
VLA-36$^*$ & 203.4 $\pm$ 24.0   & 172.5 $\pm$ 20.9        & $-$0.32 $\pm$ 0.33 \\
VLA-37$^*$ & 134.7 $\pm$ 14.1   & < 11.3                  & < $-$4.9 \\
VLA-38     & 2384.8 $\pm$ 243.7 & 1556.5 $\pm$ 160.8      & $-$0.83 $\pm$ 0.28 \\
VLA-39     & < 30.0             & 17.3 $\pm$ 2.1          & > $-$1.1 \\
VLA-40     & 65.7 $\pm$ 8.2     & 68.5 $\pm$ 8.3          & $+$0.08 $\pm$ 0.34 \\
VLA-41$^*$ & 59.2 $\pm$ 6.5     & < 61.9                  & < $+$0.1 \\
VLA-42     & 70.1 $\pm$ 7.1     & 35.3 $\pm$ 3.6          & $-$1.35 $\pm$ 0.28 \\
VLA-43     & 34.7 $\pm$ 3.7     & < 45.2                  & < $+$0.5 \\
VLA-44     & 59.1 $\pm$ 6.4     & 26.4 $\pm$ 2.7          & $-$1.58 $\pm$ 0.29 \\
VLA-45     & < 16.9             & 17.4 $\pm$ 2.2          & > $+$0.1 \\
VLA-46$^*$ & 98.1 $\pm$ 9.9     & 137.7 $\pm$ 15.5        & $+$0.66 $\pm$ 0.3 \\
VLA-47     & < 15.9             & 21.2 $\pm$ 2.2          & > $+$0.6 \\
VLA-48$^*$ & < 14.7             & 45.1 $\pm$ 4.6          & > $+$2.2  \\
VLA-49     & < 30.7             & 33.9 $\pm$ 3.4          & > $+$0.2 \\
VLA-50     & 215.7 $\pm$ 21.6   & 135.2 $\pm$ 13.6        & $-$0.91 $\pm$ 0.28 \\
VLA-51     & 157.4 $\pm$ 16.2   & 105.2 $\pm$ 10.6        & $-$0.79 $\pm$ 0.28 \\
VLA-52     & < 17.3             & 14.1 $\pm$ 1.4          & > $-$0.4 \\
VLA-53     & 44.5 $\pm$ 4.5     & 17.7 $\pm$ 1.8          & $-$1.81 $\pm$ 0.28 \\
VLA-54$^*$ & < 23.8             & 32.3 $\pm$ 3.7          & > $+$0.6 \\
VLA-55     & < 39.2             & 43.0 $\pm$ 4.3          & > $+$0.2 \\
VLA-56     & < 8.2              & 16.7 $\pm$ 1.7          & > $+$1.4 \\
VLA-57     & 20.5 $\pm$ 2.2     & < 22.9                  & < $+$0.2 \\
VLA-58     & < 8.6              & 25.7 $\pm$ 2.6          & > $+$1.4 \\
VLA-59     & < 35.7             & 42.3 $\pm$ 4.3          & > $+$0.3 \\
VLA-60$^*$ & < 52.2             & 54.4 $\pm$ 5.5          & > $+$0.1 \\
VLA-61     & 31.1 $\pm$ 3.2     & 22.8 $\pm$ 2.3          & $-$0.61 $\pm$ 0.28 \\
VLA-62     & < 21.8             & 34.1 $\pm$ 3.4          & > $+$0.9 \\
VLA-63     & 249.3 $\pm$ 25.0   & < 645.3                 & < $+$1.9 \\
VLA-64     & 72.3 $\pm$ 7.5     & < 33.9                  & < $-$1.5 \\
VLA-65     & < 31.8             & 33.7 $\pm$ 3.4          & > $+$0.1 \\
VLA-66     & 66.4 $\pm$ 6.7     & 176.4 $\pm$ 18.0        & $+$1.91 $\pm$ 0.28 \\
\hline
\end{tabular}
\tablefoot{
VLA-32, VLA-36, VLA-37, VLA-41, VLA-46, VLA-48, VLA-54 and VLA-60 are variable sources and therefore their spectral index may be inaccurate.
}
\end{small}
\end{table}

It is worth noting that there are some factors that may affect the accuracy of the spectral index estimation. First, the C- and X-band observations were not carried out simultaneously, so any variation in the emission could have led to an inaccurate spectral index. We also have to take into account that the fluxes have been calculated selecting the same region, and the differences in the spatial emission of the two bands may have introduced errors in the calculation. It should be remarked that the absence of detection in one of the frequency bands does not necessarily imply that the continuum radio source has a featureless spectrum. This can be due to the source being faint at that frequency, or its signal may be masked by background noise. This highlights the importance of having observations at multiple frequencies to infer the origin of the emission of the sources.

As listed in Table~\ref{t:IndexN}, we found 14 centimeter sources  (corresponding to $\approx$44\% of the radio sources) in G14.2-N with spectral indices clearly smaller than $-0.1$. There is one source (VLA-33) which has a positive spectral index. For some sources, the spectral index is very close to the $-0.1$ limit but we cannot classify them unambiguously, due to the uncertainty in the measurements. Having this in mind, we consider that sources with an spectral index between $-0.3$ and $+0.1$ are expected to show a nearly flat spectrum, indicating emission that remains relatively constant or that slightly varies with frequency. This behavior is also commonly associated with thermal emission. Accordingly, in G14.2-N there are two sources (VLA-11 and VLA-22), corresponding to $\approx$6\% of the radio sources, with a flat spectrum that can be considered as thermal candidates. On the other hand, and as listed in Table~\ref{t:IndexS}, in G14.2-S, we found 8 sources ($\approx$24\%) with an spectral index smaller than $-0.1$. There are 12
sources ($\approx$35\%) that have a spectral index larger than $-0.1$. There is one source (VLA-40) with a nearly flat spectrum and therefore considered as a thermal candidate. The rest of the sources are either variable, and therefore have been excluded because the spectral index is considered unreliable, or the origin of the emission could not be determined due to the uncertainty or derived limits. 

\begin{figure}
    \centering
    \includegraphics[width=0.9\linewidth]{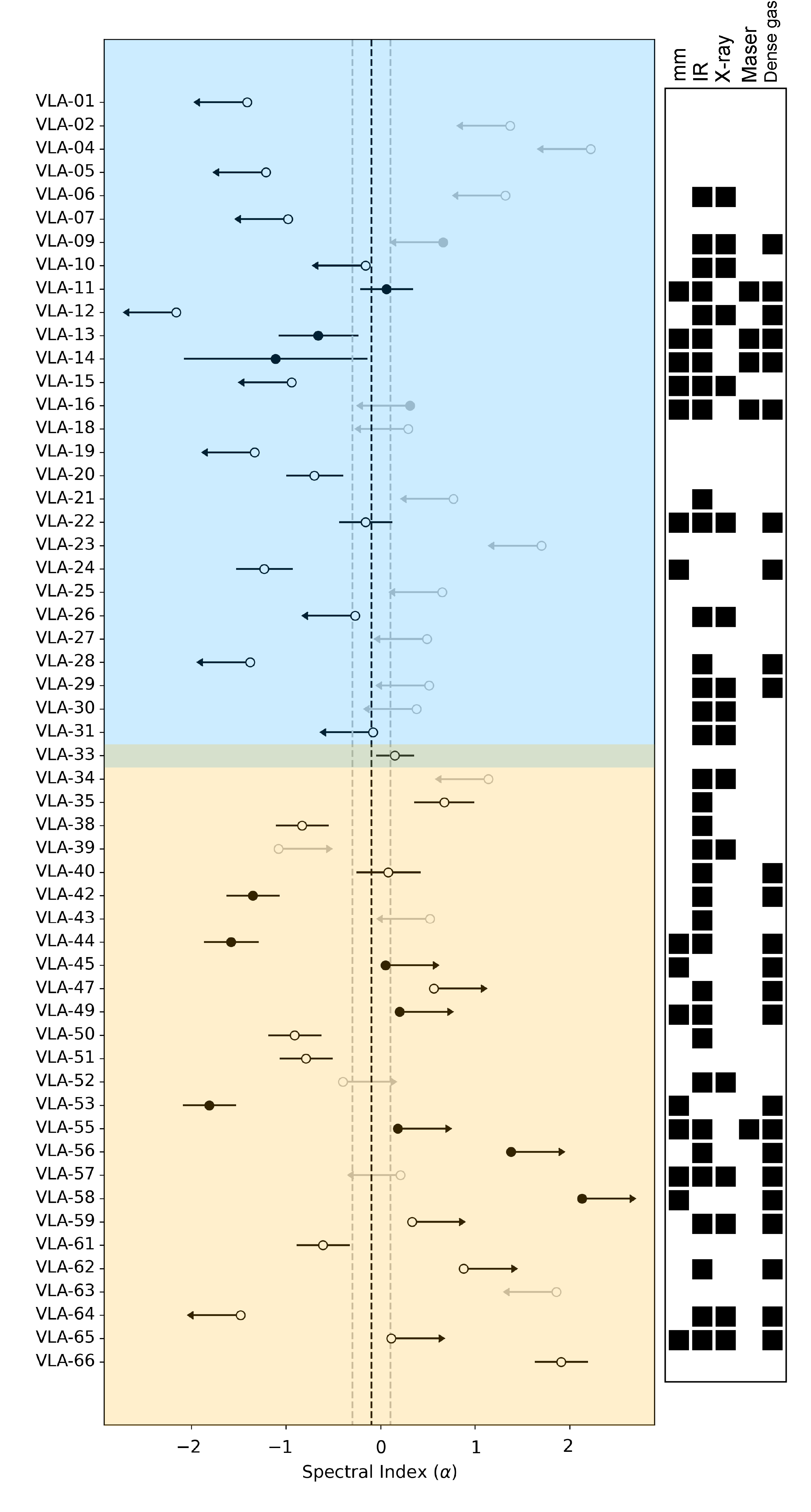}
    \caption{Spectral index of the sources detected in G14.2. Filled and open symbols denote sources located within the 0.4~pc inner region (\ie\ hubs) or outside the 0.4~pc inner region, respectively. Black symbols represent those sources for which it has been possible to determine the origin of the radio continuum emission. Grey symbols represent those sources for which it has not been possible to infer its nature. Arrows denote upper or lower limits for those sources only detected in one frequency band. Variable sources have been excluded for this representation. The black dashed line at $-$0.1 draws the boundary between thermal emission ($\alpha > -0.1$) and non-thermal emission ($\alpha < -0.1$). The grey dashed lines trace the limits where sources show a nearly flat spectrum, probably associated with thermal emission. The blue- and orange-shaded regions represent G14.2-N and G14.2-S, respectively. The right black panels indicate if the sources present a reported counterpart at mm, IR and/or X-ray, associated maser emission and/or dense gas emission.}
    \label{fig:Index}
\end{figure}

\begin{figure}[!ht]
      \centering
      \includegraphics[width=1\linewidth]{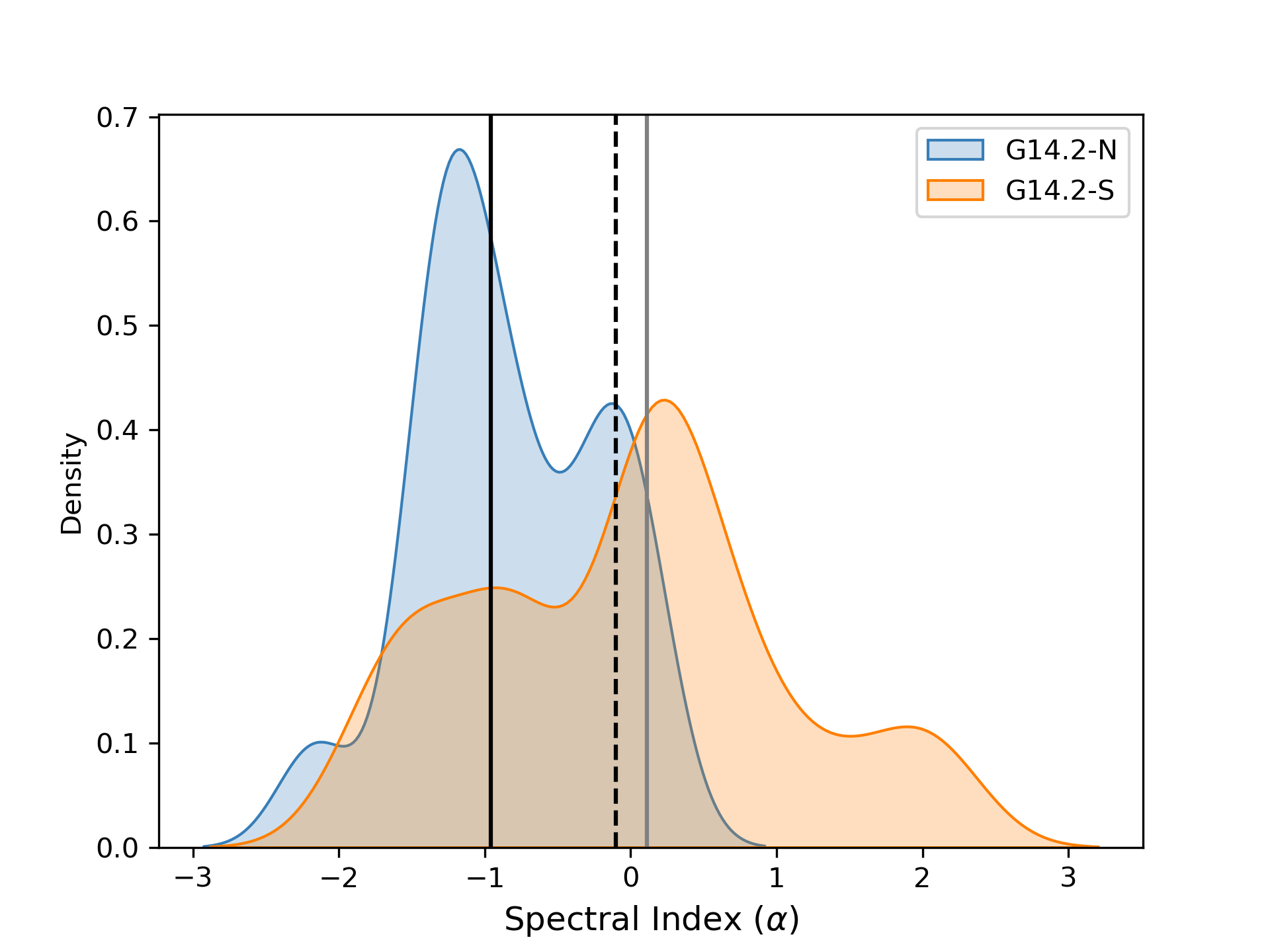}
      \caption{Kernel density estimation (KDE) showing the probability distribution of the spectral index for the sources detected in G14.2 for which it has been possible to determine the origin of the radio continuum emission. The black dashed line at $\alpha=-0.1$ draws the boundary between thermal emission and non-thermal emission. Solid lines show the median value of the spectral index for G14.2-N (black) and G14.2-S (grey).}
      \label{fig:Index-KDE}
\end{figure}

Fig.~\ref{fig:Index} shows the spectral indices and limits obtained for the sources in G14.2. In G14.2-N there are more sources, compared to G14.2-S, whose uncertainty has not allowed us to classify them as thermal or non-thermal emitters. Interestingly, for those sources for which we can unambiguously determine the origin of the radio emission, we find a vast majority of non-thermal objects in G14.2-N ($\approx$70\%) compared to G14.2-S ($\approx$40\%).
Fig.~\ref{fig:Index-KDE} displays the probability density of the spectral index of the radio sources for which it has been possible to determine the origin of the radio emission, corresponding to the black dots shown in Fig.~\ref{fig:Index}. As we can see, G14.2-N is dominated by non-thermal sources. In G14.2-S, we see a wider range of spectral indices, although with a tendency towards positive values. We note that in this figure, the values of the upper and lower limits have been taken as true values. Since in G14.2-N we have mainly upper limits, while in G14.2-S we have more lower limits, the difference between non-thermal and thermal populations in the two regions would be more pronounced if accurate spectral indices, instead of limits, could be derived for all objects.

\subsection{Counterparts at other wavelengths}\label{s:count}

The study of the counterparts at other wavelengths can give us more information about the evolutionary stage and properties of the radio continuum sources detected in G14.2. We have searched for counterparts at millimeter, infrared and X-ray, as well as presence of dense gas \citep{busquet2013} and maser emission \citep{palagi1993,wang2006,Green_2010,sugiyama2017}. For this, we have used the millimeter source catalogues from \cite{busquet2016} and \cite{ohashi2016}, which have been completed by new high-resolution data at 1.3~mm from ALMA (Q.\ Zhang, priv.\ communication); as well as the catalogue of infrared and X-ray sources from \cite{povich2016}. We established a radius of 2$^{\prime\prime}$ around every source to consider sources at different wavelengths to be counterparts. It should be taken into account that we have different fields of view for the observations at different wavelengths. We note that while the IR and X-ray observations cover the whole area of G14.2 ($\sim17^{\prime}$), the mm observations focus only on smaller regions around the central hubs. The radio observations presented in this work cover two large pointings ($\sim7'$ and $4.2'$ for C and X-band, respectively). Therefore, there may be additional counterparts with mm sources that cannot be identified with the current catalogues. Table~\ref{tab:Counter} lists all the identified counterparts for the radio sources detected in this work. The stage-system classification taken from \cite{povich2016} and introduced by \cite{robitaille2006} is based on the physical parameters of the spectral energy distribution (SED) models, with Stage~0/I objects modeled as an SED with an infalling envelope and Stage~II objects modeled as an SED with only circumstellar disks. 

In \citet{povich2016}, some of the objects were classified as \textit{diskless}, referring to IR point sources detected in X-rays but with no infrared excess emission above a normally-reddened stellar photosphere. Most of these are intermediate-mass pre-main-sequence stars with strong magneto-coronal X-ray emission but lacking inner dust disks. Therefore, YSOs established by \citet{povich2016} as diskless are most likely sources in the process of clearing up the circumstellar disk material. 
As shown in Table~\ref{tab:Counter}, most of these sources present variability, which is usually found in sources in a more advanced evolutionary stage. Moreover, 4 out of the 5 diskless sources detected in our observations present non-thermal emission. Thus, our results confirm that the objects classified as diskless by \cite{povich2016} could be equivalent to Stage~III YSOs.

\begin{figure*}[htb]
\centering
\includegraphics[width=0.8\linewidth]{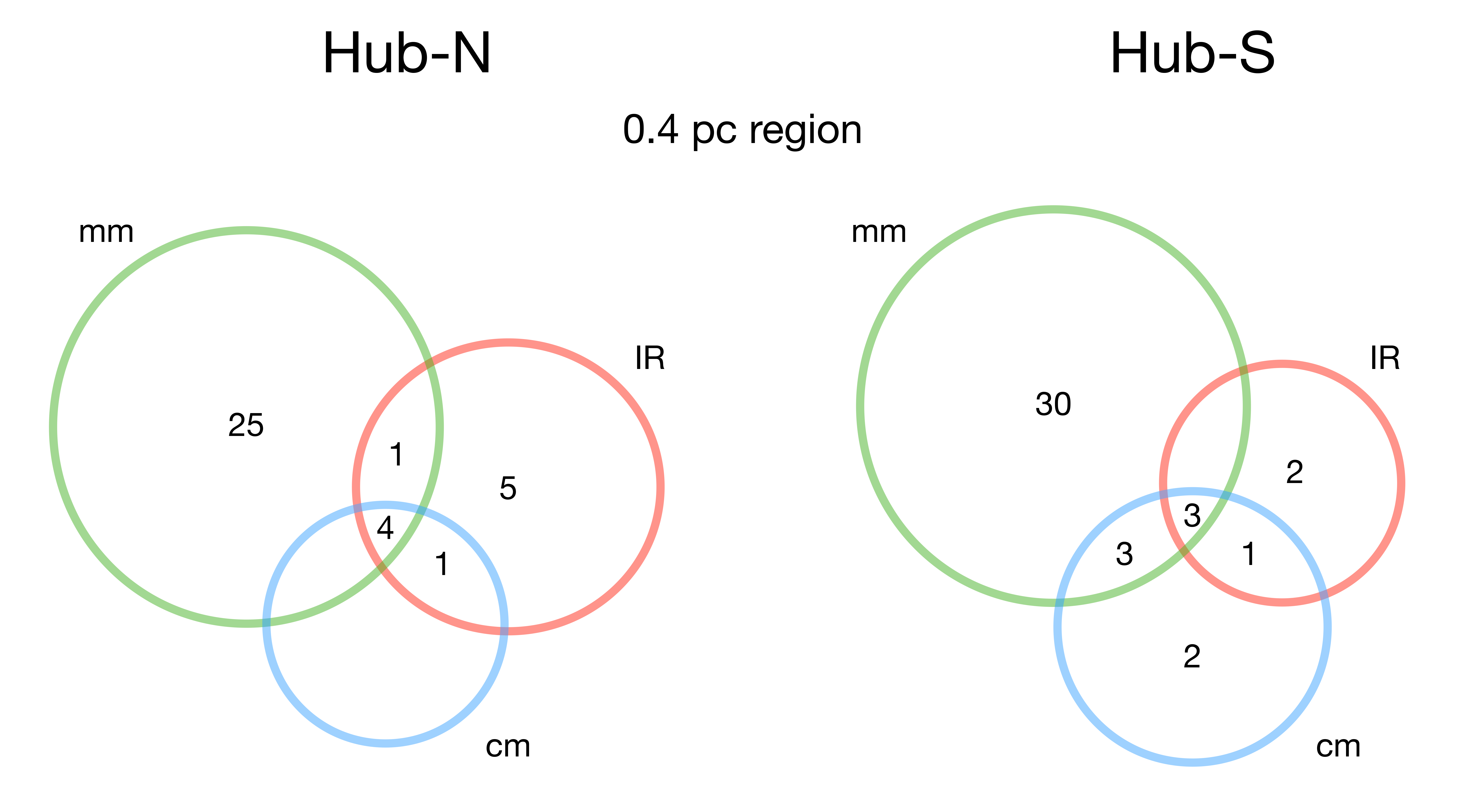}
    \caption{Schematic representation of the sources detected in the 0.4~pc region around each hub and their counterparts at different wavelengths. Millimeter sources correspond to the ones detected in \citet{busquet2016}, \citet{ohashi2016} and Zhang et al. (private communication). Infrared sources correspond to those detected in \citet{povich2016}. Centimeter sources correspond to those detected in this work.}
    \label{fig:counterparts}
\end{figure*}

\cite{busquet2016} found that the ratio between the number of infrared sources without a millimeter counterpart and the total number of sources, within a region of about 0.4~pc in diameter around the center of each hub, is 4 times larger in G14.2-N than in G14.2-S, suggesting a more evolved population in the northern hub. We have expanded this analysis to include the radio continuum emission reported in this work as well as the X-ray sources. When evaluating the inner 0.4~pc region, we have 5 and 9 radio sources in the G14.2-N and G14.2-S hubs, respectively. Fig.~\ref{fig:counterparts} summarizes the number of sources detected at each wavelength in each hub. Similar to the study by \cite{busquet2016}, we list in Table~\ref{tab:number} the number of sources at each wavelengths. Computing the ratio of IR sources without a millimeter and/or centimeter counterpart in each hub, we obtained $N_\text{IR}/N_\text{radio} = 0.2$ in G14.2-N and $N_\text{IR}/N_\text{radio} \simeq 0.05$ in G14.2-S. Thus, the relative number of infrared sources with respect to the radio sources is larger in the northern hub by a factor of approximately 4, similar to the results found by \citet{busquet2016} using observations at millimeter wavelengths.

\begin{table*}[h!]
\centering
\caption{Counterparts of the radio continuum sources detected in G14.2.} 
\label{tab:Counter}
\begin{scriptsize}
\begin{tabular}{lcccccccccll}
\hline\hline \noalign{\smallskip}
Source
& Dense gas
& mm
& IR
& X-rays
& Maser
& Radio\tablefootmark{a}
& \multicolumn{2}{c}{Object type\tablefootmark{b}}
& Stage\tablefootmark{c}
& Catalogue name
& References
\\
\hline \noalign{\smallskip}
\multicolumn{6}{l}{ Northern region G14.2-N} \\
\hline \noalign{\smallskip}

VLA-03 & \checkmark & --- & \checkmark & \checkmark & --- & V &  YSO$^{\dagger}$ & U & Stage III & G014.2405-00.4991, CXO J181811.60-164830.9 &  (1), (2) (3) \\

VLA-06 & --- & --- & \checkmark & \checkmark & ---  & U & YSO$^{\dagger}$ & U & Stage III & CXO J181824.90-164901.1  &  (1) \\

VLA-08 & \checkmark & --- & \checkmark & \checkmark & ---  & V & YSO$^{\dagger}$ & U & Stage III & G014.2360-00.5146, CXO J181814.48-164911.6 &  (1), (2) (3) \\

VLA-09$^*$ & \checkmark & --- & \checkmark & \checkmark  & --- & U & YSO & J &  Stage II  & [PW2016] 572, G014.2341-00.5067, & (1), (2), (3) \\ 
& & & & & & & & & & CXO J181812.54-164904.6&  \\

VLA-10 & --- & --- & \checkmark & \checkmark & ---  & non-thermal & YSO$^{\dagger}$ & G & Stage III & CXO J181807.28-164916.8 &  (1) \\

VLA-11$^*$ & \checkmark & \checkmark & \checkmark & --- & \checkmark   & flat & YSO & J  & Stage 0/I & [OSC2016]N-4, N14, [PW2016] 579,  & (1), (2), (3), \\
 & & & & & & & & & & G014.2300-00.5097 & (4), (8), (9), \\
& & & & & & & & & & & (10)\\

VLA-12 & \checkmark & --- & \checkmark & \checkmark & --- &  non-thermal & YSO$^{\dagger}$ & U & Stage III & CXO J181806.99-164903.3 &  (1), (3) \\

VLA-13$^*$ & \checkmark & \checkmark & \checkmark & --- & \checkmark  & non-thermal & YSO & J & Stage 0/I &  [PW2016] 578, G014.2290-00.5102, & (1), (2), (3), \\

& & & & & & & & & & [OSC2016] N-19, [BEP2016] MM1g, N12  & (4), (5), (7), \\

& & & & & & & & &  & & (8),(9), (10)  \\

VLA-14$^*$ & \checkmark & \checkmark & \checkmark & ---  & \checkmark & non-thermal &  YSO & J & Stage 0/I &  [OSC2016] N-16, [BEP2016] Hub-N MM1, N24 & (1), (2), (3),  \\

& & & & & & & & & & [PW2016] 564, G014.2286-00.5088,& (4), (5), (6), \\

& & & & & & & & & & & (10) \\
VLA-15 & --- & \checkmark & \checkmark & \checkmark & ---  & non-thermal &  YSO$^{\dagger}$ & U & Stage III & [OSC2016] N-25, CXO J181814.95-164927.5  &  (1), (4) \\

VLA-16$^*$ & \checkmark & \checkmark & \checkmark & --- & \checkmark & U & YSO & J & Stage 0/I & [PW2016] 575, G014.2281-00.5105, N16 & (1), (2), (3)\\
& & & & & & & & & & & (7), (10)  \\

VLA-21 & --- & --- & \checkmark & --- & --- & U & YSO & J & Stage II & [PW2016] 404, G014.1996-00.4766 & (1), (2)\\

VLA-22 & \checkmark & \checkmark & \checkmark & \checkmark & --- & flat & YSO & G & Stage 0/I & [OSC2016] N-2, G014.2156-00.5151,  & (1), (2), 10)\\
& & & & & & & & & & [PW2016] 563, CXO J181812.21-165017.2, N28 & \\

VLA-24 & \checkmark & \checkmark & --- & --- & ---  & non-thermal & YSO$^{\dagger}$ & --- & --- & [OSC2016] N-7, [BEP2016] MM2, & (3), (4), (5) \\

VLA-26  & --- & --- & \checkmark & \checkmark & ---  & non-thermal &  YSO & G & Stage II & [PW2016] 709, G014.2379-00.5654  &  (1), (2)\\
& & & & & & & & & & CXO J181825.96-165037.4 &   \\

VLA-28 & \checkmark & --- & \checkmark & --- & --- & non-thermal & YSO & J & Stage 0/I & [PW2016] 465, G014.1901-00.5044 & (1), (2), (3) \\

VLA-29 & \checkmark & --- & \checkmark & \checkmark & ---  & U & YSO$^{\dagger}$ & U & Stage III & CXO J181812.33-165126.4 & (1), (3) \\

VLA-30 & --- & --- & \checkmark & \checkmark & ---  & U & YSO$^{\dagger}$ & U & Stage III & G014.1750-00.5220, CXO J181808.88-165237.7 &  (1), (2) \\

VLA-31 & --- & --- & \checkmark & \checkmark & --- & non-thermal & YSO & G & Stage 0/I & [PW2016] 409, G014.1667-00.4990,  & (1), (2)\\
 & & & & & & & & & & CXO J181802.60-165245.9 & \\

VLA-32  & --- & \checkmark & \checkmark & \checkmark & ---  & V &  YSO$^{\dagger}$ & U & Stage III & [OSC2016] N-17, N7, & (1), (2)\\
 & & & & & & & & & & G014.1642-00.5131, CXO J181805.66-165257.0 &  (4), (10)\\

\hline \noalign{\smallskip}
\multicolumn{6}{l}{ Southern region G14.2-S} \\
\hline \noalign{\smallskip}

VLA-34 & ---   & ---  & \checkmark & \checkmark  & --- & U & YSO$^{\dagger}$ & U & Stage III & G014.1554-00.5541, CXO J181813.63-165435.1 & (1), (2) \\

VLA-35 & --- & --- & \checkmark & --- & --- & thermal & YSO & J &  Stage II & [PW2016] 407, G014.1327-00.5170 & (1), (2) \\

VLA-36 & ---  & --- & \checkmark & \checkmark & --- & V & YSO$^{\dagger}$ & U & Stage III & CXO J181810.07-165508.0  & (1) \\

VLA-37 & \checkmark   & ---  & \checkmark & \checkmark  & --- & V & YSO$^{\dagger}$ & U & Stage III & G014.1310-00.5609, CXO J181812.23-165602.8 & (1), (2) \\

VLA-38 & --- & --- & \checkmark & --- & --- & non-thermal & YSO & J & Stage II & [PW2016] 377, G014.1008-00.5132 & (1), (2) \\

VLA-39 & ---  & --- & \checkmark & \checkmark & --- & U & YSO$^{\dagger}$ & U & Stage III & CXO J181811.53-165627.1 & (1)\\

VLA-40 & \checkmark  & --- & \checkmark & --- & --- & flat &  YSO & J & Stage II & [PW2016] 460 G014.1123-00.5449 & (1), (2), (3) \\

VLA-42$^*$ & \checkmark   & ---  & \checkmark & ---  & --- &non-thermal & YSO & J & A & [PW2016] 561, G014.1114-00.5711 & (1), (2) \\

VLA-43 & --- & --- & \checkmark & --- & --- & U &YSO & J & Stage 0/I & [PW2016] 451, G014.0945-00.5518 & (1), (2) \\

VLA-44$^*$ & \checkmark & \checkmark & \checkmark & --- & --- & non-thermal & YSO & J & Stage 0/I &  [OSC2016] S-18, [BEP2016] MM4 & (1), (2), (3) \\

& & & & & & & & & &  [PW2016] 584, G014.1132-00.5745  & (4),(5) \\

VLA-45${^*}$ & \checkmark & \checkmark & --- & --- & --- & thermal &  YSO$^{^\dagger}$ & U & --- &  [OSC2016] S-3, S6 &  (3), (4), (10) \\

VLA-46 & \checkmark  & --- & \checkmark & \checkmark & --- & V & YSO$^{\dagger}$ & U & Stage III & G014.1074-00.5575, CXO J181808.70-165712.0 & (1), (2), (3) \\

VLA-47 & \checkmark   & --- & \checkmark & --- & --- & thermal & YSO & J & Stage 0/I & [PW2016] G014.1101-00.5626 & (1), (2) \\

VLA-48 & \checkmark   & \checkmark  & \checkmark & \checkmark  & --- & V & YSO${^\dagger}$  & U & --- & [OSC2016] S-17, CXO J181810.98-165708.9 & (1), (3), (4) \\

VLA-49$^*$ & \checkmark & \checkmark & \checkmark & --- & --- & thermal & YSO & J & Stage 0/I &  [BEP2016] MM7a, S3, [PW2016] 594 & (1), (2), (3), \\

& & & & & & & & & &  G014.1126-00.5780  &  (5), (10) \\

VLA-50 & --- & --- & \checkmark & --- & --- & non-thermal & YSO & J & Stage II & [PW2016] 400, G014.0948-00.5298 & (1), (2) \\

VLA-52 & --- & --- & \checkmark & \checkmark & --- & U & YSO$^{\dagger}$ & U & Stage III &  CXO J181820.84-165720.7 &  (1)\\

VLA-53$^*$ & \checkmark & \checkmark & --- & --- & --- & non-thermal &YSO${^\dagger}$ & --- & --- &  [OSC2016] S-2, [BEP2016] MM2b, S20 &(3), (4), (5), \\
& & & & & & & & & & & (10)\\

VLA-54$^*$ & \checkmark & \checkmark & \checkmark & --- & --- & V & YSO & J & Stage 0/I &  [BEP2016] MM5b, [PW2016] 587, S14 & (1), (2), (3),  \\

& & & & & & & & &  &  G014.1142-00.5743  & (5), (10) \\

VLA-55$^*$ & \checkmark   & \checkmark  & \checkmark & ---  & \checkmark & thermal & YSO & J & Stage 0/I & [OSC2016] S-1, [BEP2016] MM5a, S8  & (1), (2), (3), \\

& & & & & & & & & &  [PW2016] 588, G014.1157-00.5737 & (4), (5), (6), \\
& & & & & & & & & & & (10)\\

VLA-56$^*$ & \checkmark & --- & \checkmark & --- & --- & thermal &  YSO & J & A &  [PW2016] 599, G014.1147-00.5783 &  (1), (2), (3)\\

VLA-57 & \checkmark  & \checkmark & \checkmark & \checkmark & --- & U & YSO${^\dagger}$ & U & --- & [OSC2016] S-9, CXO J181809.15-165728.0  & (1), (3), (4) \\

VLA-58$^*$ & \checkmark & \checkmark & --- & --- & --- & thermal & YSO${^\dagger}$ & J & --- &  [OSC2016] S-4, [BEP2016] MM2a & (3), (4), (5) \\

VLA-59 & \checkmark & --- & \checkmark & \checkmark & --- & thermal & YSO$^{\dagger}$ & U & Stage III & CXO J181812.03-165701.0 & (1), (3) \\

VLA-60 & \checkmark  & --- & \checkmark & \checkmark & --- & V & YSO$^{\dagger}$ & U & Stage III & G014.1000-00.5634, CXO J181809.17-165746.3 & (1), (2), (3) \\

VLA-62 & \checkmark  & --- & \checkmark & --- & --- & thermal & YSO & J & Stage II & [PW2016] 475, G014.1056-00.5537 & (1), (2), (3) \\

VLA-64 & \checkmark  & --- & \checkmark & \checkmark & --- & non-thermal & YSO$^{\dagger}$ & G & Stage III & G014.0893-00.5548, CXO J181805.90-165804.4 & (1), (2) (3) \\

VLA-65 & \checkmark  & \checkmark & \checkmark & \checkmark & --- & thermal & YSO$^{\dagger}$  & U & Stage III & [OSC2016] S-13, CXO J181805.64-165812.6 & (1), (3), (4) \\
\hline
\end{tabular}  

\tablebib{(1)~\cite{povich2016}; (2)~\cite{Spitzer2009}; (3)~\cite{busquet2013}; (4)~\cite{ohashi2016}; (5)~\cite{busquet2016}; (6)~\cite{wang2006}; (7)~\cite{palagi1993}; (8)~\cite{Green_2010}; (9)~\cite{sugiyama2017}; (10)~Q.\ Zhang (priv.\ communication).}
\tablefoot{
Sources marked with * next to their names correspond to radio continuum sources located within the inner 0.4~pc region around the center of each hub.
\tablefoottext{a}{Radio continuum emission classified as thermal, non-thermal, flat, variable (V) and unclassified (U).} 
\tablefoottext{b}{Object type: YSO, young stellar object; YSO$^{\dagger}$, YSO candidates proposed in this work; J, compatible with radio jet; G, compatible with gyrosynchrotron emission.}
\tablefoottext{c}{Evolutionary stage classification \citep{povich2016}: Stage~0/I = SED model with infalling envelope; Stage~II = SED model with only circumstellar disks; Stage III = X-ray sources with no mid-infrared excess but detected as IR point sources;  A = Ambiguous (no dominant SED model type)}
}
\end{scriptsize}
\end{table*}

\begin{table}[htb]
\caption{Number of sources detected at different wavelengths}
\begin{footnotesize}
\begin{tabular}{l c c c c c c}
\hline\hline \noalign{\smallskip}
Region
& ${N}_{\mathrm{mm}}$
& ${N}_{\mathrm{cm}}$
& ${N}_{\mathrm{IR}}$
& ${N}_{\mathrm{IR+}}$
& ${N}_{\mathrm{IR}}/{N}_{\mathrm{radio}}$
& ${N}_{\mathrm{IR^+}}/{N}_{\mathrm{radio}}$
\\ 
\hline \noalign{\smallskip}
G14.2-N & 25 & 5 & 5 & 10  & 0.2 & 0.3 \\
G14.2-S & 30 & 9 & 2 & 6  & 0.05 & 0.2 \\
\hline
\end{tabular}
\tablefoot{
Number of sources within the inner 0.4~pc around each hub.
$N_{\mathrm{mm}}$ is the number of millimeter sources without a centimeter and/or IR counterpart.
$N_{\mathrm{cm}}$ is the number of centimeter sources detected in this work.
$N_{\mathrm{IR}}$ is the number of IR sources without a millimeter and/or centimeter counterpart.
$N_{\mathrm{IR^+}}$ is the total number of IR sources.
${N}_{\mathrm{radio}}$ is the number of radio sources, that is, $N_{\mathrm{mm}}$+$N_{\mathrm{cm}}$.
}
\end{footnotesize}
\label{tab:number}
\end{table}

\section{Analysis}\label{s:analysis}

As explained in Sect.~\ref{s:intro}, the radio continuum emission from YSOs can have a thermal or non-thermal nature and can be originated in different processes (\eg\ free-free emission from thermal radio jets or young \hii\ regions, non-thermal gyrosynchrotron and synchrotron emission in magnetically-active YSOs). In this section, we analyze in detail the origin of the radio continuum emission for the 37 sources with well constrained spectral indices (see Sect.~\ref{s:spec})  by studying the well-known correlation between the radio luminosity and the bolometric luminosity for thermal radio jets \citep[see][for a review]{anglada2018} and the connection between the radio and X-ray luminosities, expected for non-thermal radio sources with active coronal activity.

\subsection{Thermal free-free emission: radio jets or \hii\ regions?} \label{sect:hii}

In this section we investigate whether the thermal radio emitters detected in G14.2 can be explained in terms of photoionization (\ie\ \hii\ regions) or ionization through shocks associated with outflows and jets. 
For this, we computed the number of Lyman-continuum photons per second, $N_{\rm{Ly}}$ \citep[see][]{sanchez-monge2013}, using the flux densities at 3.6 cm, which for 19 thermal emitters, the radio luminosities are in the range of $0.04$--$0.45$~mJy~kpc$^2$, with a mean value of $\sim0.16$~mJy~kpc$^{2}$. Adopting an electron temperature of $T_{\rm{e}}=10^4$~K, we obtained values $N_{\rm{Ly}}\sim3\times10^{42}$--$3\times10^{43}$~s$^{-1}$, which translates to spectral types B3–B4 
assuming as a ionization source a single zero-age main sequence (ZAMS) star \citep{panagia1973,thompson84,vacca96,diaz-miller98,martins2005}.  In G14.2, the YSO population detected in the IR present luminosities much lower, 10--100~$L_\odot$, so we expect  $N_{\rm{Ly}}\ll10^{42}$~s$^{-1}$. Therefore, the emission is likely due shock-induced ionization for most of the sources.

\begin{figure}[htb]
    \centering
    \includegraphics[width=1\linewidth]{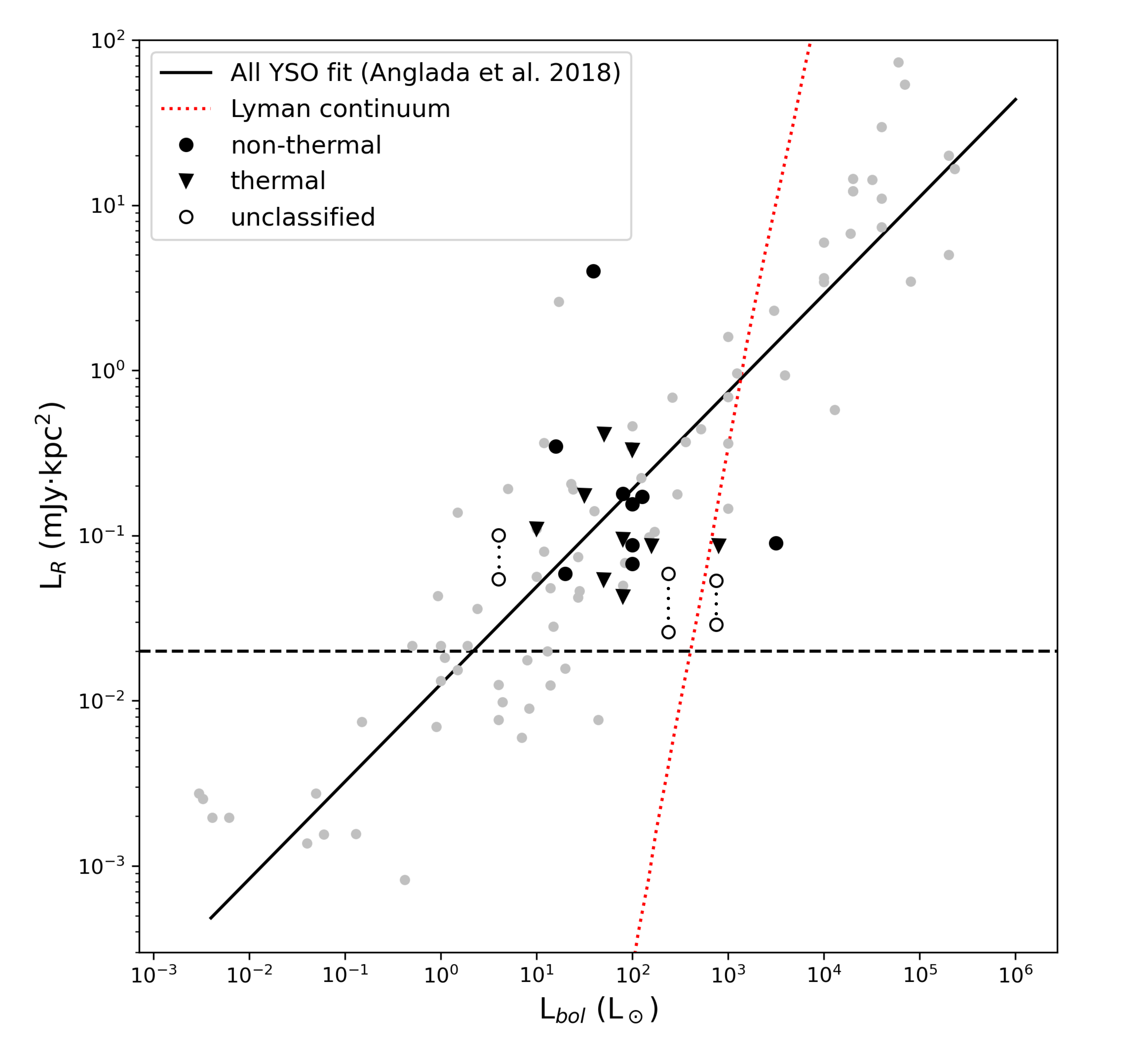}
    \caption{Radio luminosity as a function of bolometric luminosity for the sources with an IR counterpart and a measured bolometric luminosity \citep{povich2010,povich2016}. Triangles represent thermal sources, filled dots depict non-thermal sources, and open circles represent radio sources with an unclassified origin of the radio continuum emission in G14.2. For unclassified radio sources not detected at 3.6~cm we extrapolated the 6~cm flux density to 3.6~cm adopting two values of the spectral index, $\alpha=-0.7$ and $\alpha=+0.5$ (see main text). Grey dots show the thermal radio jets compiled by \cite{anglada2018}. The solid black line corresponds to the fit done to all radio jets in \cite{anglada2018}. The red dotted line depicts the radio luminosity at 3.6~cm associated with the Lyman continuum flux of \hii\ regions powered by stars of different luminosities \citep[see Fig.~4 from][]{sanchez-monge2013}. The black dashed line depicts our $5\sigma$ sensitivity limit in radio luminosity ($\sim0.02$~mJy kpc$^2$).} 
    \label{fig:lum-bol}
\end{figure}

Fig.~\ref{fig:lum-bol} presents the relation between the radio luminosity and the bolometric luminosity for 9 out of the 19 thermal sources, including flat-spectrum sources, identified in G14.2. We compare them with the sample of radio jets compiled by \citet{anglada2018} as reference. Although our sample comprises a relatively narrow range of luminosities ($\sim$10--800~\lsun), there is an excess of radio emission compared to what is expected for an \hii\ region, and hence the radio emission is compatible with the well-known correlation for thermal radio jets. Hence, we can discard that these sources are \hii\ regions. The morphology of these sources appear, in most of the cases, elongated indicating that they are potential thermal radio jets. In fact, 6 out of the 9 thermal radio sources in G14.2 with measured bolometric luminosities, have been classified by \citet{povich2016} as Stage~0/I YSOs, 3 of them are Stage~II  and 1 source is classified as Ambiguous.  
For the Stage~III YSOs we do not have the measured bolometric luminosities. Moreover, the centimeter sources found in association with H$_2$O and CH$_3$OH masers \citep{palagi1993,wang2006,Green_2010,sugiyama2017} are classified as Stage 0/I, suggesting that they are potential radio jets.

We additionally explored whether the non-thermal radio sources, as well as the unclassified radio sources, 
with an infrared counterpart, hence with measured bolometric luminosities, follow the empirical correlation for radio jets (see Fig.~\ref{fig:lum-bol}). For those sources only detected at 6~cm, we estimated the 3.6~cm radio luminosity assuming a spectral index of $+0.5$ (following the approach of \citealt{anglada2018}). Since we know, however, that most of these radio sources present a negative spectral index, for well-classified non-thermal sources, we adopted $\alpha=-0.7$ to extrapolate the flux density at 3.6~cm. 
Our sample contains 12 radio sources, 6 of them have been classified as Stage~0/I, 5 correspond to Stage~II objects and only 1 is classified as Ambiguous according the classification of \citet{povich2016} . 

In order to discard that the sources detected in G14.2 could be \hii\ regions, we also calculated the expected flux density and thus, the luminosity, from the number of Lyman-continuum photons per second that are expected for \hii\ regions \citep{panagia1973,thompson84}. As can be seen in Fig.~\ref{fig:lum-bol}, with the exception of some sources, the rest of them do not follow the expected relation. In contrast, our sample follows the expected relation between the radio luminosity and the bolometric luminosity found by \citet{anglada2018}, suggesting that these sources are also potential radio jets. In fact, several works find that radio jets can present both thermal and non-thermal emission. The central and powering source is usually associated with thermal emission whereas the jet lobes/knots are associated with non-thermal synchrotron emission from relativistic electrons accelerated in strong shocks \citep[\eg][]{carrasco-gonzalez2010,marti1993,rodriguez1989,rodriguez2005,sanna2019}. However, there are some cases in which the radio emission from jets, at the current resolution, seems to be dominated by a non-thermal origin \citep[\eg][]{reid1995,moscadelli2016,Kavak2021}. Therefore, the sample of radio sources in G14.2 with negative spectral indices or the unclassified sources are compatible with radio emission arising from radio jets although further observations spanning a wider range of frequencies and in polarization mode would be necessary to fully confirm their nature.

\subsection{The radio--X-ray relation}

Previous VLA surveys of nearby star-forming regions have reported a correlation between the radio emission of YSOs and their associated X-ray emission \citep[see \eg][]{pech2016}. Several findings suggest that YSOs adhere to the empirically Güdel–Benz relation \citep{Gudel1993, benz1994} for magnetically active stars:
\begin{equation}
    \dfrac{L_X}{L_R} = \kappa\,10^{15.5 \pm 0.5},
\end{equation} 
with $\kappa \leq 1$, depending on the type of stars. From our VLA observations, 25 out of 66 sources (\ie\ $\sim38$\% of our sample) present a reported X-ray counterpart but only 9 of them have measured luminosities \citep{povich2016} and do not present variability. 
The absorption-corrected luminosities measure the total X-ray band (0.5--8~keV) and the hard X-ray band (2–-8~keV). For this work, we use the hard X-ray band since it is less affected by absorption.

\begin{figure}[!ht]
    \centering
    \includegraphics[width=1\linewidth]{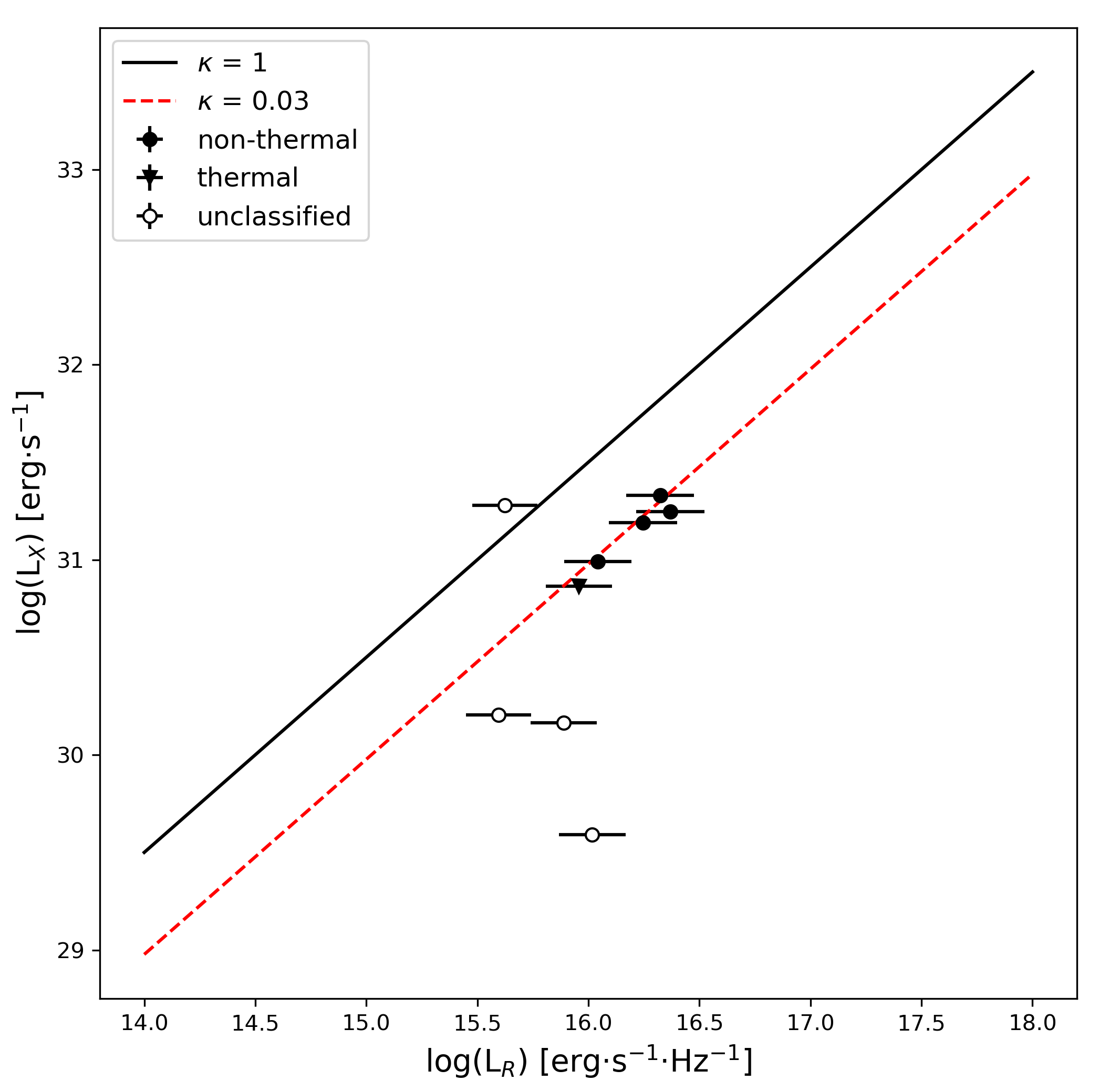}
    \caption{X-ray luminosity as a function of radio luminosity for the sources in G14.2 with a measured X-ray luminosity \citep{povich2016}. The black line corresponds to the Güdel-Benz relation with $\kappa = 1$. The red dashed line corresponds to the Güdel-Benz relation but with $\kappa = 0.03$. Black dots depict non-thermal sources, black triangles represent the thermal sources, and unfilled black dots represent the radio sources with an unclassified origin of radio continuum emission. Variable sources have been excluded for this representation.}
    \label{fig:lum-all}
\end{figure}

Fig.~\ref{fig:lum-all} shows the X-ray luminosities and our derived radio luminosities for thermal, non-thermal, and unclassified radio sources. For simplicity in the representation, we considered flat sources as thermal emitters. Particularly for the unclassified sources, our data is poorly correlated to what we expected and presents a large dispersion, similar to the results found in M17 \citep{yanza22} and in the Orion Nebula Cluster \citep{forbrich2016}. As explained in \cite{yanza22}, the lack of correlation between X-ray and radio observations can be due to the presence of potential thermal sources in the data sample. Moreover, the different timescales and high intrinsic variability of gyrocoronal flares may affect the results since simultaneous X-ray and radio observations are needed to properly study this relation. In fact, for most of the radio sources in G14.2 that present an X-ray counterpart it has not been possible to determine the origin of the emission from the spectral index. The Güdel-Benz relation is valid for non-thermal sources, so it might not apply to most of the sources. Moreover, our sample size is small and thus we cannot infer robust conclusions from the results obtained.

However, if only non-thermal radio sources are considered (i.e., filled black dots in Fig.~\ref{fig:lum-all}), our observations seem to reproduce the Güdel-Benz relation with $\kappa = 0.03$. This suggests that the radio emission in those sources is probably produced by gyrosynchrotron radiation from the mildly relativistic electrons that are responsible for the X-ray emission. The only thermal source presented in Fig.~\ref{fig:lum-all} is VLA-22, which was originally classified as flat source. This source is therefore likely to be more compatible with non-thermal emission and also produced by gyrosynchrotron radiation.

Similar results were found in nearby region such as Ophiuchus \citep{dzib2013}, Taurus-Aurgia \citep{dzib2015} and Perseus \citep{pech2016}, while in Orion and Serpens it was found that the X-ray emission of YSOs was underluminous compared to the Güdel–Benz relation with $\kappa = 1$ \citep{kounkel2014,ortiz-leon2015,forbrich2016}. Despite these promising similarities between G14.2 and other nearby star-forming complex, with only 5 sources in G14.2, we cannot draw firm conclusions regarding the expected radio-X-ray correlation. 

Finally, in G14.2 we have identified 22 non-thermal radio emitters, 36\% of them remain unresolved with our angular resolution ($\sim0\farcs3$) and, with the exception of VLA-19 whose emission is very extended, the remaining sources have sizes $<0\farcs7$. Therefore, based on the compactness of these radio sources, we suggest that the radio emission of most of the non-thermal radio population in G14.2 is most likely associated with gyrosynchrotron radiation from the very active stellar magnetosphere, typically found in Class~II/III YSOs \citep{Feigelson1985}. However, in order to fully confirm the gyrosynchrotron origin, follow-up polarization studies are needed to investigate whether these radio sources present some degree of circular polarization.

\section{Discussion}\label{s:dis}

\subsection{Levels of fragmentation in the G14.2 hubs}

Previous observations of the two hubs in G14.2 with the SMA, at an angular resolution of $\sim1\farcs5$ revealed different levels of fragmentation, with G14.2-S being more fragmented than G14.2-N \citep[see][]{busquet2016}. Despite these differences in fragmentation, the physical properties of both hubs such as the density and temperature profiles, the level of turbulence (Mach number $\sim5.6$--6.4), the Alfvén Mach number ($\sim0.4$--0.3), the rotational-to-gravitational energy ratio ($\beta_{\mathrm{rot}}\sim0.016$--0.015), the mass (979--717~\msun), and the luminosity (995--531~\lsun) are remarkable similar \citep[see Tables~5 and 6 in][for further details]{busquet2016}.
As explained in \citet{busquet2016}, the different levels of fragmentation may be due to different reasons. The first one is the difference in the magnetic field strength, with G14.2-N having a stronger magnetic field compared to G14.2-S \citep[see][]{A_ez_L_pez_2020}. The second potential cause is the presence of the luminous IRAS\,18153$-$1651 source, with a luminosity of $\sim1.1\times10^4$~$L_{\sun}$ and strong UV radiation, in G14.2-N. This suggests that the UV radiation from IRAS\,18153$-$1651, as well as from the larger number of IR sources in the northern hub compared to the southern sibling, might be suppressing fragmentation.

However, with our VLA data we do not see significant differences in the number of sources (or level of fragmentation) in the two hubs as previously studied in \cite{busquet2016}. In the current work, 32 centimeter sources were detected in G14.2-N and 34 in G14.2-S. While the detection or non-detection of radio continuum sources might be related to evolutionary effects, interestingly, the latest ALMA data at 1.3~mm, with an angular resolution comparable to the VLA observations (Q.\ Zhang, priv.\ communication, see also Figs.~\ref{fig:SMA-hubN} and \ref{fig:SMA-hubS}), do not reveal statistical differences in terms of fragmentation: with 25 millimeter sources without a centimeter and/or IR counterpart in G14.2-N and 30 millimeter sources in G14.2-S. Therefore, we conclude that both hubs show similar levels of fragmentation based on the observations with the VLA and ALMA. Hence, it seems that the different fragmentation levels reported in \citet{busquet2016} may have been due to poor sensitivity in previous SMA observations, or to different effects controlling fragmentation at different scales. Therefore, although the magnetic field and UV radiation (from the bright IRAS source) could determine the level of fragmentation at intermediate scales (i.e.\ 0.03~pc scale), the fragmentation at smaller scales (i.e.\ 0.005~pc) does not seem to be affected anymore by these effects. Thus, thanks to the new results at high-angular resolutions in the cm and mm regimes, it is very feasible that G14.2-N and G14.2-S are twin hubs in terms of fragmentation, as proposed in \citet{busquet2016} regarding their large-scale physical properties.

\subsection{Radio properties of the stellar population}

The high sensitivity VLA observations allowed us to detect 66 radio sources in the IRDC\,G14.225.
Our analysis of the spectral index in the 6--3.6~cm range reveals that in G14.2 there are 22 sources ($\approx$33\%) that clearly present non-thermal emission and 13 ($\approx$20\%) are thermal emitters. There are also 3 sources ($\approx$5\%) presenting a nearly flat emission spectrum, most likely associated with thermal emission. 

One aspect that should be taken into account when examining the origin of the radio continuum emission based on the spectral index analysis is the variability of the sources. As mentioned in previous sections, we found ten sources that are clearly variable at short-time scale (see Tables~\ref{VarC} and \ref{VarX} and Figs.~\ref{fig:var-g14n-xband} and ~\ref{fig:var-g14s-xband} ), but our observations were not designed to carefully characterize radio variability, and therefore, other sources may be also variable even if not detected as such in the current observations. Follow-up simultaneous multi-frequency observations with the VLA, similarly to \citet{Liu2014} and \citet{coutens2019}, might provide a more detailed insight into the variability of the radio sources in G14.2 and thus, a better estimation of their spectral indices and origin of the radio emission.

\begin{figure}[t!]
    \centering
    \begin{tabular}{c}
    \includegraphics[width=1\linewidth]{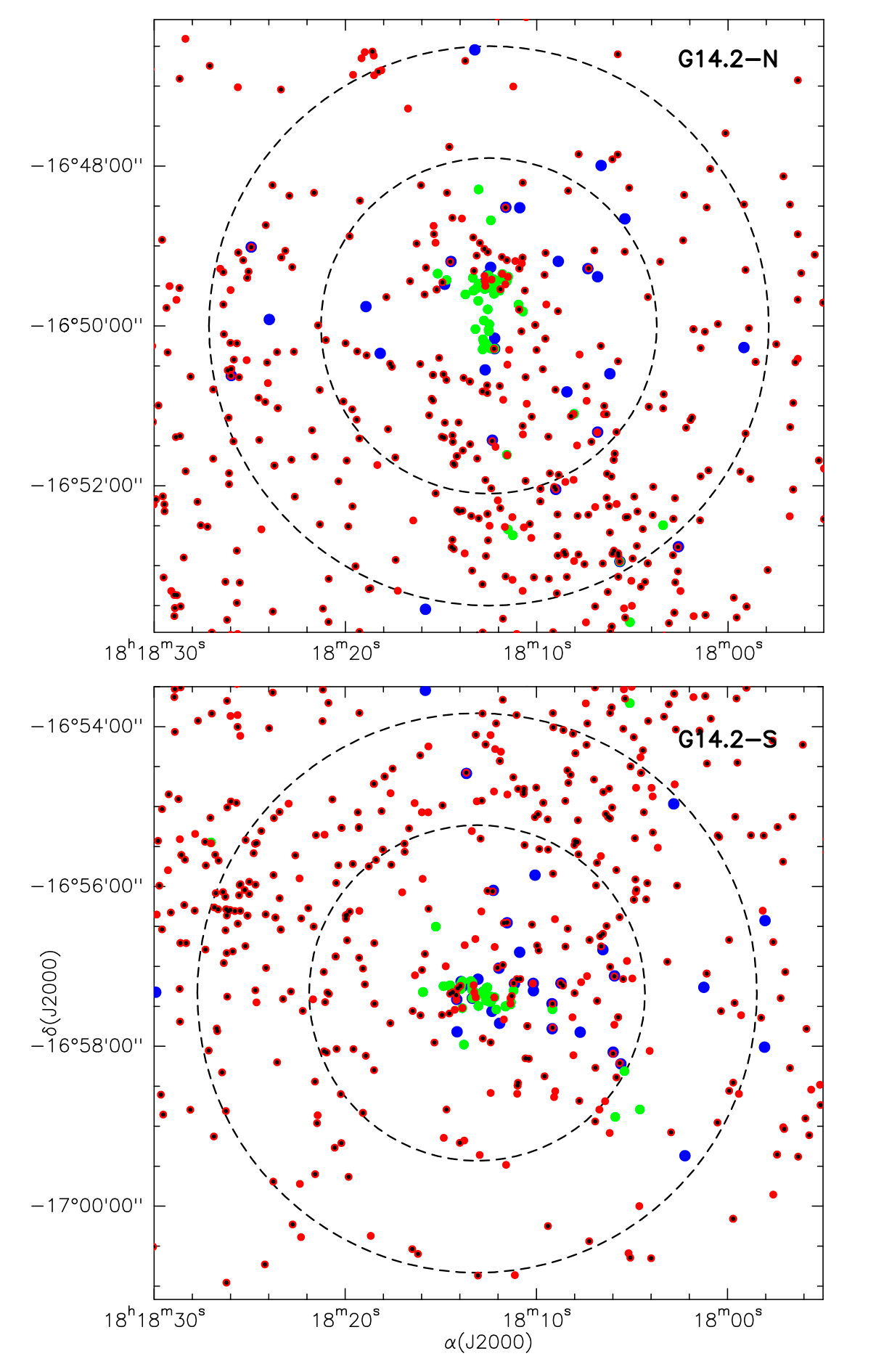} \\
    \end{tabular}
    \caption{Spatial distribution of the stellar population in G14.2. Top panel shows the northern region of G14.2, while bottom panel shows the southern region of G14.2 (G14.2-N and G14.2-S, respectively). Blue symbols indicate centimeter sources (this work). Green symbols indicate millimeter sources \citep[][Zhang et al. private communication]{ohashi2016,busquet2016}. Red symbols indicate infrared sources \citep{povich2016}. Black symbols indicate X-ray sources \citep{povich2016}. The symbol sizes do not correspond to the respective angular resolution. The outer and inner dashed circles represent the field of view at 6~cm ($\sim7'$ at 6~GHz) and 3.6~cm ($\sim4.2'$ at 10~GHz), respectively.}
    \label{fig:detections-g14}
\end{figure}

Despite the high sensitivity of VLA observations, the fraction of radio detections is low in comparison with the IR and X-ray stellar population \citep{povich2016}. Fig.~\ref{fig:detections-g14} shows the location of the four different populations in G14.2-N (top panel) and G14.2-S (bottom panel). In each region, there are between 300 to 400 sources detected at IR and/or X-rays, and only 44 have a radio counterpart. The IR/X-ray sources with no radio counterpart, could be rather evolved objects (Class II/III) with quiet corona activity, and hence with no thermal radio jet and no gyrosynchrotron emission.

Regarding the millimeter population \citep[][Zhang et al. private communication]{ohashi2016,busquet2016}, Fig.~\ref{fig:detections-g14} makes more noticeable the differences in the fields of view, since millimeter observations are centered on smaller regions around the central hubs. From our study of the counterparts, we found that four radio sources were only associated with mm emission without any other counterpart at another wavelength. We have proposed these four millimeter sources associated with centimeter emission as new YSOs candidates and it is very likely that these objects are Class~0 or deeply embedded Class~I objects. Since our study of the mm counterparts is limited only to the central region, we are likely to have more mm sources outside the hubs.

\subsubsection{Comparison with other nearby star-forming regions}

We compare now the properties of the radio sources in G14.2 to other star-forming complexes from the Gould's Belt VLA survey where their radio-source population has been studied in detail, reaching similar sensitivities and spatial resolutions as for G14.2. In particular, by comparing the radio spectral indices, which serve as indicators of the emission characteristics, we can investigate their properties across the different evolutionary stages of the YSOs. A comparative study of G14.2 with other complexes may unveil potential differences and shed light on the main characteristics of G14.2.

\begin{figure}[t!]
    \centering
    \includegraphics[width=1\linewidth]{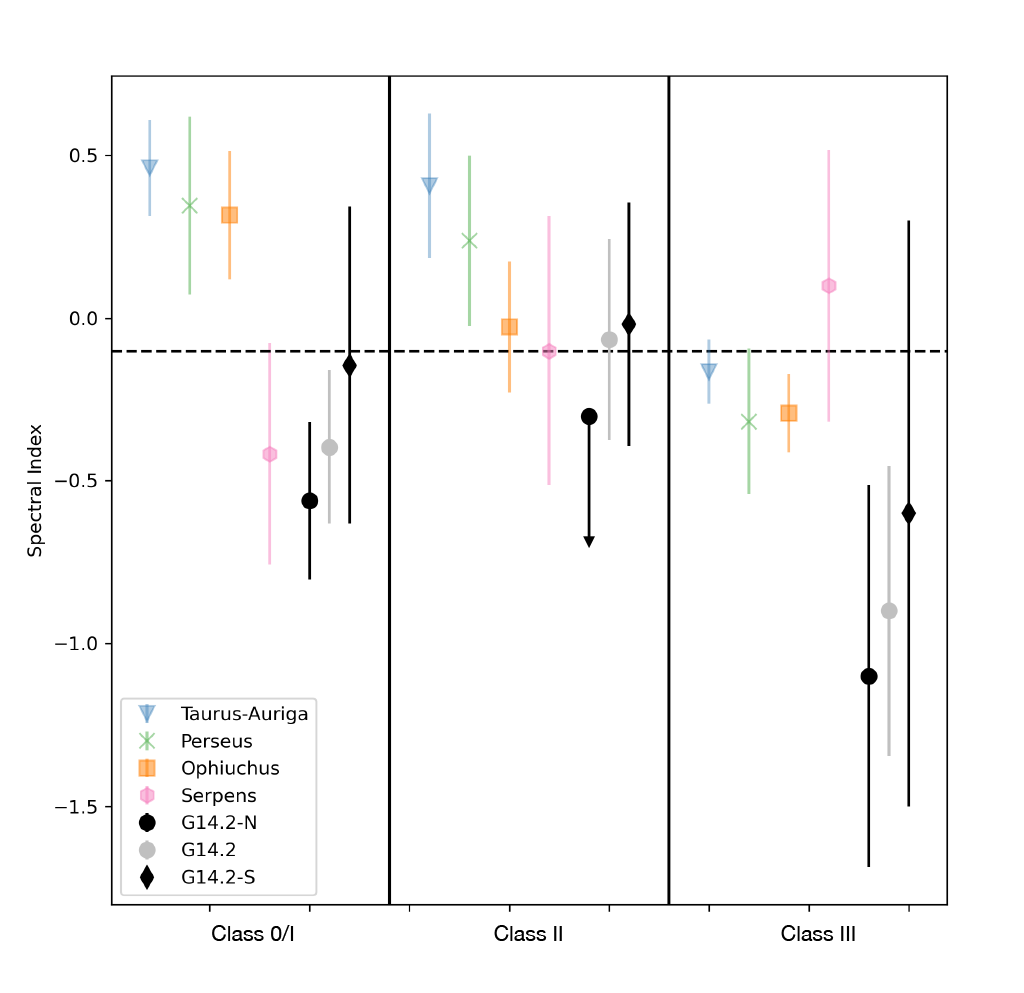}
    \caption{Mean spectral index as a function of the YSO evolutionary stages in different star-forming regions (Blue triangles: Taurus-Auriga \citep{dzib2015}; Green cross: Perseus \citep{pech2016}; Orange square: Ophiuchus \citep{dzib2013}; Pink dot: Serpens \citep{ortiz-leon2015}; Grey dot: IRDC\,G14.2, separated in G14.2-N (black dot)  and G14.2-S (black diamond)). Note that black and grey symbols are SED-based Stage classifications \citep{povich2016} and the groups could be different if a different classification scheme was employed. The black dashed line at $\alpha=-0.1$ draws the boundary between thermal emission and non-thermal emission.}
    \label{fig:YSO-comparison}
\end{figure}

As previously discussed, we adopted the stage categorization in which YSOs were classified as Stage~0/I (SED modelled with infalling envelopes), Stage~II (SED modelled with only circumstellar disks) or Stage~III (X-ray sources with no mid-IR excess from circumstellar disks) used by \cite{povich2016} \citep[see also][]{robitaille2006}. However, the standard classification of YSOs is the class categorization based on the spectral index at infrared wavelengths \citep{lada1987,andre1993,gutermuth2009}. For comparison with other nearby regions we equate Stage 0/I to Class 0/I, Stage II to Class II and Stage III to Class III. Figure~\ref{fig:YSO-comparison} shows the spectral index for different YSOs in different star forming complexes, with the YSOs classified according to their evolutionary stage. We compare the results of G14.2 with Ophiucus \citep{dzib2013}, Serpens \citep{ortiz-leon2015}, Taurus-Auriga \citep{dzib2015} and Perseus \citep{pech2016}.  

We find that for the detected YSOs in Taurus-Auriga, Ophiuchus and Perseus, the more evolved objects have a more negative spectral index. Based on this, it has been proposed that the radio emission towards Class~0/I objects, with spectral indices between +0.3 and +0.5, is likely dominated by partially optically thick free-free emission (from thermal radio jets). On the other hand, Class~II and III objects present radio emission consistent with either optically thin free-free emission or (gyro-~)synchrotron radiation \citep[see e.g.][]{dzib2013}. This is in agreement with the idea that, for more evolved sources, we are no longer able to detect the thermal emission from the surrounding material, since they have already expelled most of it. \citep[\eg][]{Forbrich_2007, Dzib_2010}.

Nevertheless, the Serpens star-forming region \citep{ortiz-leon2015} and the IRDC\,G14.2 (this work) do not follow this trend, since younger objects are associated with non-thermal spectral indices.  
This might be due to the fact that these regions are composed of more massive YSOs in which non-thermal emission may be more dominant \citep[\eg][]{carrasco-gonzalez2010,rodriguez-kamenetzk2017, Kavak2021}. An alternative explanation for the detection of non-thermal emission in the less evolved objects might be due to geometrical effects rather than the mass of the YSOs \citep[\eg][]{ortiz-leon2015}. According to this scenario, if the star is seen nearly pole-on or nearly edge-on, the non-thermal emission originating in the corona might be less absorbed by the surrounding material and can be more easily observed \citep{Forbrich_2007}. This effect could also be obtained through tidal clearing of circumstellar material in a tight binary system \citep{Dzib_2010}.

Considering that, statistically, one does not expect a preferential orientation for YSOs, the trend found for the spectral index towards the YSOs of G14.2 might be explained by the presence of more massive YSOs compared to regions such as Ophiuchus, Taurus-Auriga or Perseus. This scenario is plausible for both Serpens and G14.2, since recent studies have confirmed mass segregation effects for both regions \citep[see][]{povich2016,plunkett2018}, with more massive YSOs located in the central regions of the star-forming complex and corresponding to those preferentially studied in the radio observations. Note also that more massive YSOs evolve more quickly, which could explain why Stage III objects present more negative spectral indices in G14.2 in comparison with other regions, since they should have stronger magnetic flaring activity. As shown in Fig.~\ref{fig:YSO-comparison}, the dominating population of non-thermal emitters within the less evolved objects is likely to come mainly from the northern region G14.2-N. By examining Table~\ref{tab:Counter}, we can see that in G14.2-S only one Stage~0/I object clearly shows non-thermal emission. Thus, our results point to G14.2-N likely containing more massive objects.

\subsection{Evolution and development of G14.2}\label{sec:ev}

The molecular cloud environment in G14 extends more than $1\degr$ to the southwest of the \hii\ region, parallel to the galactic mid-plane \citep{elmegreen1976,elmegreen1979}. These authors suggested a sequential massive star formation 
from the north-eastern side with OB stars in NGC\,6618 to the M17 southwest extension, or M17 SWex \citep{povich2009,povich2010}. We discuss now on the possible evolutionary stage of the IRDC\,G14.2  
in relation to the more developed M17 star-forming complex. For this, we highlight different aspects regarding the stellar population and physical properties across the IRDC\,G14.2. 

First, the already-developed and large \hii\ region associated with the bright IRAS~18153$-$1651 source is located to the northeast of G14.2 (see Fig.~\ref{fig:radiosources}), while the southern region of the cloud appears more quiescent. This suggests a certain evolutionary gradient from 
southwest (less evolved) to northeast (more evolved), in agreement with the large scale age evolution proposed by \citet{elmegreen1976}. 

A second aspect refers to the stellar population across the IRDC\,G14.2. The counterparts of the radio sources at other wavelengths (see Table~\ref{tab:number}) provide precious information on the stellar population in both regions, their properties and evolutionary stages.  
The number of infrared sources relative to radio sources in G14.2-N suggests that the northern hub harbours a stellar population in a more advanced evolutionary stage but still having a deeply embedded population of protostellar cores (\eg\ the case of VLA-14/MM1, see Appendix \ref{app:vla-14}). On the other hand, in G14.2-S, we have identified more millimeter sources without an infrared counterpart, suggesting that there is a larger population of objects at an earlier evolutionary stage. 
The large fraction of non-thermal emitters in G14.2-N could be due to the presence of relatively evolved YSOs (Class~II and/or Class~III), consistent with the ratio of IR versus mm sources. We note however that G14.2-N harbours several sources at an early evolutionary stage (\ie\ classified as Stage~0/I by \citealt{povich2016}; see Table~\ref{tab:Counter}), and thus the non-thermal radio emission might also result from strong shocks produced by radio jets powered by intermediate-/high-mass objects \citep{carrasco-gonzalez2010,rodriguez-kamenetzk2017, Kavak2021}. Both analyses lead us to the conclusion that there are differences in the evolutionary stages of the two regions in the IRDC\,G14.2, and hint towards G14.2-N being more evolved compared to G14.2-S, and likely containing more massive objects.

The differences, in age and mass, seem to be in agreement with the `filament-halo' gradient observed by \cite{povich2016}. The proposed scenario to explain the gradient combines two interrelated star formation processes, filament-driven star formation with dynamical relaxation \citep[\eg][]{bate2003} coupled with global hierarchical filament collapse \citep{vazquez-semadeni2019}. However, the importance of each process in producing the observed distribution is still unclear \citep[see][for further discussion]{povich2016}.

One of the possibilities causing these differences could be related to the proximity of G14.2-N to the brightest IRAS\,18153$-$1651 source in the field, located at about $1.5'$ south-east from G14.2-N \citep{busquet2013, Gvaramadze2017}. As mentioned previously, it hosts two B-type stars (B1 and B3). The discovery of an optical arc near the centre of the nebula associated with the IRAS source by \citet{Gvaramadze2017} led to the hypothesis that it might represent a bubble blown by the wind of a young massive star. This could be an evidence that the northern region is a bit more evolved compared to Hub-S, which lacks of similar structures.

Finally, we compared our results to the previous work by \cite{yanza22} towards M17 (located north to G14.2 and associated with a bright and well-developed \hii\ region). In \cite{yanza22}, the M17 region is studied using VLA observations in X-band with the most extended A configuration. They find a
median source size of $\approx200$~mas.
In G14.2,
our observations at X-band result in a median size of $\approx320$~mas for G14.2-S and $\approx230$~mas for G14.2-N. Therefore, sources tend to get smaller from G14.2-S to G14.2-N, and from G14.2-N to M17. This is consistent with a sequence where centimeter sources are progressively more compact as they tend further north. When connected with evolution,
this would point to more evolved objects having a more compact radio continuum emission compared to early-stage objects. 
Early-stage Class~0/I sources are usually dominated by radio jets, which are elongated and often resolved at sub-arcsecond resolution, whereas more evolved Class~II/III YSOs are typically associated with very compact and unresolved radio emission \citep[see][]{anglada2018}.

Moreover, the compact radio continuum sources in the M17 region are mainly dominated by non-thermal emission. For the sources for which the in-band spectral index could be obtained 
(see Table~4 from \citealt{yanza22} for further information), more than 75\% of them present spectral indices lower than $-0.1$. This is in agreement with the results found for G14.2-S and G14.2-N, where most of the non-thermal radio emitters are found in the more evolved G14.2-N region.
All together, our results combined with the results of \cite{yanza22}, confirm an evolutionary sequence starting with G14.2-S, following with G14.2-N and ending in M17 as it was first proposed by \cite{elmegreen1976}. However, in our work we do not found evidences that support their claim that star formation in M17 SWex is triggered by the presence of the M17 \hii\ region.

\subsection{Can the IRDC G14.2 form massive stars in the future?}\label{sec:massive_stars}

Previous works by \citet{povich2010,povich2016} have pointed out the remarkable star formation activity in the IRDC\,G14.2, characterized by a high star formation rate (SFR) of $\dot{M}=0.0072$~\msun\,yr$^{-1}$. This value is even higher than that of the Orion Nebula Cluster (ONC) and NGC\,6618, the cluster ionizing the bright M17 \hii\ region ($\dot{M}=0.005$~\msun\,yr$^{-1}$). Interestingly, despite the high SFR, G14.2 lacks of O-type stars ($M>20$~\msun), which is difficult to explain in the context of a standard Initial Mass Function (IMF; \citealt{salpeter1955,kroupa2001}.

Using the N(H$_2$) column density map of G14.2 \citep{lin2017} and adopting a distance of  $d \sim 1.6$~kpc to the cloud, we estimate the total mass of G14.2 to be approximately of 12000~\msun. This differs from the previous estimation of $\sim$ 20000~\msun\ \citep{lin2017} due to the different distances adopted. Taking this updated mass estimate into account, and incorporating it into an analytical model for the cloud's evolution presented in \cite{camacho2020} (see their Figure~13 for further details),
G14.2 would be located closer to the trace of the initial accretion rate of $2.9 \times 10^3$ \msun Myr$^{-1}$ and thus, it would correspond to a age of 6-7~Myr, younger than the previously estimated.

These findings shed light on the question raised by \citet{povich2016} regarding the late birth of massive stars in the IRDC\,G14.2. The cloud's slightly younger age suggests that it could continue to evolve and potentially form massive stars in the future. Furthermore, the estimated mass reservoir is lower than what was previously thought, challenging our previous understanding of the cloud's star-forming potential. Assuming a total mass of 12000~\msun and a global star formation efficiency (SFE) of 30\% \citep{bontemps2010}, we would have a total of 3600~\msun available to be in the stellar cluster. Considering the IMF described by \cite{kroupa2001} and 3600~\msun, we would get that the typical total number of stars in the cluster would be about 8300 $\pm$ 300 stars\footnote{We have used the python tool \texttt{imf} available at \url{https://github.com/keflavich/imf}}, with a maximum of 84 $\pm$ 22 \msun for the most massive star in the cluster. These values come from 1000 different generations of clusters, following the IMF from \cite{kroupa2001}, and they correspond to the median of the number of stars and maximum total mass. The uncertainty in the values stand for the standard deviation of the 1000 runs. Therefore, our results suggest that, even considering that the total mass reservoir estimation is lower, G14.2 could end up forming massive stars in the future.

\section{Summary and Conclusions}\label{s:conc}

We have carried out VLA observations in its most extended configurations of the radio continuum emission at 6~cm (C-band: 4--8~GHz) and 3.6~cm (X-band: 8--12~GHz) towards the IRDC G14.225-0.506, an infrared dark cloud that is forming stars in two main hubs (G14.2-N and G14.2-S) with similar masses and luminosities. Our study allowed us to identify a hidden radio-continuum population of compact sources and relate their properties to G14.2 as a whole. The main findings obtained in this work can be summarized as follows: 

\begin{itemize}

    \item We detected 66 sources, 32 of which are located in the G14.2-N region and 34 in the G14.2-S region, with two sources detected in both fields. Most of the detected sources in the IRDC have flux densities around 50~$\mu$Jy and are compact ($\approx200$--300~mas, or 320-480~au), specially in G14.2-N.

    \item The number of radio continuum sources in both hubs is similar, suggesting similar levels of fragmentation and consistent with the latest mm data obtained with ALMA, which suggests that the two regions are twin hubs. 

    \item We have identified 10 sources (3 located in G14.2-N and 7 in G14.2-S) to have a significant variable flux at radio wavelengths over periods of a few days. We note that variability may be an important factor to consider when detecting and characterizing these faint and compact objects in future studies.

    \item We looked for the counterparts at other wavelengths of the detected centimeter sources. We found that 5 of the radio sources are associated with H$_2$O and CH$_3$OH maser emission. 23 of the sources were already known YSOs and we classified 25 sources as YSOs candidates. In the inner 0.4~pc region around the two main hubs, the relative number of IR sources in front of the radio sources is larger in G14.2-N by a factor of 4, suggesting that the northern part is in a more advanced evolutionary stage. 

    \item By examining the spectral index, when possible, we determined the origin of the radio continuum emission. In G14.2-N we found 14 sources with a non-thermal origin and only one thermal emitter. Two sources present a flat spectrum, most likely associated with thermal emission. In G14.2-S we found 8 non-thermal emitters, 11 thermal emitters and 2 sources with a flat spectrum. The dominant continuum emission in G14.2-N could be an evidence of the formation of more massive YSOs, resulting in non-thermal emission due to strong shocks.

    \item By comparing the bolometric luminosity with the radio luminosity, we found that the studied sources are compatible with thermal radio jets, and excluded the presence of embedded \hii\ regions in G14.2 by comparing our observations with the expected relation for \hii\ regions.

    \item When comparing the radio sources with their counterparts in X-rays, we found that most of the sources are underluminous with respect to the Güdel-Benz relation with $\kappa = 1$. When examining only the sources classified as non-thermal emitters, we find that they follow the Güdel-Benz relation with $\kappa = 0.03$, similarly to other star-forming regions. This suggests that radio and X-ray emission are probably caused by magnetic reconnection in the stellar coronae.

    \item We compared the radio properties of the stellar population in G14.2 with other nearby star-forming complexes, such as Taurus-Auriga, Perseus, Ophiuchus and Serpens. The objects in G14.2 follow a similar trend as found in Serpens, with Stage~0/I objects being associated with more non-thermal emission than Stage~II YSOs. Similar to Serpens, this may point to our region being composed of more massive objects compare to other low-mass star forming complexes. 

    \item A comparison of our results of G14.2 with M17, a more evolved star-forming region to the north-east of G14.2, confirms a wider evolutionary sequence starting in G14.2-S and onwards to the most evolved region M17.

    \item Based on the new distance estimations, G14.2 is slightly younger and harbors a lower mass reservoir than what was previously though. Our analysis point out that the complex has the potential to form massive stars in the future.

\end{itemize}

Currently, conducting deep radio surveys is highly time-consuming, taking $\sim11$~hour for a single pointing at X-band, and hence they are limited to relatively nearby ($d<2$~kpc) star-forming regions. However, the next generation of radio interferometers, such as the Square Kilometre Array (SKA) and the Next Generation Very Large Array (ngVLA) are expected to significantly improve the observations. These advanced radio telescopes will reach the same sensitivity achieved in G14.2 with just 1 hour of telescope time, revolutionizing the study of young stellar clusters at radio wavelengths by performing systematic surveys across the Milky Way. In addition, these new instruments will allow for systematic studies of short- and long-term variability in radio emissions. This capability is essential for disentangle the nature of the radio emission and, when combined with deep X-ray observations, for investigating coronal-type magnetic activity across a wide range of stellar masses and evolutionary stages. Finally, it is crucial to conduct polarization observations to better understand the radio properties of YSOs. Gyrosynchrotron emission exhibits circular polarization, while synchrotron emission is linearly polarized. Although there are a few instances where linear polarization has been detected in radio jets, such as in the case of HH80-81 with a relatively high degree of polarization \citep{carrasco-gonzalez2010}, in most cases, the polarization degree is lower than in HH80-81, only accessible with the next generation of radio interferometers.

\begin{acknowledgements}
We are grateful to the anonymous referee for the valuable comments and suggestions.
We are grateful to Vanessa Yanza for useful discussions about the M17-\hii\ properties. Project supported by a 2022 Leonardo Grant for Researchers in Physics from the BBVA Foundation. The BBVA Foundation accepts no responsibility for the opinions, statements and contents included in the project and/or the results thereof, which are entirely the responsibility of the authors.
EDM, GB, JMG, ASM and RE acknowledge support from the PID2020-117710GB-I00 grant funded by MCIN/ AEI /10.13039/501100011033.  ASM acknowledges support from the RyC2021-032892-I grant funded by
MCIN/AEI/10.13039/501100011033 and by the European Union
‘Next GenerationEU’/PRTR, as well as the program Unidad de
Excelencia María de Maeztu CEX2020-001058-M. AP acknowledges financial support from the UNAM-PAPIIT IG100223 grant, the Sistema Nacional de
Investigadores of CONAHCyT, and from the CONAHCyT project number 86372 of the `Ciencia de
Frontera 2019’ program, entitled `Citlalc\'oatl: A multiscale study at the new frontier of the formation and
early evolution of stars and planetary systems’, M\'exico. NAL acknowledges support from the European
Research Council synergy grant ECOGAL (Grant :
855130). HBL is supported by the National Science and Technology Council (NSTC) of Taiwan (Grant Nos. 111-2112-M-110-022-MY3).
\end{acknowledgements}

\bibliographystyle{aa}
\bibliography{g14}

\begin{thebibliography}{99}
\expandafter\ifx\csname natexlab\endcsname\relax\def\natexlab#1{#1}\fi

\bibitem[{{Ainsworth} {et~al.}(2014){Ainsworth}, {Scaife}, {Ray}, {Taylor},
  {Green}, \& {Buckle}}]{Ainsworth2014}
{Ainsworth}, R.~E., {Scaife}, A. M.~M., {Ray}, T.~P., {et~al.} 2014, \apjl,
  792, L18

\bibitem[{{Andre} {et~al.}(1993){Andre}, {Ward-Thompson}, \&
  {Barsony}}]{andre1993}
{Andre}, P., {Ward-Thompson}, D., \& {Barsony}, M. 1993, \apj, 406, 122

\bibitem[{A{\~{n}}ez-L{\'{o}}pez {et~al.}(2020)A{\~{n}}ez-L{\'{o}}pez, Busquet,
  Koch, Girart, Liu, Santos, Chapman, Novak, Palau, Ho, \&
  Zhang}]{A_ez_L_pez_2020}
A{\~{n}}ez-L{\'{o}}pez, N., Busquet, G., Koch, P.~M., {et~al.} 2020, \aap, 644,
  A52

\bibitem[{{Anglada} {et~al.}(2018){Anglada}, {Rodr{\'\i}guez}, \&
  {Carrasco-Gonz{\'a}lez}}]{anglada2018}
{Anglada}, G., {Rodr{\'\i}guez}, L.~F., \& {Carrasco-Gonz{\'a}lez}, C. 2018,
  \aapr, 26, 3

\bibitem[{Anglada {et~al.}(1998)Anglada, Villuendas, Estalella, Beltr{\'{a}}n,
  Rodr{\'{\i}}guez, Torrelles, \& Curiel}]{Anglada_1998}
Anglada, G., Villuendas, E., Estalella, R., {et~al.} 1998, The Astronomical
  Journal, 116, 2953

\bibitem[{{Ballering} {et~al.}(2023){Ballering}, {Cleeves}, {Haworth}, {Bally},
  {Eisner}, {Ginsburg}, {Boyden}, {Fang}, \& {Kim}}]{ballering2023}
{Ballering}, N.~P., {Cleeves}, L.~I., {Haworth}, T.~J., {et~al.} 2023, \apj,
  954, 127

\bibitem[{{Bate} {et~al.}(2003){Bate}, {Bonnell}, \& {Bromm}}]{bate2003}
{Bate}, M.~R., {Bonnell}, I.~A., \& {Bromm}, V. 2003, \mnras, 339, 577

\bibitem[{{Benz} \& {Guedel}(1994)}]{benz1994}
{Benz}, A.~O. \& {Guedel}, M. 1994, \aap, 285, 621

\bibitem[{{Bontemps} {et~al.}(2010){Bontemps}, {Motte}, {Csengeri}, \&
  {Schneider}}]{bontemps2010}
{Bontemps}, S., {Motte}, F., {Csengeri}, T., \& {Schneider}, N. 2010, \aap,
  524, A18

\bibitem[{{Busquet} {et~al.}(2016){Busquet}, {Estalella}, {Palau}, {Liu},
  {Zhang}, {Girart}, {de Gregorio-Monsalvo}, {Pillai}, {Anglada}, \&
  {Ho}}]{busquet2016}
{Busquet}, G., {Estalella}, R., {Palau}, A., {et~al.} 2016, \apj, 819, 139

\bibitem[{{Busquet} {et~al.}(2013){Busquet}, {Zhang}, {Palau}, {Liu},
  {S{\'a}nchez-Monge}, {Estalella}, {Ho}, {de Gregorio-Monsalvo}, {Pillai},
  {Wyrowski}, {Girart}, {Santos}, \& {Franco}}]{busquet2013}
{Busquet}, G., {Zhang}, Q., {Palau}, A., {et~al.} 2013, \apjl, 764, L26

\bibitem[{{Camacho} {et~al.}(2020){Camacho}, {V{\'a}zquez-Semadeni}, {Palau},
  {Busquet}, \& {Zamora-Avil{\'e}s}}]{camacho2020}
{Camacho}, V., {V{\'a}zquez-Semadeni}, E., {Palau}, A., {Busquet}, G., \&
  {Zamora-Avil{\'e}s}, M. 2020, \apj, 903, 46

\bibitem[{{Carey} {et~al.}(2009){Carey}, {Noriega-Crespo}, {Mizuno}, {Shenoy},
  {Paladini}, {Kraemer}, {Price}, {Flagey}, {Ryan}, {Ingalls}, {Kuchar},
  {Pinheiro Gon{\c{c}}alves}, {Indebetouw}, {Billot}, {Marleau}, {Padgett},
  {Rebull}, {Bressert}, {Ali}, {Molinari}, {Martin}, {Berriman}, {Boulanger},
  {Latter}, {Miville-Deschenes}, {Shipman}, \& {Testi}}]{carey2009}
{Carey}, S.~J., {Noriega-Crespo}, A., {Mizuno}, D.~R., {et~al.} 2009, \pasp,
  121, 76

\bibitem[{{Carrasco-Gonz{\'a}lez} {et~al.}(2010){Carrasco-Gonz{\'a}lez},
  {Rodr{\'\i}guez}, {Anglada}, {Mart{\'\i}}, {Torrelles}, \&
  {Osorio}}]{carrasco-gonzalez2010}
{Carrasco-Gonz{\'a}lez}, C., {Rodr{\'\i}guez}, L.~F., {Anglada}, G., {et~al.}
  2010, Science, 330, 1209

\bibitem[{{Chen} {et~al.}(2019){Chen}, {Zhang}, {Wright}, {Busquet}, {Lin},
  {Liu}, {Olguin}, {Sanhueza}, {Nakamura}, {Palau}, {Ohashi}, {Tatematsu}, \&
  {Liao}}]{chen2019}
{Chen}, H.-R.~V., {Zhang}, Q., {Wright}, M.~C.~H., {et~al.} 2019, \apj, 875, 24

\bibitem[{{Chini} {et~al.}(1980){Chini}, {Elsaesser}, \& {Neckel}}]{chini1980}
{Chini}, R., {Elsaesser}, H., \& {Neckel}, T. 1980, \aap, 91, 186

\bibitem[{{Cornwell} {et~al.}(2008){Cornwell}, {Golap}, \&
  {Bhatnagar}}]{Cornwell2008}
{Cornwell}, T.~J., {Golap}, K., \& {Bhatnagar}, S. 2008, IEEE Journal of
  Selected Topics in Signal Processing, 2, 647

\bibitem[{{Coutens} {et~al.}(2019){Coutens}, {Liu}, {Jim{\'e}nez-Serra},
  {Bourke}, {Forbrich}, {Hoare}, {Loinard}, {Testi}, {Audard}, {Caselli},
  {Chac{\'o}n-Tanarro}, {Codella}, {Di Francesco}, {Fontani}, {Hogerheijde},
  {Johansen}, {Johnstone}, {Maddison}, {Pani{\'c}}, {P{\'e}rez}, {Podio},
  {Punanova}, {Rawlings}, {Semenov}, {Tazzari}, {Tobin}, {van der Wiel}, {van
  Langevelde}, {Vlemmings}, {Walsh}, \& {Wilner}}]{coutens2019}
{Coutens}, A., {Liu}, H.~B., {Jim{\'e}nez-Serra}, I., {et~al.} 2019, \aap, 631,
  A58

\bibitem[{Cutri {et~al.}(2003)Cutri, Skrutskie, Dyk, Beichman, Carpenter,
  Chester, Cambresy, Evans, Fowler, Gizis, Howard, Huchra, Jarrett, Kopan,
  Kirkpatrick, Light, Marsh, McCallon, Schneider, \& Zacarias}]{Cutri2003}
Cutri, R., Skrutskie, M., Dyk, S., {et~al.} 2003, -1

\bibitem[{{Cutri} {et~al.}(2013){Cutri}, {Wright}, {Conrow}, {Fowler},
  {Eisenhardt}, {Grillmair}, {Kirkpatrick}, {Masci}, {McCallon}, {Wheelock},
  {Fajardo-Acosta}, {Yan}, {Benford}, {Harbut}, {Jarrett}, {Lake}, {Leisawitz},
  {Ressler}, {Stanford}, {Tsai}, {Liu}, {Helou}, {Mainzer}, {Gettings},
  {Gonzalez}, {Hoffman}, {Marsh}, {Padgett}, {Skrutskie}, {Beck}, {Papin}, \&
  {Wittman}}]{Cutri2013}
{Cutri}, R.~M., {Wright}, E.~L., {Conrow}, T., {et~al.} 2013, {Explanatory
  Supplement to the AllWISE Data Release Products}, Explanatory Supplement to
  the AllWISE Data Release Products

\bibitem[{{De Pree} {et~al.}(2014){De Pree}, {Peters}, {Mac Low}, {Wilner},
  {Goss}, {Galv{\'a}n-Madrid}, {Keto}, {Klessen}, \& {Monsrud}}]{depree2014}
{De Pree}, C.~G., {Peters}, T., {Mac Low}, M.~M., {et~al.} 2014, \apjl, 781,
  L36

\bibitem[{{Deller} {et~al.}(2013){Deller}, {Forbrich}, \&
  {Loinard}}]{deller2013}
{Deller}, A.~T., {Forbrich}, J., \& {Loinard}, L. 2013, \aap, 552, A51

\bibitem[{{Diaz-Miller} {et~al.}(1998){Diaz-Miller}, {Franco}, \&
  {Shore}}]{diaz-miller98}
{Diaz-Miller}, R.~I., {Franco}, J., \& {Shore}, S.~N. 1998, \apj, 501, 192

\bibitem[{Dzib {et~al.}(2010)Dzib, Loinard, Mioduszewski, Boden, Rodr{\'{\i}
  }guez, \& Torres}]{Dzib_2010}
Dzib, S., Loinard, L., Mioduszewski, A.~J., {et~al.} 2010, The Astrophysical
  Journal, 718, 610

\bibitem[{{Dzib} {et~al.}(2013){Dzib}, {Loinard}, {Mioduszewski},
  {Rodr{\'\i}guez}, {Ortiz-Le{\'o}n}, {Pech}, {Rivera}, {Torres}, {Boden},
  {Hartmann}, {Evans}, {Brice{\~n}o}, \& {Tobin}}]{dzib2013}
{Dzib}, S.~A., {Loinard}, L., {Mioduszewski}, A.~J., {et~al.} 2013, \apj, 775,
  63

\bibitem[{{Dzib} {et~al.}(2015){Dzib}, {Loinard}, {Rodr{\'\i}guez},
  {Mioduszewski}, {Ortiz-Le{\'o}n}, {Kounkel}, {Pech}, {Rivera}, {Torres},
  {Boden}, {Hartmann}, {Evans}, {Brice{\~n}o}, \& {Tobin}}]{dzib2015}
{Dzib}, S.~A., {Loinard}, L., {Rodr{\'\i}guez}, L.~F., {et~al.} 2015, \apj,
  801, 91

\bibitem[{{Dzib} {et~al.}(2023){Dzib}, {Yang}, {Urquhart}, {Medina},
  {Brunthaler}, {Menten}, {Wyrowski}, {Cotton}, {Dokara}, {Ortiz-Le{\'o}n},
  {Rugel}, {Nguyen}, {Gong}, {Chakraborty}, {Beuther}, {Billington},
  {Carrasco-Gonzalez}, {Csengeri}, {Hofner}, {Ott}, {Pandian}, {Roy}, \&
  {Yanza}}]{dzib2023}
{Dzib}, S.~A., {Yang}, A.~Y., {Urquhart}, J.~S., {et~al.} 2023, \aap, 670, A9

\bibitem[{{Elmegreen} \& {Lada}(1976)}]{elmegreen1976}
{Elmegreen}, B.~G. \& {Lada}, C.~J. 1976, \aj, 81, 1089

\bibitem[{{Elmegreen} {et~al.}(1979){Elmegreen}, {Lada}, \&
  {Dickinson}}]{elmegreen1979}
{Elmegreen}, B.~G., {Lada}, C.~J., \& {Dickinson}, D.~F. 1979, \apj, 230, 415

\bibitem[{{Feigelson} \& {Montmerle}(1985)}]{Feigelson1985}
{Feigelson}, E.~D. \& {Montmerle}, T. 1985, \apjl, 289, L19

\bibitem[{Forbrich {et~al.}(2007)Forbrich, Massi, Ros, Brunthaler, \&
  Menten}]{Forbrich_2007}
Forbrich, J., Massi, M., Ros, E., Brunthaler, A., \& Menten, K.~M. 2007, \aap,
  469, 985

\bibitem[{{Forbrich} {et~al.}(2016){Forbrich}, {Rivilla}, {Menten}, {Reid},
  {Chandler}, {Rau}, {Bhatnagar}, {Wolk}, \& {Meingast}}]{forbrich2016}
{Forbrich}, J., {Rivilla}, V.~M., {Menten}, K.~M., {et~al.} 2016, \apj, 822, 93

\bibitem[{Green {et~al.}(2010)Green, Caswell, Fuller, Avison, Breen, Ellingsen,
  Gray, Pestalozzi, Quinn, Thompson, \& Voronkov}]{Green_2010}
Green, J.~A., Caswell, J.~L., Fuller, G.~A., {et~al.} 2010, Monthly Notices of
  the Royal Astronomical Society, 409, 913

\bibitem[{{G{\"u}del}(2002)}]{Gudel2002}
{G{\"u}del}, M. 2002, \araa, 40, 217

\bibitem[{{Guedel} \& {Benz}(1993)}]{Gudel1993}
{Guedel}, M. \& {Benz}, A.~O. 1993, \apjl, 405, L63

\bibitem[{{Gutermuth} {et~al.}(2009){Gutermuth}, {Megeath}, {Myers}, {Allen},
  {Pipher}, \& {Fazio}}]{gutermuth2009}
{Gutermuth}, R.~A., {Megeath}, S.~T., {Myers}, P.~C., {et~al.} 2009, \apjs,
  184, 18

\bibitem[{{Gvaramadze} {et~al.}(2017){Gvaramadze}, {Mackey}, {Kniazev},
  {Langer}, {Chen{\'e}}, {Castro}, {Haworth}, \& {Grebel}}]{Gvaramadze2017}
{Gvaramadze}, V.~V., {Mackey}, J., {Kniazev}, A.~Y., {et~al.} 2017, \mnras,
  466, 1857

\bibitem[{{Hoffmeister} {et~al.}(2008){Hoffmeister}, {Chini}, {Scheyda},
  {Schulze}, {Watermann}, {N{\"u}rnberger}, \& {Vogt}}]{hoffmeister2008}
{Hoffmeister}, V.~H., {Chini}, R., {Scheyda}, C.~M., {et~al.} 2008, \apj, 686,
  310

\bibitem[{{Hofner} {et~al.}(2017){Hofner}, {Cesaroni}, {Kurtz}, {Rosero},
  {Anderson}, {Furuya}, {Araya}, \& {Molinari}}]{hofner2017}
{Hofner}, P., {Cesaroni}, R., {Kurtz}, S., {et~al.} 2017, \apj, 843, 99

\bibitem[{{Irabor} {et~al.}(2023){Irabor}, {Hoare}, {Burton}, {Cotton},
  {Diamond}, {Dougherty}, {Ellingsen}, {Fender}, {Fuller}, {Garrington},
  {Goldsmith}, {Green}, {Gunn}, {Jackson}, {Kurtz}, {Lumsden}, {Marti},
  {McDonald}, {Molinari}, {Moore}, {Mutale}, {Muxlow}, {O'Brien}, {Oudmaijer},
  {Paladini}, {Pandian}, {Paredes}, {Richards}, {Sanchez-Monge}, {Spencer},
  {Thompson}, {Umana}, {Urquhart}, {Wieringa}, \& {Zijlstra}}]{irabor2023}
{Irabor}, T., {Hoare}, M.~G., {Burton}, M., {et~al.} 2023, \mnras, 520, 1073

\bibitem[{{Ishihara} {et~al.}(2010){Ishihara}, {Onaka}, {Kataza}, {Salama},
  {Alfageme}, {Cassatella}, {Cox}, {Garc{\'\i}a-Lario}, {Stephenson}, {Cohen},
  {Fujishiro}, {Fujiwara}, {Hasegawa}, {Ita}, {Kim}, {Matsuhara}, {Murakami},
  {M{\"u}ller}, {Nakagawa}, {Ohyama}, {Oyabu}, {Pyo}, {Sakon}, {Shibai},
  {Takita}, {Tanab{\'e}}, {Uemizu}, {Ueno}, {Usui}, {Wada}, {Watarai},
  {Yamamura}, \& {Yamauchi}}]{akari2010}
{Ishihara}, D., {Onaka}, T., {Kataza}, H., {et~al.} 2010, \aap, 514, A1

\bibitem[{{Jaffe} {et~al.}(1981){Jaffe}, {Guesten}, \& {Downes}}]{jaffe1981}
{Jaffe}, D.~T., {Guesten}, R., \& {Downes}, D. 1981, \apj, 250, 621

\bibitem[{{Kavak} {et~al.}(2021){Kavak}, {S{\'a}nchez-Monge},
  {L{\'o}pez-Sepulcre}, {Cesaroni}, {van der Tak}, {Moscadelli}, {Beltr{\'a}n},
  \& {Schilke}}]{Kavak2021}
{Kavak}, {\"U}., {S{\'a}nchez-Monge}, {\'A}., {L{\'o}pez-Sepulcre}, A.,
  {et~al.} 2021, \aap, 645, A29

\bibitem[{{Kounkel} {et~al.}(2014){Kounkel}, {Hartmann}, {Loinard},
  {Mioduszewski}, {Dzib}, {Ortiz-Le{\'o}n}, {Rodr{\'\i}guez}, {Pech}, {Rivera},
  {Torres}, {Boden}, {Evans}, {Brice{\~n}o}, \& {Tobin}}]{kounkel2014}
{Kounkel}, M., {Hartmann}, L., {Loinard}, L., {et~al.} 2014, \apj, 790, 49

\bibitem[{{Kroupa}(2001)}]{kroupa2001}
{Kroupa}, P. 2001, \mnras, 322, 231

\bibitem[{{Kurtz} {et~al.}(1994){Kurtz}, {Churchwell}, \& {Wood}}]{Kurtz1994}
{Kurtz}, S., {Churchwell}, E., \& {Wood}, D.~O.~S. 1994, \apjs, 91, 659

\bibitem[{{Lada}(1987)}]{lada1987}
{Lada}, C.~J. 1987, in Star Forming Regions, ed. M.~{Peimbert} \& J.~{Jugaku},
  Vol. 115, 1

\bibitem[{{Lada} \& {Lada}(2003)}]{lada2003}
{Lada}, C.~J. \& {Lada}, E.~A. 2003, \araa, 41, 57

\bibitem[{{Lin} {et~al.}(2017){Lin}, {Liu}, {Dale}, {Li}, {Busquet}, {Zhang},
  {Ginsburg}, {Galv{\'a}n-Madrid}, {Kov{\'a}cs}, {Koch}, {Qian}, {Wang},
  {Longmore}, {Chen}, \& {Walker}}]{lin2017}
{Lin}, Y., {Liu}, H.~B., {Dale}, J.~E., {et~al.} 2017, \apj, 840, 22

\bibitem[{{Liu} {et~al.}(2014){Liu}, {Galv{\'a}n-Madrid}, {Forbrich},
  {Rodr{\'\i}guez}, {Takami}, {Costigan}, {Manara}, {Yan}, {Karr}, {Chou},
  {Ho}, \& {Zhang}}]{Liu2014}
{Liu}, H.~B., {Galv{\'a}n-Madrid}, R., {Forbrich}, J., {et~al.} 2014, \apj,
  780, 155

\bibitem[{{Marti} {et~al.}(1993){Marti}, {Rodriguez}, \&
  {Reipurth}}]{marti1993}
{Marti}, J., {Rodriguez}, L.~F., \& {Reipurth}, B. 1993, \apj, 416, 208

\bibitem[{{Martins} {et~al.}(2005){Martins}, {Schaerer}, \&
  {Hillier}}]{martins2005}
{Martins}, F., {Schaerer}, D., \& {Hillier}, D.~J. 2005, \aap, 436, 1049

\bibitem[{{Medina} {et~al.}(2018){Medina}, {Dzib}, {Tapia}, {Rodr{\'\i}guez},
  \& {Loinard}}]{medina2018}
{Medina}, S. N.~X., {Dzib}, S.~A., {Tapia}, M., {Rodr{\'\i}guez}, L.~F., \&
  {Loinard}, L. 2018, \aap, 610, A27

\bibitem[{{Meng} {et~al.}(2019){Meng}, {S{\'a}nchez-Monge}, {Schilke},
  {Padovani}, {Marcowith}, {Ginsburg}, {Schmiedeke}, {Schw{\"o}rer}, {DePree},
  {Veena}, \& {M{\"o}ller}}]{meng2019}
{Meng}, F., {S{\'a}nchez-Monge}, {\'A}., {Schilke}, P., {et~al.} 2019, \aap,
  630, A73

\bibitem[{{Moscadelli} {et~al.}(2016){Moscadelli}, {S{\'a}nchez-Monge},
  {Goddi}, {Li}, {Sanna}, {Cesaroni}, {Pestalozzi}, {Molinari}, \&
  {Reid}}]{moscadelli2016}
{Moscadelli}, L., {S{\'a}nchez-Monge}, {\'A}., {Goddi}, C., {et~al.} 2016,
  \aap, 585, A71

\bibitem[{{Motte} {et~al.}(2018){Motte}, {Bontemps}, \& {Louvet}}]{motte2018}
{Motte}, F., {Bontemps}, S., \& {Louvet}, F. 2018, \araa, 56, 41

\bibitem[{{Ohashi} {et~al.}(2016){Ohashi}, {Sanhueza}, {Chen}, {Zhang},
  {Busquet}, {Nakamura}, {Palau}, \& {Tatematsu}}]{ohashi2016}
{Ohashi}, S., {Sanhueza}, P., {Chen}, H.-R.~V., {et~al.} 2016, \apj, 833, 209

\bibitem[{{Ortiz-Le{\'o}n} {et~al.}(2015){Ortiz-Le{\'o}n}, {Loinard},
  {Mioduszewski}, {Dzib}, {Rodr{\'\i}guez}, {Pech}, {Rivera}, {Torres},
  {Boden}, {Hartmann}, {Evans}, {Brice{\~n}o}, {Tobin}, {Kounkel}, \&
  {Gonz{\'a}lez-L{\'o}pezlira}}]{ortiz-leon2015}
{Ortiz-Le{\'o}n}, G.~N., {Loinard}, L., {Mioduszewski}, A.~J., {et~al.} 2015,
  \apj, 805, 9

\bibitem[{{Osorio} {et~al.}(2017){Osorio}, {D{\'\i}az-Rodr{\'\i}guez},
  {Anglada}, {Megeath}, {Rodr{\'\i}guez}, {Tobin}, {Stutz}, {Furlan},
  {Fischer}, {Manoj}, {G{\'o}mez}, {Gonz{\'a}lez-Garc{\'\i}a}, {Stanke},
  {Watson}, {Loinard}, {Vavrek}, \& {Carrasco-Gonz{\'a}lez}}]{osorio2017}
{Osorio}, M., {D{\'\i}az-Rodr{\'\i}guez}, A.~K., {Anglada}, G., {et~al.} 2017,
  \apj, 840, 36

\bibitem[{{Padovani} {et~al.}(2019){Padovani}, {Marcowith},
  {S{\'a}nchez-Monge}, {Meng}, \& {Schilke}}]{padovani2019}
{Padovani}, M., {Marcowith}, A., {S{\'a}nchez-Monge}, {\'A}., {Meng}, F., \&
  {Schilke}, P. 2019, \aap, 630, A72

\bibitem[{{Palagi} {et~al.}(1993){Palagi}, {Cesaroni}, {Comoretto}, {Felli}, \&
  {Natale}}]{palagi1993}
{Palagi}, F., {Cesaroni}, R., {Comoretto}, G., {Felli}, M., \& {Natale}, V.
  1993, \aaps, 101, 153

\bibitem[{{Panagia}(1973)}]{panagia1973}
{Panagia}, N. 1973, \aj, 78, 929

\bibitem[{{Pech} {et~al.}(2016){Pech}, {Loinard}, {Dzib}, {Mioduszewski},
  {Rodr{\'\i}guez}, {Ortiz-Le{\'o}n}, {Rivera}, {Torres}, {Boden}, {Hartmann},
  {Kounkel}, {Evans}, {Brice{\~n}o}, {Tobin}, \& {Zapata}}]{pech2016}
{Pech}, G., {Loinard}, L., {Dzib}, S.~A., {et~al.} 2016, \apj, 818, 116

\bibitem[{{Plunkett} {et~al.}(2018){Plunkett}, {Fern{\'a}ndez-L{\'o}pez},
  {Arce}, {Busquet}, {Mardones}, \& {Dunham}}]{plunkett2018}
{Plunkett}, A.~L., {Fern{\'a}ndez-L{\'o}pez}, M., {Arce}, H.~G., {et~al.} 2018,
  \aap, 615, A9

\bibitem[{{Povich} {et~al.}(2009){Povich}, {Churchwell}, {Bieging}, {Kang},
  {Whitney}, {Brogan}, {Kulesa}, {Cohen}, {Babler}, {Indebetouw}, {Meade}, \&
  {Robitaille}}]{povich2009}
{Povich}, M.~S., {Churchwell}, E., {Bieging}, J.~H., {et~al.} 2009, \apj, 696,
  1278

\bibitem[{{Povich} {et~al.}(2016){Povich}, {Townsley}, {Robitaille}, {Broos},
  {Orbin}, {King}, {Naylor}, \& {Whitney}}]{povich2016}
{Povich}, M.~S., {Townsley}, L.~K., {Robitaille}, T.~P., {et~al.} 2016, \apj,
  825, 125

\bibitem[{{Povich} \& {Whitney}(2010)}]{povich2010}
{Povich}, M.~S. \& {Whitney}, B.~A. 2010, \apjl, 714, L285

\bibitem[{{Pudritz}(2002)}]{pudritz2002}
{Pudritz}, R.~E. 2002, Science, 295, 68

\bibitem[{Purcell {et~al.}(2013)Purcell, Hoare, Cotton, Lumsden, Urquhart,
  Chandler, Churchwell, Diamond, Dougherty, Fender, Fuller, Garrington,
  Gledhill, Goldsmith, Hindson, Jackson, Kurtz, Mart{\'{\i} }, Moore, Mundy,
  Muxlow, Oudmaijer, Pandian, Paredes, Shepherd, Smethurst, Spencer, Thompson,
  Umana, \& Zijlstra}]{Purcell_2013}
Purcell, C.~R., Hoare, M.~G., Cotton, W.~D., {et~al.} 2013, The Astrophysical
  Journal Supplement Series, 205, 1

\bibitem[{{Purser} {et~al.}(2021){Purser}, {Lumsden}, {Hoare}, \&
  {Kurtz}}]{purser2021}
{Purser}, S.~J.~D., {Lumsden}, S.~L., {Hoare}, M.~G., \& {Kurtz}, S. 2021,
  \mnras, 504, 338

\bibitem[{{Reid} {et~al.}(1995){Reid}, {Argon}, {Masson}, {Menten}, \&
  {Moran}}]{reid1995}
{Reid}, M.~J., {Argon}, A.~L., {Masson}, C.~R., {Menten}, K.~M., \& {Moran},
  J.~M. 1995, \apj, 443, 238

\bibitem[{{Rivilla} {et~al.}(2015){Rivilla}, {Chandler}, {Sanz-Forcada},
  {Jim{\'e}nez-Serra}, {Forbrich}, \& {Mart{\'\i}n-Pintado}}]{rivilla2015}
{Rivilla}, V.~M., {Chandler}, C.~J., {Sanz-Forcada}, J., {et~al.} 2015, \apj,
  808, 146

\bibitem[{{Robitaille} {et~al.}(2006){Robitaille}, {Whitney}, {Indebetouw},
  {Wood}, \& {Denzmore}}]{robitaille2006}
{Robitaille}, T.~P., {Whitney}, B.~A., {Indebetouw}, R., {Wood}, K., \&
  {Denzmore}, P. 2006, \apjs, 167, 256

\bibitem[{{Rodr{\'\i}guez}(1999)}]{rodriguez1999}
{Rodr{\'\i}guez}, L.~F. 1999, in Star Formation 1999, ed. T.~{Nakamoto},
  257--262

\bibitem[{{Rodriguez} {et~al.}(1989){Rodriguez}, {Curiel}, {Moran}, {Mirabel},
  {Roth}, \& {Garay}}]{rodriguez1989}
{Rodriguez}, L.~F., {Curiel}, S., {Moran}, J.~M., {et~al.} 1989, \apjl, 346,
  L85

\bibitem[{{Rodr{\'\i}guez} {et~al.}(2005){Rodr{\'\i}guez}, {Garay}, {Brooks},
  \& {Mardones}}]{rodriguez2005}
{Rodr{\'\i}guez}, L.~F., {Garay}, G., {Brooks}, K.~J., \& {Mardones}, D. 2005,
  \apj, 626, 953

\bibitem[{{Rodr{\'\i}guez} {et~al.}(2012){Rodr{\'\i}guez}, {Gonz{\'a}lez},
  {Montes}, {Asiri}, {Raga}, \& {Cant{\'o}}}]{rodriguez2012}
{Rodr{\'\i}guez}, L.~F., {Gonz{\'a}lez}, R.~F., {Montes}, G., {et~al.} 2012,
  \apj, 755, 152

\bibitem[{{Rodr{\'\i}guez-Kamenetzky}
  {et~al.}(2017){Rodr{\'\i}guez-Kamenetzky}, {Carrasco-Gonz{\'a}lez}, {Araudo},
  {Romero}, {Torrelles}, {Rodr{\'\i}guez}, {Anglada}, {Mart{\'\i}}, {Perucho},
  \& {Valotto}}]{rodriguez-kamenetzk2017}
{Rodr{\'\i}guez-Kamenetzky}, A., {Carrasco-Gonz{\'a}lez}, C., {Araudo}, A.,
  {et~al.} 2017, \apj, 851, 16

\bibitem[{{Rodr{\'\i}guez-Kamenetzky}
  {et~al.}(2016){Rodr{\'\i}guez-Kamenetzky}, {Carrasco-Gonz{\'a}lez}, {Araudo},
  {Torrelles}, {Anglada}, {Mart{\'\i}}, {Rodr{\'\i}guez}, \&
  {Valotto}}]{rodriguez-kamenetzky2016}
{Rodr{\'\i}guez-Kamenetzky}, A., {Carrasco-Gonz{\'a}lez}, C., {Araudo}, A.,
  {et~al.} 2016, \apj, 818, 27

\bibitem[{{Rodr{\'\i}guez-Kamenetzky}
  {et~al.}(2019){Rodr{\'\i}guez-Kamenetzky}, {Carrasco-Gonz{\'a}lez},
  {Gonz{\'a}lez-Mart{\'\i}n}, {Araudo}, {Rodr{\'\i}guez}, {Vig}, \&
  {Hofner}}]{rodriguez-kamenetzky2019}
{Rodr{\'\i}guez-Kamenetzky}, A., {Carrasco-Gonz{\'a}lez}, C.,
  {Gonz{\'a}lez-Mart{\'\i}n}, O., {et~al.} 2019, \mnras, 482, 4687

\bibitem[{{Rosero} {et~al.}(2016){Rosero}, {Hofner}, {Claussen}, {Kurtz},
  {Cesaroni}, {Araya}, {Carrasco-Gonz{\'a}lez}, {Rodr{\'\i}guez}, {Menten},
  {Wyrowski}, {Loinard}, \& {Ellingsen}}]{rosero2016}
{Rosero}, V., {Hofner}, P., {Claussen}, M., {et~al.} 2016, \apjs, 227, 25

\bibitem[{{Rosero} {et~al.}(2019){Rosero}, {Hofner}, {Kurtz}, {Cesaroni},
  {Carrasco-Gonz{\'a}lez}, {Araya}, {Rodr{\'\i}guez}, {Menten}, {Wyrowski},
  {Loinard}, {Ellingsen}, \& {Molinari}}]{rosero2019}
{Rosero}, V., {Hofner}, P., {Kurtz}, S., {et~al.} 2019, \apj, 880, 99

\bibitem[{{Salpeter}(1955)}]{salpeter1955}
{Salpeter}, E.~E. 1955, \apj, 121, 161

\bibitem[{{S{\'a}nchez-Monge} {et~al.}(2013{\natexlab{a}}){S{\'a}nchez-Monge},
  {Beltr{\'a}n}, {Cesaroni}, {Fontani}, {Brand}, {Molinari}, {Testi}, \&
  {Burton}}]{sanchez-monge2013}
{S{\'a}nchez-Monge}, {\'A}., {Beltr{\'a}n}, M.~T., {Cesaroni}, R., {et~al.}
  2013{\natexlab{a}}, \aap, 550, A21

\bibitem[{{S{\'a}nchez-Monge} {et~al.}(2013{\natexlab{b}}){S{\'a}nchez-Monge},
  {Kurtz}, {Palau}, {Estalella}, {Shepherd}, {Lizano}, {Franco}, \&
  {Garay}}]{sanchez-monge2013b}
{S{\'a}nchez-Monge}, {\'A}., {Kurtz}, S., {Palau}, A., {et~al.}
  2013{\natexlab{b}}, \apj, 766, 114

\bibitem[{{Sanna} {et~al.}(2019){Sanna}, {Moscadelli}, {Goddi}, {Beltr{\'a}n},
  {Brogan}, {Caratti o Garatti}, {Carrasco-Gonz{\'a}lez}, {Hunter}, {Massi}, \&
  {Padovani}}]{sanna2019}
{Sanna}, A., {Moscadelli}, L., {Goddi}, C., {et~al.} 2019, \aap, 623, L3

\bibitem[{{Spitzer Science}(2009)}]{Spitzer2009}
{Spitzer Science}, C. 2009, VizieR Online Data Catalog, II/293

\bibitem[{{Sugiyama} {et~al.}(2017){Sugiyama}, {Nagase}, {Yonekura}, {Momose},
  {Yasui}, {Saito}, {Motogi}, {Honma}, {Hachisuka}, {Matsumoto}, {Uchiyama}, \&
  {Fujisawa}}]{sugiyama2017}
{Sugiyama}, K., {Nagase}, K., {Yonekura}, Y., {et~al.} 2017, \pasj, 69, 59

\bibitem[{{Thompson}(1984)}]{thompson84}
{Thompson}, R.~I. 1984, \apj, 283, 165

\bibitem[{{Vacca} {et~al.}(1996){Vacca}, {Garmany}, \& {Shull}}]{vacca96}
{Vacca}, W.~D., {Garmany}, C.~D., \& {Shull}, J.~M. 1996, \apj, 460, 914

\bibitem[{{Vargas-Gonz{\'a}lez} {et~al.}(2021){Vargas-Gonz{\'a}lez},
  {Forbrich}, {Dzib}, \& {Bally}}]{vargas-gonzalez2021}
{Vargas-Gonz{\'a}lez}, J., {Forbrich}, J., {Dzib}, S.~A., \& {Bally}, J. 2021,
  \mnras, 506, 3169

\bibitem[{{V{\'a}zquez-Semadeni} {et~al.}(2017){V{\'a}zquez-Semadeni},
  {Gonz{\'a}lez-Samaniego}, \& {Col{\'\i}n}}]{vazquez-semadeni2017}
{V{\'a}zquez-Semadeni}, E., {Gonz{\'a}lez-Samaniego}, A., \& {Col{\'\i}n}, P.
  2017, \mnras, 467, 1313

\bibitem[{{V{\'a}zquez-Semadeni} {et~al.}(2019){V{\'a}zquez-Semadeni}, {Palau},
  {Ballesteros-Paredes}, {G{\'o}mez}, \&
  {Zamora-Avil{\'e}s}}]{vazquez-semadeni2019}
{V{\'a}zquez-Semadeni}, E., {Palau}, A., {Ballesteros-Paredes}, J.,
  {G{\'o}mez}, G.~C., \& {Zamora-Avil{\'e}s}, M. 2019, \mnras, 490, 3061

\bibitem[{{Wang} {et~al.}(2006){Wang}, {Zhang}, {Rathborne}, {Jackson}, \&
  {Wu}}]{wang2006}
{Wang}, Y., {Zhang}, Q., {Rathborne}, J.~M., {Jackson}, J., \& {Wu}, Y. 2006,
  \apjl, 651, L125

\bibitem[{{Wu} {et~al.}(2014){Wu}, {Sato}, {Reid}, {Moscadelli}, {Zhang}, {Xu},
  {Brunthaler}, {Menten}, {Dame}, \& {Zheng}}]{wu2014}
{Wu}, Y.~W., {Sato}, M., {Reid}, M.~J., {et~al.} 2014, \aap, 566, A17

\bibitem[{{Xu} {et~al.}(2011){Xu}, {Moscadelli}, {Reid}, {Menten}, {Zhang},
  {Zheng}, \& {Brunthaler}}]{xu2011}
{Xu}, Y., {Moscadelli}, L., {Reid}, M.~J., {et~al.} 2011, \apj, 733, 25

\bibitem[{{Yanza} {et~al.}(2022){Yanza}, {Masqu{\'e}}, {Dzib},
  {Rodr{\'\i}guez}, {Medina}, {Kurtz}, {Loinard}, {Trinidad}, {Menten}, \&
  {Rodr{\'\i}guez-Rico}}]{yanza22}
{Yanza}, V., {Masqu{\'e}}, J.~M., {Dzib}, S.~A., {et~al.} 2022, \aj, 163, 276

\bibitem[{{Zapata} {et~al.}(2004){Zapata}, {Rodr{\'\i}guez}, {Kurtz}, \&
  {O'Dell}}]{zapata2004}
{Zapata}, L.~A., {Rodr{\'\i}guez}, L.~F., {Kurtz}, S.~E., \& {O'Dell}, C.~R.
  2004, \aj, 127, 2252

\bibitem[{{Zucker} {et~al.}(2020){Zucker}, {Speagle}, {Schlafly}, {Green},
  {Finkbeiner}, {Goodman}, \& {Alves}}]{Zucker2020}
{Zucker}, C., {Speagle}, J.~S., {Schlafly}, E.~F., {et~al.} 2020, \aap, 633,
  A51

\end{thebibliography}

\begin{appendix}

\section{Variable source subtraction}
\label{appendix:AppA}

The presence of the highly variable sources during observations generates artifacts with a psf-like pattern that cannot be properly removed. In order to use the \textit{tclean} task correctly, the substraction of the sources had to be done before. One of the sources is located in G14.2-N at the C-band, while the other is in G14.2-S at the X-band. The following subsections detail the two subtraction processes.

\subsection{G14.2-N C-band}
The variable source in G14.2-N is located at R.A.(J2000)=18$^{\mathrm{h}}$18$^{\mathrm{m}}$05.651$^{\mathrm{s}}$ and Dec.(J2000)=$-$16\grau52$'$56.87$''$. The flux of this source changes from $0.235$~mJy to $1.686$~mJy from the data collected on September 25 to the one collected on September 26, which is a variation of a factor of 6. 
Therefore, we need to subtract the source before combining the visibilities of both days to make the final map.
First of all, we tried to subtract the source using the CASA \textit{uvsub} command, which worked correctly for 25th September but had some significant residual flux for the 26th of September. 
We applied the self-calibration to the data of 26th September and then we used the same command to subtract the source.
After that we combined both datasets (the 25th of September and the 26th of September) to obtain the final image.

\subsection{G14.2-S X-band}

The source we need to subtract in G14.2-S is located at R.A.(J2000)=18$^{\mathrm{h}}$18$^{\mathrm{m}}$02.910$^{\mathrm{s}}$ and Dec.(J2000)=$-$17\grau05$'$40.73$''$. We had to use a different method for this source since is located outside the field of view of the VLA observations. The first step was to generate an image with a larger field of view which had to include the source. This process needed too much computer processing time, so we used an alternative way, which consisted of 
creating a small image around the source, using the CASA \textit{tclean} option called \textit{outlier}. For this small image, we defined a mask around the source, avoiding the lobes of the dirty beam (psf), that is used to constrain the cleaning algorithm where the true emission is located. 

At the X-band observations, the source had a variability of more than one order of magnitude and therefore we generated images for the different days independently. 
The next step was to subtract the source using the CASA \textit{uvsub} command like in the G14.2-N C-band variable source, but as this source is far from the phase center, a geometric correction for the curvature had to be done before. This correction takes into account the non-coplanarity of the baselines as function of the distance from the phase center which results in a non zero w-term. This results in a phase term that limits the dynamic range of the final image. This correction projects the sky curvature onto many smaller planes which is known as w-projection \citep[see][]{Cornwell2008}. The source was subtracted for every observation day and then we combined all the days to get the final image.

\section{X Band recentering}
\label{appendix:AppB}

For the G14.2-N X-band data, it was necessary to realign the data for each day by recentering it due to the discrepancy in the positions of all the sources. This mismatch was detected while viewing the map of every day of observation individually.

In order to correct for this mismatch all the data had to be recentered. The first step was to get the positions of bright sources that had a counterpart in our G14.2-N C-band image. With these positions, the next step was to check which day was the one where the measured positions were the closest to the peak position of the C-band counterparts.
Then we used the clean components of the brightest sources from the selected image as a model for self-calibrating the other days data. Phase corrections obtained for each day corrects for atmospheric changes during observations but also correct astrometry to fit the value used as a model. Finally the new self-calibrated visibilities were used to obtain a new and corrected image. 
\section{Parameters of the detected sources}\label{app_fluxes}

Each source's position, peak intensity, flux density and deconvolved size were determined by selecting a small region around each source. The uncertainties in the fluxes are the result of the quadratic sum of the statistical error of the region, i.e., the rms of the region and the uncertainty in the determination of the flux itself, which has been considered to be 10\% of it.
The sources have been sorted by declination, showing the separation between G14.2-N and G14.2-S. Table~\ref{t:g14Cband} shows the sources that have been detected at C-band, and Table \ref{t:g14Xband} shows the sources that have been detected at X-band. For unresolved point-like sources or for sources having extended and non-Gaussian emission, the peak position and integrated flux were determined by selecting a region around the emitting region. All fluxes and intensities presented are corrected for primary beam attenuation.

\begin{table*}[!ht]
\centering
\caption{Parameters of the radio sources detected at C-band.}
\label{t:g14Cband}
\begin{tabular}{lcccccc}
\hline\hline \noalign{\smallskip}

& \multicolumn{2}{c}{Position}
& \multicolumn{1}{c}{$I_{\rm peak}$}
& \multicolumn{1}{c}{$S_{\nu}$}
& \multicolumn{1}{c}{Size\tablefootmark{$a$}}
& \multicolumn{1}{c}{PA}
\\
\cline{2-3} 
Source
& R.A.(J2000) 18$^{\mathrm{h}}$
& Dec.(J2000) $-16^\circ$
& ($\mu$Jy/beam)
& \multicolumn{1}{c}{($\mu$Jy)}
& (mas)
& ($^\circ$)
\\ 
\hline \noalign{\smallskip}
\multicolumn{4}{l}{Northern region G14.2-N} \\
\hline \noalign{\smallskip}
VLA-01            & 18:13.2411 $\pm$ 0.0003 & 46:32.830 $\pm$ 0.010 & 87.7 $\pm$ 8.8 & 162.0 $\pm$ 16.4 & 446 $\pm$ 45 & 8 $\pm$ 2 \\
VLA-02            & 18:06.6380 $\pm$ 0.0030 & 47:59.598 $\pm$ 0.013 & 20.6 $\pm$ 2.3 & 21.9 $\pm$ 2.8 & 104 $\pm$ 10 & 87 $\pm$ 62 \\
VLA-03            & 18:11.6160 $\pm$ 0.0010 & 48:30.997 $\pm$ 0.006 & 296.6 $\pm$ 30.2 & 305.2 $\pm$ 32.3 & point source & \\
VLA-04            & 18:10.8910 $\pm$ 0.0070 & 48:31.332 $\pm$ 0.027 & 20.7 $\pm$ 2.1 & 17.3 $\pm$ 1.9 & point source & \\
VLA-05            & 18:05.3939 $\pm$ 0.0002 & 48:39.643 $\pm$ 0.007 & 19.8 $\pm$ 2.0 & 19.3 $\pm$ 2.0 & point source & \\
VLA-06            & 18:24.9224 $\pm$ 0.0005 & 49:00.898 $\pm$ 0.014 & 24.2 $\pm$ 2.6 & 31.7 $\pm$ 3.6 & point source & \\
VLA-07            & 18:08.8731 $\pm$ 0.0005 & 49:11.557 $\pm$ 0.015 & 32.3 $\pm$ 3.4 & 31.1 $\pm$ 3.6 & point source & \\
VLA-08            & 18:14.4820 $\pm$ 0.0010 & 49:11.684 $\pm$ 0.018 & 25.5 $\pm$ 2.6 & 42.1 $\pm$ 4.6 & 376 $\pm$ 38 & 121 $\pm$ 24 \\
VLA-09            & 18:12.4110 $\pm$ 0.0010 & 49:16.070 $\pm$ 0.029 & 13.6 $\pm$ 1.4 & 16.2 $\pm$ 1.7 & 207 $\pm$ 21 & 165 $\pm$ 86 \\
VLA-10            & 18:07.3270 $\pm$ 0.0010 & 49:17.014 $\pm$ 0.030 & 20.7 $\pm$ 2.6 & 45.3 $\pm$ 6.6 & 509 $\pm$ 51 & 109 $\pm$ 85 \\
VLA-11            & 18:12.5702 $\pm$ 0.0002 & 49:22.128 $\pm$ 0.006 & 121.3 $\pm$ 12.2 & 125.5 $\pm$ 12.7 & point source \\
VLA-12            & 18:06.8200 $\pm$ 0.0070 & 49:23.119 $\pm$ 0.022 & 27.0 $\pm$ 3.0 & 57.4 $\pm$ 6.8 & 513 $\pm$ 52 & 172 $\pm$ 14 \\
VLA-13            & 18:12.7140 $\pm$ 0.0010 & 49:27.384 $\pm$ 0.021 & 74.9 $\pm$ 9.5 & \phn98.0 $\pm$ 15.7 & point source & \\
VLA-14            & 18:12.2910 $\pm$ 0.0020 & 49:28.235 $\pm$ 0.035 & \phn50.9 $\pm$ 11.2 & \phn40.4 $\pm$ 14.4 & point source & \\
VLA-15            & 18:14.8110 $\pm$ 0.0003 & 49:28.778 $\pm$ 0.013 & 35.5 $\pm$ 3.8 & 38.66 $\pm$ 4.59 & 138 $\pm$ 14 & 12 $\pm$ 6 \\
VLA-16            & 18:12.5050 $\pm$ 0.0010 & 49:30.695 $\pm$ 0.034 & 15.0 $\pm$ 1.5 & 14.6 $\pm$ 1.6 & point source & \\
VLA-17            & 17:54.6870 $\pm$ 0.0010 & 49:42.697 $\pm$ 0.015 & 252.1 $\pm$ 25.8 & 359.1 $\pm$ 37.9 & point source & \\
VLA-18            & 18:18.9245 $\pm$ 0.0002 & 49:45.559 $\pm$ 0.009 & 21.4 $\pm$ 2.2 & 29.3 $\pm$ 3.0 & 291 $\pm$ 29 & 168 $\pm$ 5 \\
VLA-19            & 18:23.9770 $\pm$ 0.0070 & 49:54.278 $\pm$ 0.225 & 48.2 $\pm$ 5.2 & 107.5 $\pm$ 27.8 & 1093 $\pm$ 109 & 37 $\pm$ 10 \\
VLA-20            & 18:12.1986 $\pm$ 0.0003 & 50:09.309 $\pm$ 0.013 & 24.0 $\pm$ 2.5 & 21.7 $\pm$ 2.4 & point source & \\
VLA-21            & 17:59.1680 $\pm$ 0.0010 & 50:16.184 $\pm$ 0.018 & 29.1 $\pm$ 2.9 & 30.4 $\pm$ 3.1 & point source \\
VLA-22            & 18:12.2000 $\pm$ 0.0030 & 50:17.147 $\pm$ 0.012 & 36.4 $\pm$ 3.6 & 40.4 $\pm$ 4.0 & 134 $\pm$ 13 & 157 $\pm$ 1 \\
VLA-23            & 18:18.1790 $\pm$ 0.0002 & 50:20.696 $\pm$ 0.008 & 19.1 $\pm$ 2.0 & 12.9 $\pm$ 1.5 & point source & \\
VLA-24            & 18:12.6940 $\pm$ 0.0010 & 50:33.061 $\pm$ 0.018 & 38.9 $\pm$ 4.0 & 38.1 $\pm$ 4.1 & point source & \\
VLA-25            & 18:06.1755 $\pm$ 0.0004 & 50:35.942 $\pm$ 0.013 & 28.1 $\pm$ 2.9 & 30.6 $\pm$ 3.3 & point source & \\
VLA-26            & 18:25.9654 $\pm$ 0.0005 & 50:37.329 $\pm$ 0.008 & 71.5 $\pm$ 7.2 & 86.4 $\pm$ 8.9 & point source & \\
VLA-27            & 18:08.4200 $\pm$ 0.0010 & 50:49.543 $\pm$ 0.014 & 27.8 $\pm$ 2.9 & 37.2 $\pm$ 4.0 & 274 $\pm$ 27 & 42 $\pm$ 41 \\
VLA-28            & 18:06.8150 $\pm$ 0.0020 & 51:19.707 $\pm$ 0.033 & 22.3 $\pm$ 2.8 & 48.8 $\pm$ 7.1 & 509 $\pm$ 51 & 109 $\pm$ 85 \\
VLA-29            & 18:12.3152 $\pm$ 0.0005 & 51:25.829 $\pm$ 0.017 & 24.5 $\pm$ 2.5 & 24.7 $\pm$ 2.5 & point source & \\
VLA-30            & 18:09.0020 $\pm$ 0.0040 & 52:02.777 $\pm$ 0.017 & 39.9 $\pm$ 4.2 & 42.7 $\pm$ 5.0 & point source & \\
VLA-31            & 18:02.5890 $\pm$ 0.0010 & 52:45.716 $\pm$ 0.016 & 69.4 $\pm$ 7.2 & \phn96.0 $\pm$ 10.3 & 255 $\pm$ 26 & 34 $\pm$ 8 \\
\hdashline
VLA-32$^\text{S}$ & 18:05.6480 $\pm$ 0.0010 & 52:56.862 $\pm$ 0.024 & 1197.7 $\pm$ 121.5 & 2192.3 $\pm$ 225.7 & 610 $\pm$ 61 & 155 $\pm$ 1 \\
VLA-33$^\text{N}$ & 18:15.8140 $\pm$ 0.0002 & 53:32.749 $\pm$ 0.007 & 420.0 $\pm$ 42.2   & 554.0 $\pm$ 57.4 & 255 $\pm$ 25 & 174 $\pm$ 4 \\
VLA-33$^\text{S}$ &                         &                       & 304.3 $\pm$ 30.5   & 494.4 $\pm$ 49.6 &  605 $\pm$ 61 & 8 $\pm$ 1 \\
\hline \noalign{\smallskip}
\multicolumn{4}{l}{Southern region G14.2-S} \\
\hline \noalign{\smallskip}
VLA-34            & 18:13.6610 $\pm$ 0.0020 & 54:35.194 $\pm$ 0.041 & 33.5 $\pm$ 3.6 & 33.1 $\pm$ 4.0 & point source & \\
VLA-35            & 18:02.8200 $\pm$ 0.0020 & 54:58.004 $\pm$ 0.039 & 84.4 $\pm$ 9.3 & 114.2 $\pm$ 14.1 & 475 $\pm$ 47 & 135 $\pm$ 14 \\
VLA-36            & 18:10.0710 $\pm$ 0.0020 & 55:51.177 $\pm$ 0.086 & 66.7 $\pm$ 7.4 & 203.4 $\pm$ 24.0 & 1329 $\pm$ 133 & 7 $\pm$ 5 \\
VLA-37            & 18:12.2524 $\pm$ 0.0002 & 56:02.988 $\pm$ 0.008 & 123.0 $\pm$ 12.5 & 134.7 $\pm$ 14.1 & 279 $\pm$ 28 & 156 $\pm$ 42 \\
VLA-38            & 17:58.0390 $\pm$ 0.0020 & 56:25.562 $\pm$ 0.011 & 1439.3 $\pm$ 145.3 & 2384.8 $\pm$ 243.7 & 576 $\pm$ 58 & 107 $\pm$ 2 \\
VLA-40            & 18:06.5210 $\pm$ 0.0010 & 56:47.449 $\pm$ 0.014 & 57.7 $\pm$ 6.3 & 65.7 $\pm$ 8.2 & point source & \\
VLA-41            & 18:10.8736 $\pm$ 0.0005 & 56:49.420 $\pm$ 0.013 & 57.1 $\pm$ 5.9 & 59.2 $\pm$ 6.5 & point source & \\
VLA-42            & 18:11.9586 $\pm$ 0.0003 & 57:01.371 $\pm$ 0.012 & 56.9 $\pm$ 5.7 & 70.1 $\pm$ 7.1 & 428 $\pm$ 43 & 138 $\pm$ 5 \\
VLA-43            & 18:05.9117 $\pm$ 0.0030 & 57:07.138 $\pm$ 0.017 & 31.6 $\pm$ 3.3 & 34.7 $\pm$ 3.7 & point source & \\
VLA-44            & 18:13.0550 $\pm$ 0.0010 & 57:09.695 $\pm$ 0.012 & 21.2 $\pm$ 2.2 & 59.1 $\pm$ 6.5 & 1248 $\pm$ 125 & 7 $\pm$ 4 \\
VLA-46            & 18:08.7120 $\pm$ 0.0010 & 57:12.628 $\pm$ 0.014 & 92.8 $\pm$ 9.3 & \phn98.1 $\pm$ 10.0 & point source & \\
VLA-50            & 18:01.2390 $\pm$ 0.0020 & 57:15.700 $\pm$ 0.013 & 144.2 $\pm$ 14.4 & 215.7 $\pm$ 21.7 & 599 $\pm$ 60 & 94 $\pm$ 1 \\
VLA-51            & 18:10.1513 $\pm$ 0.0002 & 57:18.301 $\pm$ 0.006 & 141.3 $\pm$ 14.3 & 157.4 $\pm$ 16.2 & 304 $\pm$ 30 & 109 $\pm$ 33 \\
VLA-53            & 18:12.7680 $\pm$ 0.0010 & 57:19.485 $\pm$ 0.004 & 24.9 $\pm$ 2.6 & 35.4 $\pm$ 4.0 & 533 $\pm$ 53 & 47 $\pm$ 8 \\
VLA-57            & 18:09.1810 $\pm$ 0.0010 & 57:28.051 $\pm$ 0.017 & 22.0 $\pm$ 2.3 & 20.5 $\pm$ 2.2 & point source & \\
VLA-61            & 18:14.1490 $\pm$ 0.0010 & 57:49.089 $\pm$ 0.022 & 26.7 $\pm$ 2.7 & 31.1 $\pm$ 3.2 & point source & \\
VLA-63            & 17:58.0902 $\pm$ 0.0007 & 58:00.580 $\pm$ 0.010 & 35.2 $\pm$ 3.6 & 493.8 $\pm$ 49.5 & 29 $\pm$ 22 \\
VLA-64            & 18:05.9810 $\pm$ 0.0010 & 58:04.365 $\pm$ 0.016 & 68.1 $\pm$ 6.9 & 72.3 $\pm$ 7.5 & 116 $\pm$ 12 & 75 $\pm$ 11 \\
VLA-66            & 18:02.2220 $\pm$ 0.0030 & 59:22.112 $\pm$ 0.035 & 58.1 $\pm$ 5.8 & 66.4 $\pm$ 6.7 & 323 $\pm$ 32 & 74 $\pm$ 7 \\

\hline
\end{tabular}
\tablefoot{
The sources between the dashed lines correspond to those that have been detected at both regions.
\tablefoottext{$a$}{Sizes have been calculated as $\sqrt{\theta_\text{maj} \times \theta_\text{min}}$, where $\theta_\text{maj}$ and $\theta_\text{min}$ are the deconvolved major and minor axis, respectively.}
}
\end{table*}

\begin{table*}[t]
\centering
\caption{Parameters of the radio sources detected at X-band.}
\label{t:g14Xband}
\begin{tabular}{lcccccc}
\hline\hline \noalign{\smallskip}

& \multicolumn{2}{c}{Position}
& \multicolumn{1}{c}{$I_{\rm peak}$}
& \multicolumn{1}{c}{$S_{\nu}$}
& \multicolumn{1}{c}{Size\tablefootmark{$a$}}
& \multicolumn{1}{c}{PA}
\\
\cline{2-3} 
Source
& R.A.(J2000) 18$^{\mathrm{h}}$
& Dec.(J2000) $-16^\circ$
& ($\mu$Jy/beam)
& \multicolumn{1}{c}{($\mu$Jy)}
& (mas)
& ($^\circ$)
\\ 
\hline \noalign{\smallskip}
\multicolumn{4}{l}{Northern region G14.2-N} \\
\hline \noalign{\smallskip}
VLA-03            & 18:11.6166 $\pm$ 0.0005 & 48:31.007 $\pm$ 0.037 & 110.1 $\pm$ 11.3 & 243.3 $\pm$ 25.4 & 444 $\pm$ 44 & 171 $\pm$ 1 \\
VLA-08            & 18:14.4830 $\pm$ 0.0010 & 49:11.620 $\pm$ 0.019 & 37.3 $\pm$ 3.8 & 79.3 $\pm$ 8.1 & 459 $\pm$ 46 & 27 $\pm$ 1 \\
VLA-11            & 18:12.5698 $\pm$ 0.0002 & 49:22.110 $\pm$ 0.008 & 103.1 $\pm$ 10.4 & 129.6 $\pm$ 13.2 & 230 $\pm$ 23 & 3 $\pm$ 2 \\
VLA-13            & 18:12.7130 $\pm$ 0.0010 & 49:27.377 $\pm$ 0.026 & 46.7 $\pm$ 5.5 & 70.0 $\pm$ 9.7 & 262 $\pm$ 26 & 32 $\pm$ 12 \\
VLA-14            & 18:12.3130 $\pm$ 0.0020 & 49:27.937 $\pm$ 0.024 & 26.7 $\pm$ 9.7 & 22.9 $\pm$ 7.8 & point source & \\
VLA-20            & 18:12.1986 $\pm$ 0.0003 & 50:09.309 $\pm$ 0.013 & 13.7 $\pm$ 1.4 & 15.2 $\pm$ 1.6 & point source & \\
VLA-22            & 18:12.2000 $\pm$ 0.0010 & 50:17.140 $\pm$ 0.023 & 26.4 $\pm$ 2.6 & 37.2 $\pm$ 3.7 & 313 $\pm$ 31 & 8 $\pm$ 2 \\
VLA-24            & 18:12.6950 $\pm$ 0.0010 & 50:33.071 $\pm$ 0.020 & 17.0 $\pm$ 1.8 & 20.3 $\pm$ 2.2 & 205 $\pm$ 21 & 11 $\pm$ 12 \\
\hdashline
VLA-32$^\text{N}$ & 18:05.6530 $\pm$ 0.0020 & 52:56.872 $\pm$ 0.055 & 182.1 $\pm$ 29.2 & 1050.0 $\pm$ 195.0 & point source & \\
VLA-33$^\text{S}$ & 18:15.8104 $\pm$ 0.0003 & 53:32.798 $\pm$ 0.018 & 429.5 $\pm$ 43.2  & 600.2 $\pm$ 60.6 & point source & \\
\hline \noalign{\smallskip}
\multicolumn{4}{l}{Southern region G14.2-S} \\
\hline \noalign{\smallskip}
VLA-35            & 18:02.8150 $\pm$ 0.0012 & 54:57.921 $\pm$ 0.009 & 142.2 $\pm$ 14.4 & 161.0 $\pm$ 16.6 & 345 $\pm$ 34 & 163 $\pm$ 85 \\
VLA-36            & 18:10.0720 $\pm$ 0.0030 & 55:51.302 $\pm$ 0.071 & 64.6 $\pm$ 7.2 & 172.5 $\pm$ 20.9 & 1202 $\pm$ 120 & 178 $\pm$ 7 \\
VLA-38            & 17:58.0390 $\pm$ 0.0020 & 56:25.562 $\pm$ 0.011 & 1302.5 $\pm$ 132.0 & 1557.5 $\pm$ 160.8 & 392 $\pm$ 39 & 49 $\pm$ 13 \\
VLA-39            & 18:11.5230 $\pm$ 0.0010 & 56:27.282 $\pm$ 0.021 & 13.4 $\pm$ 1.5 & 17.3 $\pm$ 2.1 & point source & \\
VLA-40            & 18:06.5220 $\pm$ 0.0010 & 56:47.473 $\pm$ 0.007 & 70.0 $\pm$ 7.5 & 68.5 $\pm$ 8.3 & point source & \\
VLA-42            & 18:11.9610 $\pm$ 0.0010 & 57:01.432 $\pm$ 0.008 & 27.6 $\pm$ 2.8 & 35.3 $\pm$ 3.6 & 497 $\pm$ 50 & 31 $\pm$ 28 \\
VLA-44            & 18:13.0550 $\pm$ 0.0010 & 57:09.695 $\pm$ 0.012 & 20.1 $\pm$ 2.0 & 26.4 $\pm$ 2.7 & 396 $\pm$ 40 & 175 $\pm$ 1 \\
VLA-45            & 18:13.9380 $\pm$ 0.0010 & 57:11.438 $\pm$ 0.025 & 16.0 $\pm$ 1.8 & 17.4 $\pm$ 2.2 & point source & \\
VLA-46            & 18:08.7040 $\pm$ 0.0020 & 57:12.595 $\pm$ 0.008 & 46.2 $\pm$ 5.0 & 137.7 $\pm$ 15.6 & 1320 $\pm$ 132 & 43 $\pm$ 8 \\
VLA-47            & 18:10.1620 $\pm$ 0.0020 & 57:12.793 $\pm$ 0.019 & 13.3 $\pm$ 1.4 & 21.2 $\pm$ 2.2 & 709 $\pm$ 71 & 92 $\pm$ 10 \\
VLA-48            & 18:11.1420 $\pm$ 0.0010 & 57:13.015 $\pm$ 0.010 & 44.1 $\pm$ 4.4 & 45.1 $\pm$ 4.6 & 65 $\pm$ 7 & 93 $\pm$ 41 \\
VLA-49            & 18:13.8490 $\pm$ 0.0010 & 57:14.848 $\pm$ 0.013 & 23.9 $\pm$ 2.4 & 33.9 $\pm$ 3.4 & point source & \\
VLA-50            & 18:01.2405 $\pm$ 0.0005 & 57:15.652 $\pm$ 0.004 & 118.4 $\pm$ 11.9 & 135.2 $\pm$ 13.6 & 323 $\pm$ 32 & 74 $\pm$ 7 \\
VLA-51            & 18:10.1526 $\pm$ 0.0004 & 57:18.354 $\pm$ 0.003 & 98.8 $\pm$ 9.9 & 105.2 $\pm$ 10.6 & 232 $\pm$ 23 & 42 $\pm$ 15 \\
VLA-52            & 18:20.9130 $\pm$ 0.0020 & 57:19.476 $\pm$ 0.014 & 15.2 $\pm$ 1.5 & 14.1 $\pm$ 1.4 & point source & \\
VLA-53            & 18:12.7680 $\pm$ 0.0010 & 57:19.485 $\pm$ 0.004 & 17.1 $\pm$ 1.7 & 26.4 $\pm$ 2.3 & 672 $\pm$ 67 & 164 $\pm$ 5 \\
VLA-54            & 18:13.2020 $\pm$ 0.0060 & 57:22.155 $\pm$ 0.036 & 16.4 $\pm$ 1.8 & 32.3 $\pm$ 3.7 & point source & \\
VLA-55            & 18:13.3470 $\pm$ 0.0010 & 57:23.950 $\pm$ 0.004 & 28.3 $\pm$ 2.8 & 43.0 $\pm$4.3 & 682 $\pm$ 68 & 5 $\pm$ 1 \\
VLA-56            & 18:14.1640 $\pm$ 0.0020 & 57:24.866 $\pm$ 0.002 & 19.0 $\pm$ 1.9 & 16.7 $\pm$ 1.7 & point source & \\
VLA-58            & 18:12.3200 $\pm$ 0.0010 & 57:33.786 $\pm$ 0.008 & 25.6 $\pm$ 2.6 & 25.7 $\pm$ 2.6 & point source & \\
VLA-59            & 18:11.9210 $\pm$ 0.0010 & 57:42.746 $\pm$ 0.011 & 20.7 $\pm$ 2.1 & 42.3 $\pm$ 4.3 & 953 $\pm$ 95 & 174 $\pm$ 3 \\
VLA-60            & 18:09.1710 $\pm$ 0.0010 & 57:46.801 $\pm$ 0.007 & 54.6 $\pm$ 0.4 & 54.4 $\pm$ 5.5 & point source & \\
VLA-61            & 18:14.1480 $\pm$ 0.0010 & 57:49.143 $\pm$ 0.010 & 24.0 $\pm$ 2.4 & 22.8 $\pm$ 2.3 & point source & \\
VLA-62            & 18:07.7070 $\pm$ 0.0020 & 57:49.439 $\pm$ 0.018 & 21.2 $\pm$ 2.1 & 34.1 $\pm$ 3.4 & 697 $\pm$ 70 & 146 $\pm$ 1 \\
VLA-65            & 18:05.5760 $\pm$ 0.0010 & 58:13.171 $\pm$ 0.008 & 26.3 $\pm$ 2.6 & 33.7 $\pm$ 3.4 & 440 $\pm$ 44 & 27 $\pm$ 3 \\
VLA-66            & 18:02.2280 $\pm$ 0.0010 & 59:22.077 $\pm$ 0.006 & 115.5 $\pm$ 11.7 & 176.4 $\pm$ 18.0 & 671 $\pm$ 67 & 4 $\pm$ 3 \\

\hline
\end{tabular}
\tablefoot{
The sources between the dashed lines correspond to those that have been detected at both regions.
\tablefoottext{$a$}{Sizes have been calculated as $\sqrt{\theta_\text{maj} \times \theta_\text{min}}$, where $\theta_\text{maj}$ and $\theta_\text{min}$ are the deconvolved major and minor axis, respectively.}
}

\end{table*}

\section{Comments on individual sources}\label{app:comments}

\subsection{VLA-14} \label{app:vla-14}

\begin{figure*}[ht!]
    \centering
    \includegraphics[width=0.9\linewidth]{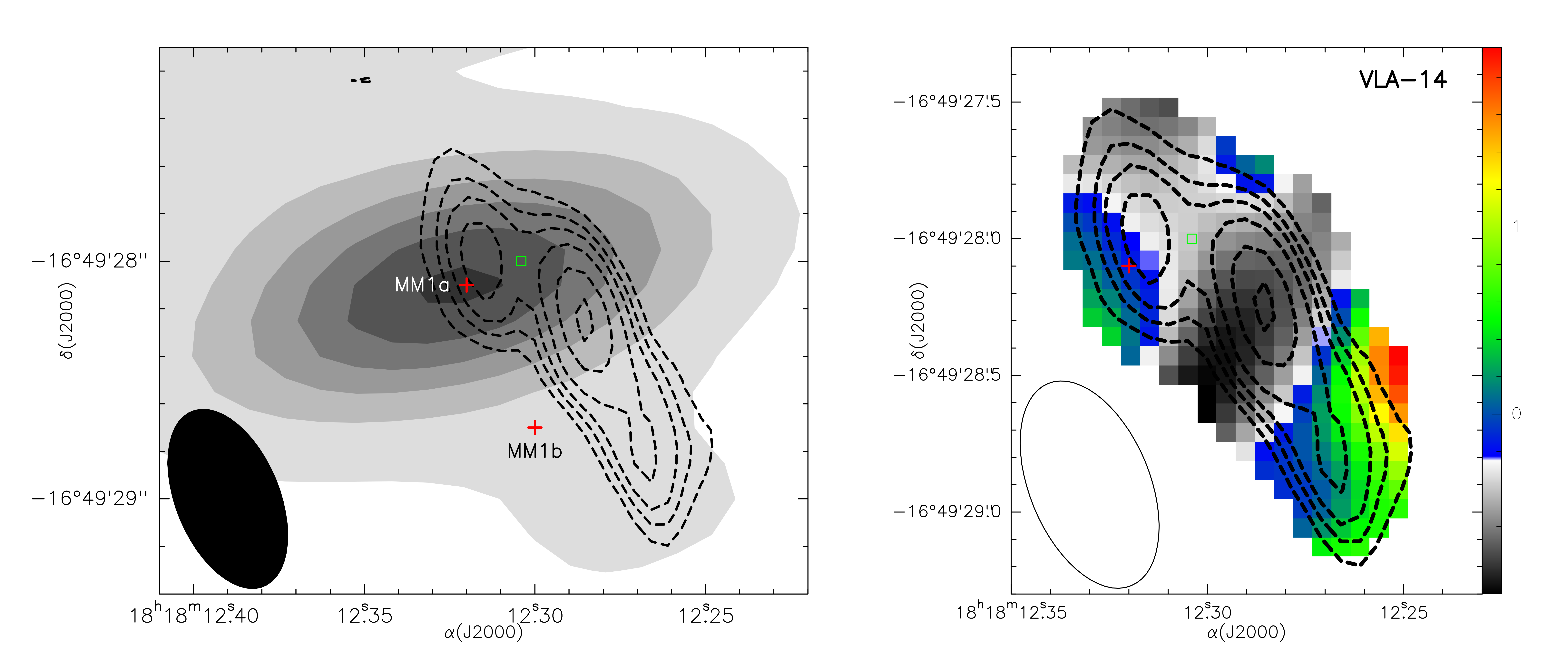}
    \caption{VLA 3.6~cm emission (contours) towards VLA-14. Left: SMA image (greyscale) at 1.3~mm \citep{busquet2016} of MM1 located in G14.2-hub-N. Right: Spectral index map (colour scale). In both images contour levels are -5, -3, 3, 5, 7, 10, 15 and 20 times the rms of the map (1.5~$\mu$Jy beam\,$^{-1}$). Red crosses show the peak position of the millimeter detections in \citep{busquet2016}. Green square shows the position of the H$_2$O maser \citep{wang2006}. The synthesized beam of the X-band is shown in the bottom-left corner of each image.}
    \label{fig:vla14}
\end{figure*}

VLA-14 is associated with the brightest millimeter dust core MM1 of G14.2-hub-N, with a derived mass $\sim13$~\msun\ \citep{busquet2016}, and with H$_2$O maser emission \citep{wang2006}. In Fig.~\ref{fig:vla14} (left panel) we present the 1.3~mm dust emission observed with the SMA with an angular resolution of $\sim1\farcs5$ overlaid on the 3.6~cm continuum emission. The radio emission of VLA-14 has an elongated morphology, extending $\sim1\farcs8\sim2800$~au, roughly in the northeast to southwest direction, reminiscent of a jet with knots. The emission consists of three peaks, with the northern one coinciding with the dust continuum peak (MM1a). Figure~\ref{fig:vla14} (right panel) shows the spectral index map obtained from the 6~cm and 3.6~cm images produced with a common $(u,v)$ range (see Sect.~\ref{s:obs}). There are significant changes in the spectral index along the radio source. While the southern knot shows a clear positive spectral index ($\alpha\sim+0.3$), suggesting partially optically thick free-free emission from thermal radio jets, the central and northern knots are clearly associated with non-thermal emission, with an spectral index around $-0.7<\alpha<-0.3$. This behavior, with the driving source having non-thermal emission, is slightly different to what has been found in well-studied radio jets, with the radio emission from the jet driving source associated with thermal free-free emission and the jet knots presenting negative spectral indices \citep{carrasco-gonzalez2010,rodriguez-kamenetzky2016,rodriguez-kamenetzk2017}. Nevertheless, we suggest that VLA-14 is tracing a collimated radio jet powered by MM1a with the central and northeast radio knots associated with synchrotron emission produced by strong shocks within the jet.

\subsection{VLA-19}

\begin{figure}[ht!]
    \centering
    \includegraphics[width=0.9\linewidth]{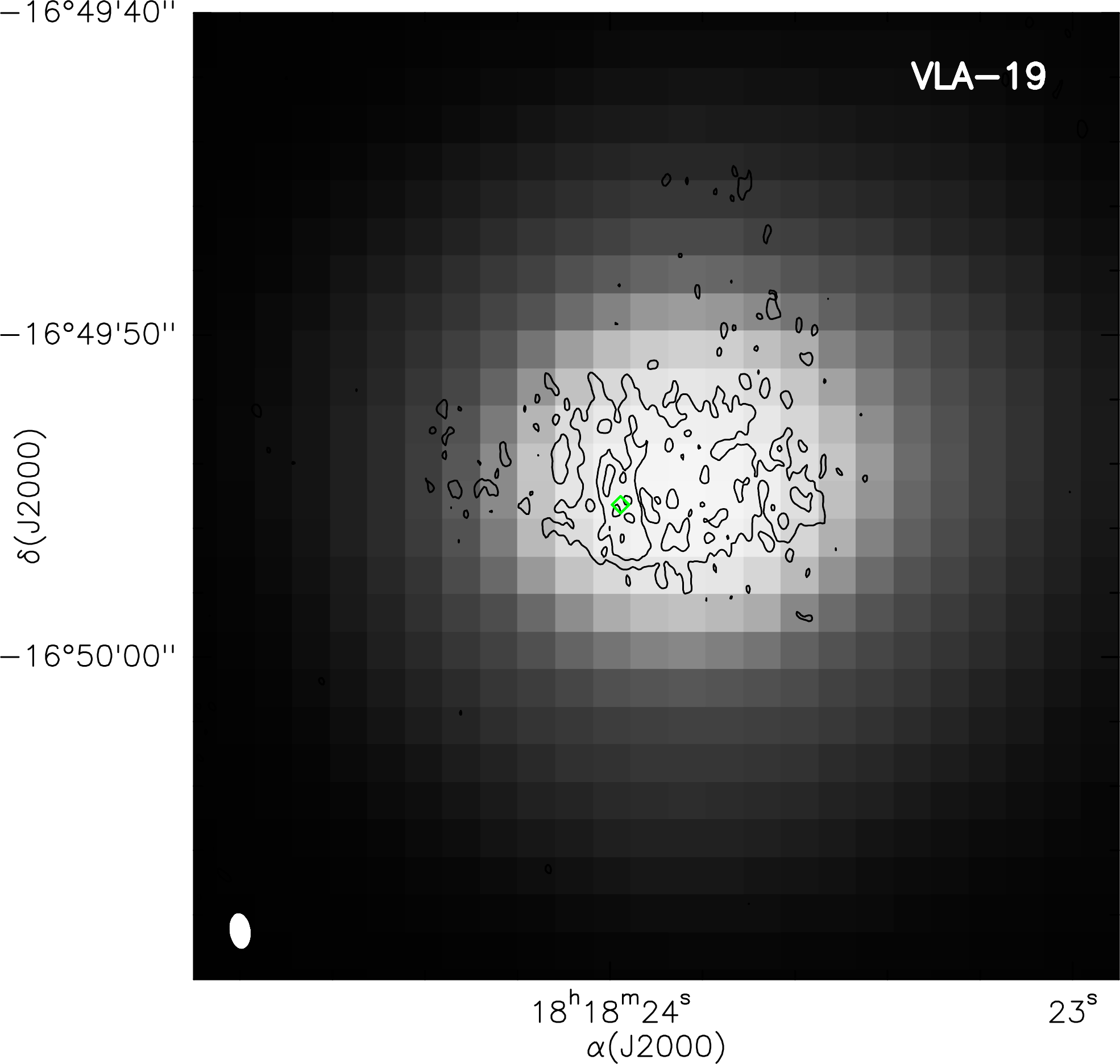}
    \caption{MIPSGAL image \citep{carey2009} at 24~\mum\ of VLA-19 overlaid on the 6~cm emission (C-band). Contour levels start at 3$\sigma$ and increase in steps of 4$\sigma$, where $\sigma$ is the rms of the map (2.2 $\mu$Jy\,beam$^{-1}$).  Green diamond show the peak position of the centimeter emission detected in this work. The synthesized beam of the C-band is shown in the bottom-left corner of the image.}
    \label{fig:vla19}
\end{figure}

VLA-19 is the most extended source detected in this work ($\sim1100$~mas, or $\approx1700$~au). It is associated with bright, extended 24 $\mu$m nebulosity (see Fig.~\ref{fig:vla19}) and also it has 8~\mum\ nebulosity (see Fig. \ref{fig:radiosources}). Interestingly, this is not in association with an IR point source but diffuse MIR emission. Unlike the ring-shaped \hii\ region bubble nearby associated with the IRAS\,18153$-$1651 source in the northern region, this seems to be a blob of MIR nebulosity. The same source has also been detected at 2~$\mu$m with 2MASS \citep{Cutri2003}, at 18 $\mu$m with AKARI \citep{akari2010}, and at 3.4, 4.6, 12, and 22 $\mu$m with WISE \citep{Cutri2013}. Normally, a combination of near and mid-infrared dust with radio emission implies a compact \hii\ region but instead, VLA-19 presents a negative spectral index consistent with non-thermal emission. We note that some \hii\ regions have been found to be associated with non-thermal emission and negative spectral indices \citep[e.g][]{meng2019,padovani2019}. However, we can not exclude that the large negative spectral index determined for VLA-19 might be due to interferometer filtering of a large source size. Further observations at other frequencies and sensitive to larger angular sizes are necessary to better understand this puzzling source.

\section{Additional images}\label{app_figures}

\begin{figure}[t]
\begin{center}
\begin{tabular}[!ht]{c} 
\includegraphics[width=0.8\linewidth]{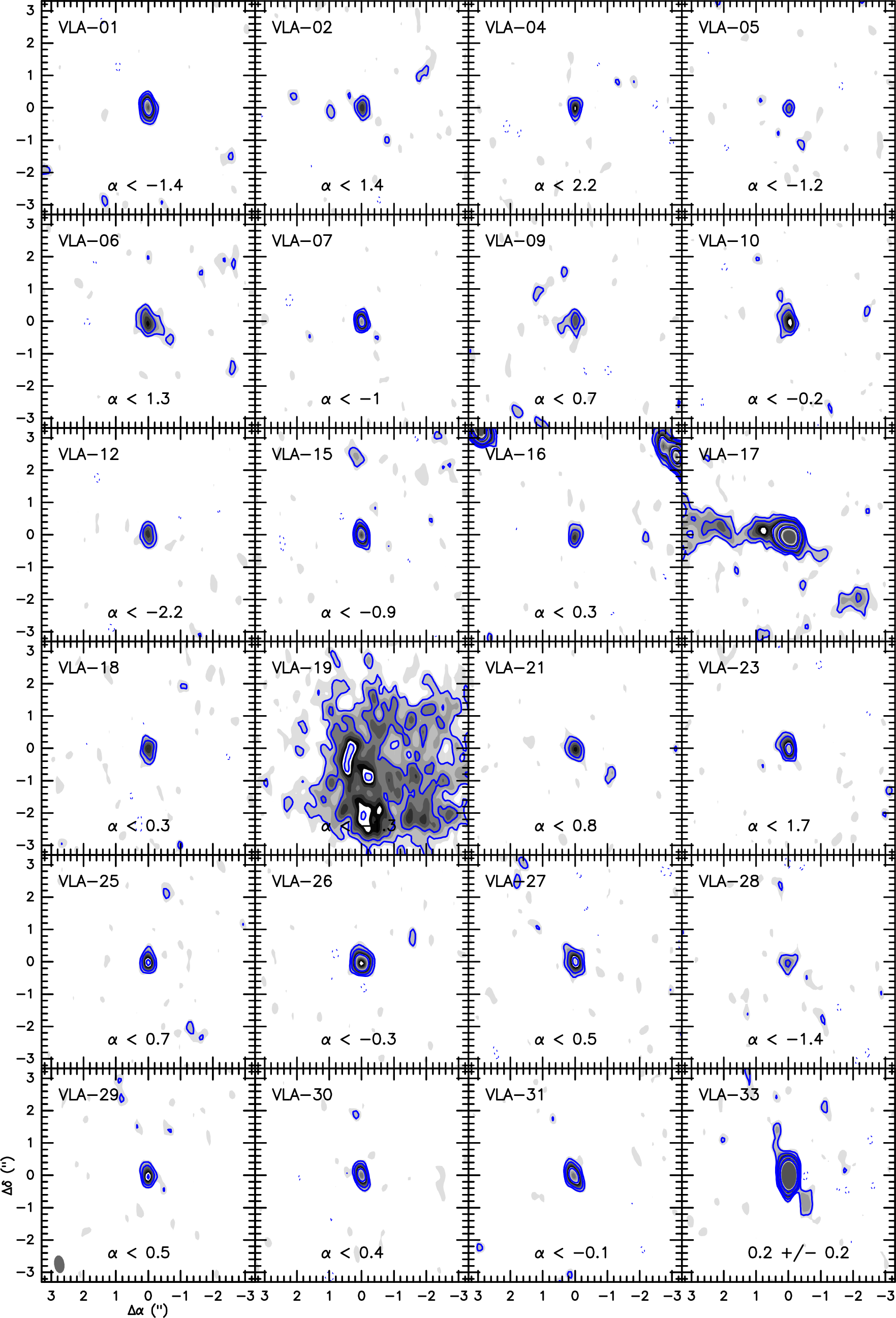} 
\end{tabular}
\caption{VLA continuum images at C-band (blue contours and grey image) of the sources detected in G14.2-N only at the C-band. Contour levels are $\pm$5, $\pm$3, 10, and 20 times the rms of the maps (2.2 ~$\mu$Jy\,beam$^{-1}$). The synthesized beam of the C-band is shown in the bottom-left corner of the bottom-left panel.}
\label{fig:g14N-Cband}
\end{center}
\end{figure}

\begin{figure}[t]
\begin{center}
\begin{tabular}[!ht]{c} 
\includegraphics[width=0.3\linewidth]{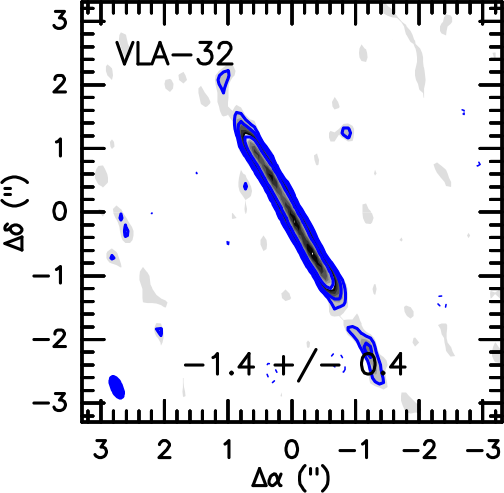}
\end{tabular}
\caption{VLA continuum image at X-band (blue contours and grey image) of the source detected in G14.2-N only at the X-band. Contour levels are $\pm$5, $\pm$3, 10, and 20 times the rms of the maps (1.5 ~$\mu$Jy\,beam$^{-1}$). The synthesized beam of the X-band image is shown in the bottom-left corner of the panel.}
\label{fig:g14N-Xband}
\end{center}
\end{figure}

\begin{figure}[t]
\begin{center}
\begin{tabular}[!ht]{c} 
\includegraphics[width=0.9\linewidth]{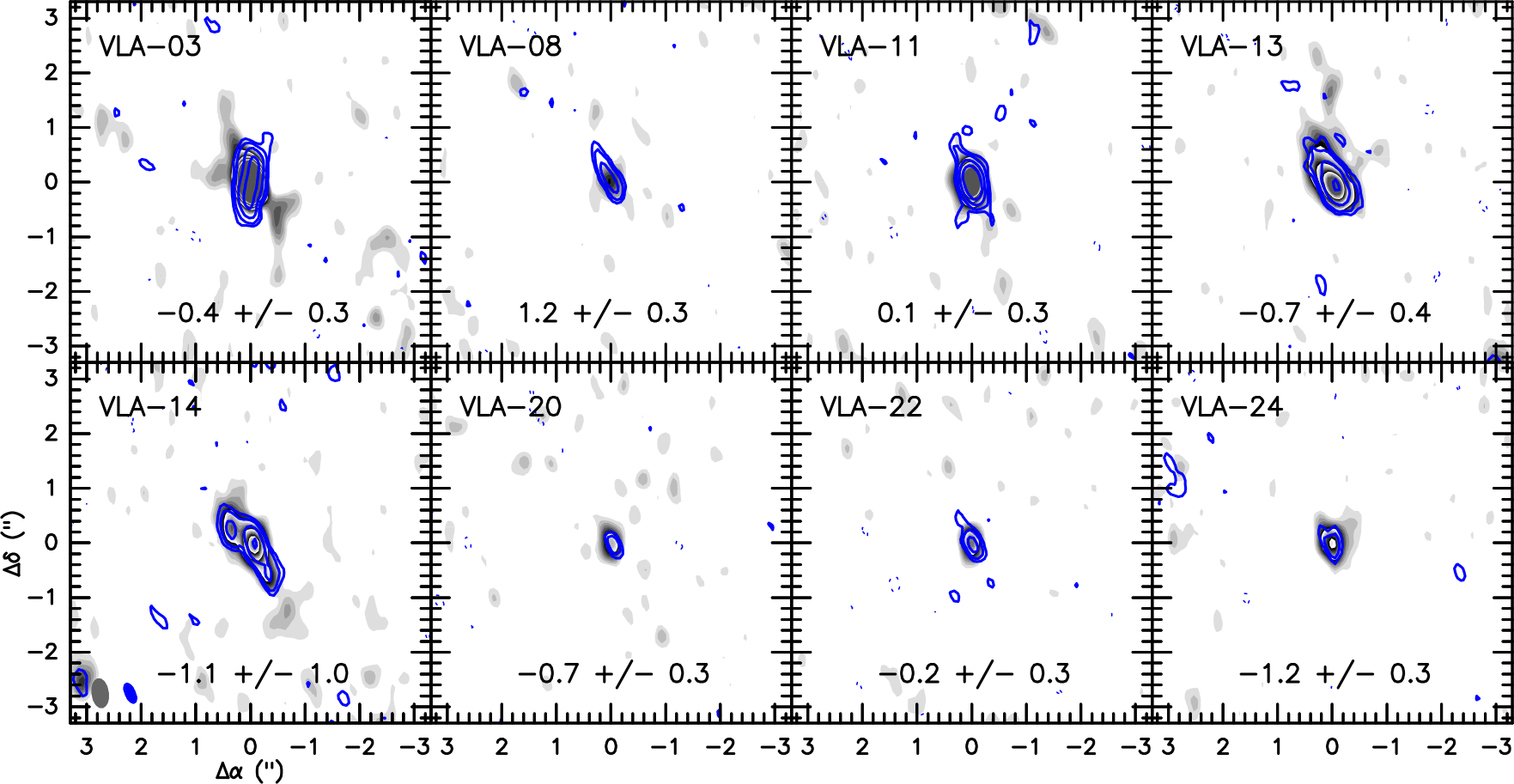}
\end{tabular}
\caption{VLA continuum images at C-band (grey image) and X-band (blue contours) of the sources detected at both frequency bands in G14.2-N. Contour levels of the grey image range from 2 to 30 times the rms of the map (2.2 ~$\mu$Jy\,beam$^{-1}$). Blue contour are $\pm$5, $\pm$3, 10, and 20 times the rms of the map (1.5 ~$\mu$Jy\,beam$^{-1}$). The synthesized beams of the two bands (grey and blue for C- and X-
band) are shown in the bottom-left corner of the bottom-left panel.}
\label{fig:g14N-CXbands}
\end{center}
\end{figure}

\begin{figure}[t]
\begin{center}
\begin{tabular}[!ht]{c} 
\includegraphics[width=0.9\linewidth]{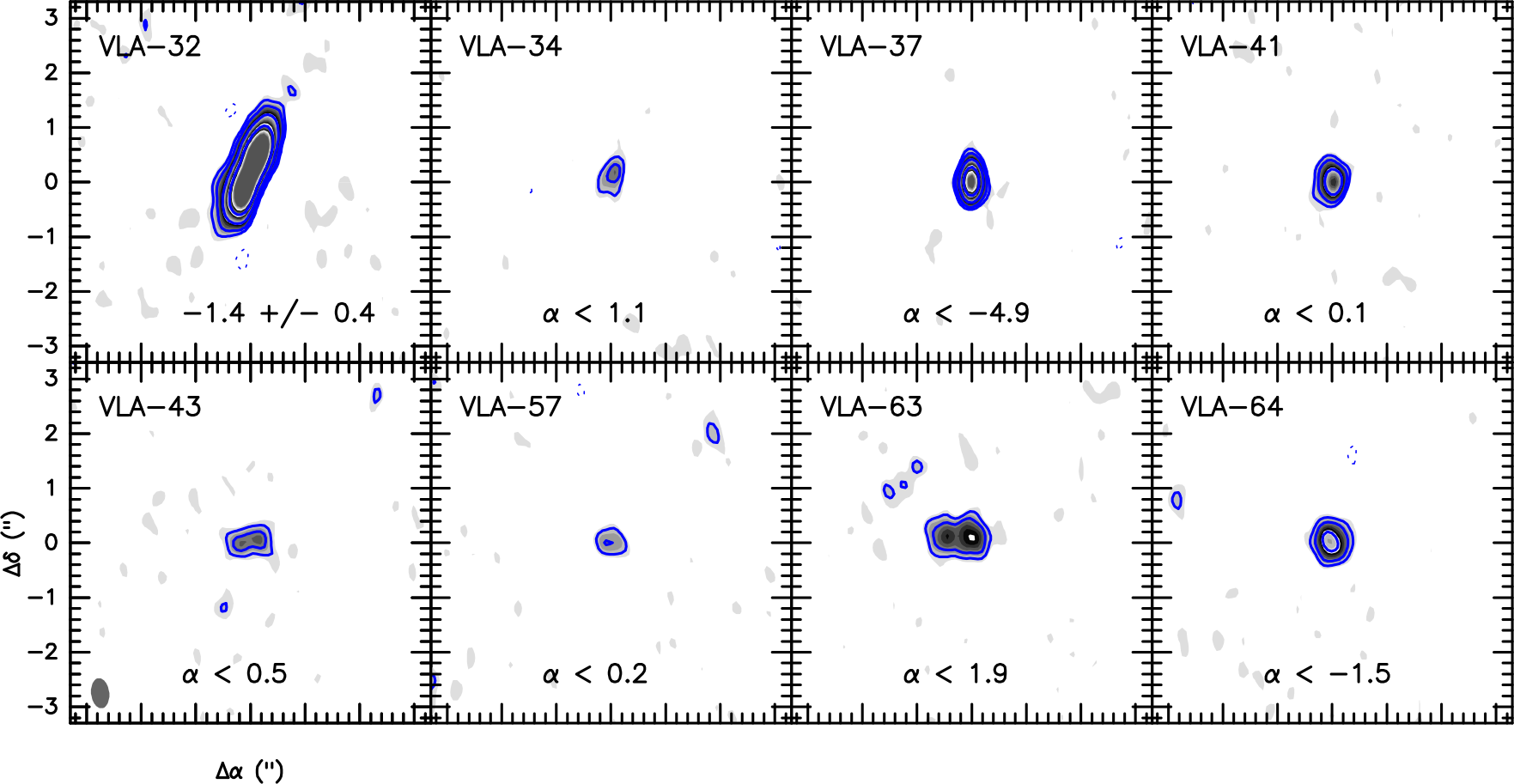}
\end{tabular}
\caption{VLA continuum images at C-band (blue contours and grey image) of the sources detected in G14.2-S only at the C-band. Contour levels are $\pm$5, $\pm$3, 10, and 20 times the rms of the map (2.9 ~$\mu$Jy\,beam$^{-1}$). The synthesized beam of the C-band is shown in the bottom-left corner of the bottom-left panel.}
\label{fig:g14S-Cband}
\end{center}
\end{figure}

\begin{figure}[t]
\begin{center}
\begin{tabular}[!ht]{c} 
\includegraphics[width=0.9\linewidth]{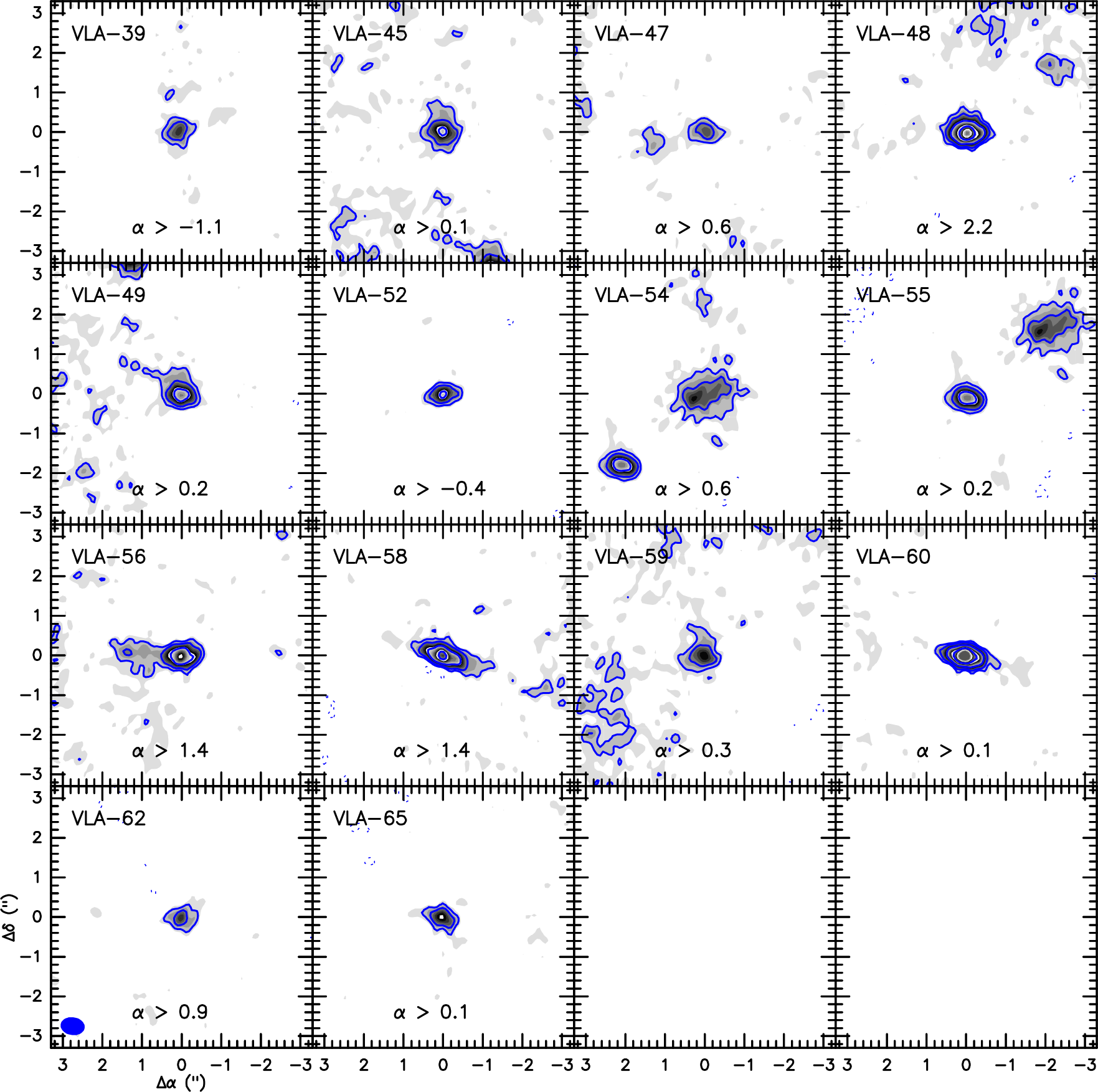} \\ \\
\end{tabular}
\caption{VLA continuum images at X-band (blue contours and grey image) of the sources detected in G14.2-S only at the X-band. Contour levels are $\pm$5, $\pm$3, 10, and 20 times the rms of the map (1.4 ~$\mu$Jy\,beam$^{-1}$). The synthesized beam of the X-band is shown in the bottom-left corner of of the bottom-left panel.}
\label{fig:g14S-Xband}
\end{center}
\end{figure}

\begin{figure}[t]
\begin{center}
\begin{tabular}[!ht]{c} 
\includegraphics[width=0.9\linewidth]{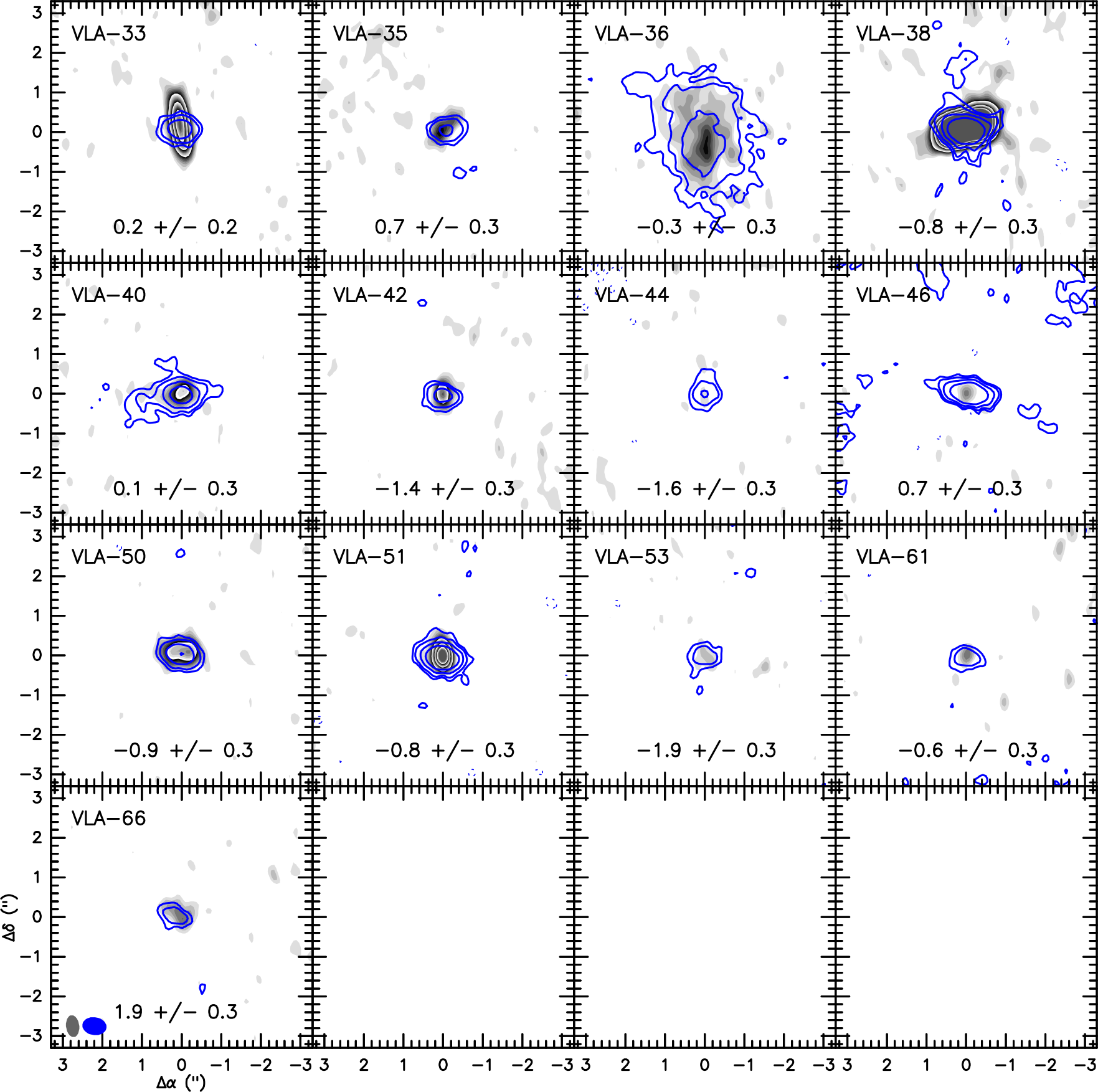}
\end{tabular}
\caption{VLA continuum images at C-band (grey image) and X-band (blue contours) of the sources detected at both frequency bands in G14.2-S. Contour levels of the grey image range from 2 to 30 times the rms of the map (2.9 ~$\mu$Jy\,beam$^{-1}$). Blue contours are $\pm$5, $\pm$3, 10, and 20 times the rms of the map (1.4 ~$\mu$Jy\,beam$^{-1}$). The synthesized beams of the two bands (grey and blue for C- and X-
band) are shown in the bottom-left corner of the bottom-left panel}
\label{fig:g14S-CXbands}
\end{center}
\end{figure}

In this appendix we present the individual images of the radio sources detected in the two fields G14.2-N and G14.2-S using the common $(u,v)$ range. For G14.2-N, in Fig.~\ref{fig:g14N-Cband} we show all sources detected only at C-band and in Fig.~\ref{fig:g14N-Xband} we show the sources detected only at X-band. Figure~\ref{fig:g14N-CXbands} presents all sources detected at both C- and X-bands. For G14.2-S, Fig.~\ref{fig:g14S-Cband} presents all centimeter sources detected only at C-band and in  Fig.~\ref{fig:g14S-Xband} we show all centimeter sources detected only at X-band. In Fig.~\ref{fig:g14S-CXbands} we present all sources detected at both C- and X-band. All images also show the spectral index value of each source, or the upper or lower limits obtained.

\onecolumn

\end{appendix}

\end{document}